\RequirePackage{fix-cm}
\documentclass{svjour3}                     
\smartqed  

\usepackage{amssymb}
\usepackage{amsmath}

\usepackage{graphicx}
\usepackage{epstopdf}
\usepackage{subfigure}
\usepackage{hyperref}
\usepackage{xcolor}
\usepackage{caption}
\usepackage{float}
\begin{document}

\title{Topology optimization of nonlinear forced response curves via reduction on spectral submanifolds}


\author{Hongming Liang \and
        Matteo Pozzi \and
        Jacopo Marconi \and
        Shobhit Jain \and
        Mingwu Li 
}


\institute{H. Liang\at
              Department of Mechanics and Aerospace Engineering, Southern University of Science and Technology, Shenzhen, 518055, China   
           \and
           M. Pozzi \and J Marconi \at
           Department of Mechanical Engineering, Politecnico di Milano, Milan, 20156, Italy
           \and
           S. Jain \at
              Delft Institute of Applied Mathematics, TU Delft, Mekelweg 4, 2628CD, Delft, The Netherlands
          \and
          M. Li\at
              Department of Mechanics and Aerospace Engineering, Southern University of Science and Technology, Shenzhen, 518055, China \\
              State Key Laboratory of Structural Analysis, Optimization and CAE Software for Industrial Equipment, Dalian University of Technology, Dalian 116024, China\\
              \email{limw@sustech.edu.cn (M. Li)}   
}

\date{Received: date / Accepted: date}

\maketitle

\begin{abstract}
Forced response curves (FRCs) of nonlinear systems can exhibit complex behaviors, including hardening/softening behavior and bifurcations. Although topology optimization holds great potential for tuning these nonlinear dynamic responses, its use in high-dimensional systems is limited by the high cost of repeated response and sensitivity analyses. To address this challenge, we employ the spectral submanifolds (SSMs) reduction theory, which reformulates the periodic response as the equilibria of an associated reduced-order model (ROM). This enables efficient and analytic evaluation of both response amplitudes and their sensitivities. Based on the SSM-based ROM, we formulate optimization problems that optimize the peak amplitude, the hardening/softening behavior, and the distance between two saddle-node bifurcations for an FRC. The proposed method is applied to the design of nonlinear MEMS devices, achieving targeted performance optimization. This framework provides a practical and efficient strategy for incorporating nonlinear dynamic effects into the topology optimization of structures.
\end{abstract}

\section{Introduction}
\label{section1}

Nonlinear dynamic responses are commonly observed in micro-electro-mechanical systems (MEMS). On the one hand, such responses may cause the systems to fail to work properly. For example, in some high-precision sensors, such as gyroscopes~\cite{su_characteristics_2020,tiwari_using_2019}, pressure sensors~\cite{LI201219}, nonlinear responses reduce the accuracy of the sensor and even cause malfunctions~\cite{s20247207}. On the other hand, nonlinearities can also be utilized to provide advantages over linear designs. For example, nonlinear dynamic responses can be used to improve the performance of energy harvesters~\cite{chen2015internal,zhang2024internal,amin_karami_powering_2012}, and their characteristic hardening or softening behavior can be utilized in the design of filters~\cite{hajjaj_mode_2017} and binary counters~\cite{yao_counter_2013}. Thus, it is essential to tailor the dynamic responses of nonlinear systems if one aims to suppress the amplitude of structural vibration or to utilize them to optimally design devices.

To utilize nonlinear dynamic responses, several studies have focused on tailoring the hardening and softening behavior of nonlinear systems. One approach involves adjusting the hardening behavior of clamped-clamped beams by maximizing or minimizing the coefficient of cubic nonlinearity in the normal form~\cite{dou_structural_2015}. Another method optimizes both the Q-factor and Duffing nonlinearity in finite element-based reduced-order models using derivative-free optimization techniques~\cite{li_finite_2025}. Additionally, backbone curves have been optimized on a reduced-order model (ROM) constructed via Lyapunov subcenter manifolds~\cite{pozzi_topology_2025,pozzi_backbone_2024} and spectral submanifolds (SSMs)~\cite{pozzi2025adjointsensitivitiesoptimizationnonlinear}. 

Tailoring the hardening and softening behavior of a nonlinear system typically involves analyzing its autonomous dynamics and does not require consideration of external forcing. However, if the objective is to suppress or enhance the vibration amplitude under harmonic excitation, it is necessary to tailor the forced response curve (FRC). For low-dimensional nonlinear systems, the FRC of nonlinear systems can be computed by the harmonic balance method~\cite{von_groll_harmonic_2001,krack_application_2019,detroux_harmonic_2015} combined with continuation techniques~\cite {ascher1995numerical,dankowicz2013recipes}. Based on these techniques, \cite{dou_optimization_2015} applied shape optimization to a clamped-clamped beam to minimize or maximize the amplitude of primary resonance. For high-dimensional nonlinear systems, such as finite element models (FEM) in topology optimization~\cite{bendsoe2013topology}, computing and tailoring the FRC becomes challenging due to the increased computational cost. \textcolor{blue}{In previous work, the moving morphable components (MMC) optimization framework has been employed to reduce the level of nonlinear vibrations in under-platform dampers~\cite{denimal_topological_2021}.}

In addition to amplitude control, tailoring the bifurcations of the FRC is also essential to nonlinear systems. Under harmonic excitation, nonlinear systems often exhibit saddle-node (SN) bifurcations on the FRC as the forcing amplitude increases. These bifurcations lead to abrupt jumps in the steady-state response, known as jump or hysteresis phenomena, which are particularly detrimental to devices intended to operate within a linear or predictable regime. Therefore, suppressing the occurrence of SN bifurcations becomes critical. In previous work, shape optimization along with numerical continuation is used to tailor the distance of two SN bifurcations on the nonlinear FRC~\cite{saghafi_nonlinear_2015}. \textcolor{blue}{Multi-parametric continuation methods have also been used to track SN bifurcations with respect to multiple system parameters, enabling the identification of parameter ranges in which isolated solutions are suppressed~\cite{grenat_multi-parametric_2019}. In addition, multi-parametric optimization frameworks have been proposed to control bifurcation behavior by regulating both the existence and the locations of different bifurcation types, including SN, branch-point, Neimark--Sacker, and period-doubling bifurcations~\cite{melot_multi-parametric_2024,melot_control_2024}.}

\textcolor{blue}{Internal resonance is another characteristic phenomenon of nonlinear systems and can lead to complex dynamic behaviors~\cite{liang_bifurcation_2025}. Topology optimization has been employed to tailor the eigenfrequencies of structures in order to avoid internal resonances~\cite{pedersen_designing_2005,sun_topology_2020}. In addition, shape optimization has been used to regulate energy transfer between modes involved in internal resonance, such as the 1:2 resonance case~\cite{dou_structural_2015}. For systems exhibiting 1:3 internal resonance, a synthesis approach has been proposed to suppress modal interactions by translating the associated nonlinear normal mode branch toward lower energy levels~\cite{detroux_tailoring_2021}}.

Here, we aim to use topology optimization to tune the FRC of high-dimensional nonlinear systems, including minimizing or maximizing the peak of FRC, manipulating the hardening/softening behavior, and suppressing the occurrence of SN bifurcations. \textcolor{blue}{In this study, internal resonance effects are not considered. Extending the proposed topology optimization framework to account for internal resonance phenomena constitutes an interesting direction for future research.} As a method to design continuum structures~\cite{jung_topology_2004}, topology optimization can produce more refined structures than shape optimization. However, the number of degrees of freedom (DOFs) of the FEM and the associated design variables in topology optimization is considerable. 

To address the challenge of high-dimensionality, reduction on SSMs has been used to analytically extract the FRC of FEM with high DOFs~\cite{haller_nonlinear_2016,howJain,haller_modeling_2025,kaszas_globalizing_2025}. In particular, SSM reduction transforms forced periodic orbits into fixed points of the associated reduced-order model (ROMs).~\textcolor{blue}{Consequently, one can obtain orders-of-magnitude speed-up gain from SSM reduction in comparison with collocation, harmonic balance, and shooting methods applied to full model~\cite{howJain,li_nonlinear_2022}.} In addition, the sensitivity of coefficients in SSM-based ROMs that govern the peak of FRC and hardening/softening behavior has been discussed in~\cite{li_explicit_2024}. These sensitivities and their efficient computation~\cite{pozzi_topology_2025} are essential in the iteration process of topology optimization. Based on these contributions, we will perform topology optimization of FRCs.


In this work, we adopt a density-based topology optimization algorithm to determine the optimal material layout. Specifically, the Solid Isotropic Material with Penalization (SIMP) scheme~\cite{bendsoe_material_1999} is used to interpolate material properties and promote 0–1 designs. To enhance numerical stability and obtain clear structural boundaries, both density filtering and density projection techniques are employed~\cite{wang_projection_2011}. The design variables are updated using the Method of Moving Asymptotes (MMA)~\cite{svanberg_method_1987}. All these techniques follow the implementation described in~\cite{pozzi_topology_2025}.
 
The rest of this paper is organized as follows. In Sec.~\ref{section2}, we introduce the SSM theory and the problem description about minimizing/maximizing the peak of FRC, tailoring the hardening/softening behavior, and suppressing the occurrence of SN bifurcations. In Sec.~\ref{section3}, we derive the sensitivity of coefficients used in the iteration process of topology optimization. Three numerical examples are presented in Sec.~\ref{section4}. Finally, the conclusions are presented in Sec.~\ref{section5}.

\section{Problem formulation}
\label{section2}

\subsection{Setup}

We consider a periodically forced nonlinear mechanical system
\begin{equation}
\label{eq:eom-second-full}
\boldsymbol{M}(\boldsymbol{\mu})\ddot{\boldsymbol{x}}+\boldsymbol{C}(\boldsymbol{\mu})\dot{\boldsymbol{x}}+\boldsymbol{K}(\boldsymbol{\mu})\boldsymbol{x}+\boldsymbol{f}_2(\boldsymbol{\mu},\boldsymbol{x},\boldsymbol{x})+\boldsymbol{f}_3(\boldsymbol{\mu},\boldsymbol{x},\boldsymbol{x},\boldsymbol{x})=\epsilon \boldsymbol{f}^{\mathrm{ext}}(\Omega t),\quad 0<\epsilon\ll1
\end{equation}
where $\boldsymbol{x}\in\mathbb{R}^n$ is the displacement vector; $\boldsymbol{\mu}\in\mathbb{R}^q$ is the design variable representing the spatial distribution of material density in the structure; $\boldsymbol{M},\boldsymbol{C},\boldsymbol{K}\in\mathbb{R}^{n\times n}$ are the mass, damping and stiffness matrices; $\boldsymbol{f}_2(\boldsymbol{x},\boldsymbol{x})$ and $\boldsymbol{f}_3(\boldsymbol{x},\boldsymbol{x},\boldsymbol{x})$ are internal nonlinear elastic forces arising from geometric nonlinearities; and $\epsilon \boldsymbol{f}^{\mathrm{ext}}(\Omega t)$ denotes external periodic forcing with small amplitude $\epsilon\ll1$. 

Solving the linear part of~\eqref{eq:eom-second-full}, we have the following generalized eigenvalue problem
\begin{equation}
	\label{eq:eom-second-eig}
	(\lambda_j^2\boldsymbol{M}+\lambda_j\boldsymbol{C}+\boldsymbol{K})\boldsymbol{\phi}_j=\boldsymbol{0},\quad (\lambda_j^2\boldsymbol{M}+\lambda_j\boldsymbol{C}+\boldsymbol{K})^{\mathrm{T}}\boldsymbol{\psi}_j=\boldsymbol{0}
\end{equation}
where $\lambda_j$ is an eigenvalue, and $\boldsymbol{\phi}_j$ and $\boldsymbol{\psi}_j$ are associated right and left eigenvectors. In this work, we do not allow for internal resonance $(\lambda_i \approx k\lambda_j,\  k \in \mathbb{Z}^+)$ such that the high-dimensional system~\eqref{eq:eom-second-full} can be reduced on a two-dimensional spectral submanifold (SSM). When internal resonance occurs during the optimization iteration process, additional constraints are introduced to eliminate the internal resonance. This will be discussed in Sec.~\ref{ssec:min-FRC}.

In this work, we consider positive definite mass and stiffness matrices as well as Rayleigh damping, namely,
\begin{equation}
	\label{eq:Rayleigh damping}
	\boldsymbol{K} \boldsymbol{\phi}_j = \omega_j^2 \boldsymbol{M} \boldsymbol{\phi}_j, \ \boldsymbol{\phi}_j^\mathrm{T}\boldsymbol{M}\boldsymbol{\phi}_j = 1, \
	\boldsymbol{C} =\alpha\boldsymbol{M} + \beta \boldsymbol{K},
\end{equation}
where $\omega_j$ is natural frequency, $\alpha$ and $\beta$ are the stiffness-proportional and  mass-proportional damping constants. The values of $\alpha$ and $\beta$ are computed by $\boldsymbol{\phi}_j^\mathrm{T}\boldsymbol{C}\boldsymbol{\phi}_j =\alpha + \beta \omega_j^2 =  2 \xi \omega_j$, where $\xi$ is the damping ratio of the system. Further, taking $j=1,2$, yields
\begin{equation}
	\label{eq:Rayleigh damping coefficient}
	\alpha = \xi \frac{2\omega_1 \omega_2}{\omega_1 + \omega_2}, \quad \beta = \xi \frac{2}{\omega_1 + \omega_2}.
\end{equation}
\textcolor{blue}{Substitution of $\boldsymbol{\phi}_1^\mathrm{T}\boldsymbol{C}\boldsymbol{\phi}_1 =  2 \xi \omega_1$ and Eq.~(3) into Eq.~(2) yields
\begin{equation}
\label{eq:lambda1}
\lambda_1^2 + 2\xi\omega_1\lambda_1 + \omega_1^2 = 0,   \implies  \lambda_1 = -\xi\,\omega_1 + \mathrm{i}\,\sqrt{1-\xi^2}\omega_1.
\end{equation}
}

\subsection{Reduction on spectral submanifolds}
\label{ssec:reduc_on_SSM}

Computing and tuning the FRC of the mechanical system~\eqref{eq:eom-second-full} is challenging for $n \gg 1$. To address this challenge, we briefly introduce SSMs reduction theory in this subsection, which transforms the full system~\eqref{eq:eom-second-full} into an SSM-based ROM.

The second-order system~\eqref{eq:eom-second-full} can be transformed into a first-order system as below
\begin{equation}
	\label{eq:full-first}
	\boldsymbol{B}\dot{\boldsymbol{z}}	=\boldsymbol{A}\boldsymbol{z}+\boldsymbol{F}(\boldsymbol{z})+\epsilon\boldsymbol{F}^{\mathrm{ext}}({\Omega t}),
\end{equation}
where
\begin{gather}
	\label{eq:zABF}
	\boldsymbol{z}=\begin{pmatrix}\boldsymbol{x}\\\dot{\boldsymbol{x}}\end{pmatrix},\,\,
	\boldsymbol{A}=\begin{pmatrix}-\boldsymbol{K} 
		& \boldsymbol{0}\\\boldsymbol{0} & \boldsymbol{M}\end{pmatrix},\,\,
	\boldsymbol{B}=\begin{pmatrix}\boldsymbol{C} 
		& \boldsymbol{M}\\\boldsymbol{M} & \boldsymbol{0}\end{pmatrix},\nonumber\\
	\boldsymbol{F}(\boldsymbol{z})=\begin{pmatrix}{-\boldsymbol{f}_2(\boldsymbol{x},\boldsymbol{x})-\boldsymbol{f}_3(\boldsymbol{x},\boldsymbol{x},\boldsymbol{x})}\\\boldsymbol{0}\end{pmatrix},\,\,
	\boldsymbol{F}^{\mathrm{ext}}(\Omega t) = \begin{pmatrix}\boldsymbol{f}^{\mathrm{ext}}(\Omega t)\\\boldsymbol{0}\end{pmatrix}.
\end{gather}
The associated generalized eigenvalue problem becomes 
\begin{equation}
	\boldsymbol{A}\boldsymbol{v}_j=\lambda_j\boldsymbol{B}\boldsymbol{v}_j,\quad \boldsymbol{u}_j^\ast \boldsymbol{A}=\lambda_j \boldsymbol{u}_j^\ast \boldsymbol{B},
\end{equation}
where $\lambda_j$ is a generalized eigenvalue and $\boldsymbol{v}_j$ and $\boldsymbol{u}_j$ are the corresponding \emph{right} and \emph{left} eigenvectors, respectively. In particular, we have
\begin{equation}
	\boldsymbol{v}_j=\begin{pmatrix}\boldsymbol{\phi}_j \\ \lambda_j\boldsymbol{\phi}_j\end{pmatrix},\quad \boldsymbol{u}_j=\begin{pmatrix}\bar{\boldsymbol{\psi}}_j \\ \bar{\lambda}_j \bar{\boldsymbol{\psi}}_j\end{pmatrix}.
\end{equation}

In this study, we construct the two-dimensional SSM associated with the master subspace $\mathcal{E}=\mathrm{span}(\boldsymbol{v}_1,\bar{\boldsymbol{v}}_1)$ for model reduction. Specifically, let $\boldsymbol{p} = (p, \bar{p})$ denote the reduced coordinates. We seek the autonomous SSM map $\boldsymbol{z} = \boldsymbol{W}(\boldsymbol{p})$ and its associated reduced dynamics $\dot{\boldsymbol{p}} = \boldsymbol{R}(\boldsymbol{p})$ at $\epsilon = 0$. When external periodic forcing $\epsilon \boldsymbol{f}^{\mathrm{ext}}(\Omega t)$ is considered, the SSM becomes time-dependent, yielding a periodic SSM map $\boldsymbol{W}_\epsilon(\boldsymbol{p},\Omega t)$ and reduced dynamics $\dot{\boldsymbol{p}}=\boldsymbol{R}_\epsilon(\boldsymbol{p},\Omega t)$. The map and reduced dynamics on the periodic SSM can be expressed as~\cite{li_nonlinear_2022,li_explicit_2024}
\begin{equation}
	\begin{aligned}
		\label{eq:red-reps}
		\boldsymbol{W}_\epsilon(\boldsymbol{p},\Omega t)&=\boldsymbol{W}(\boldsymbol{p})+\epsilon\boldsymbol{X}(\boldsymbol{p},\Omega t)+\mathcal{O}(\epsilon^2), \\
		\dot{\boldsymbol{p}}=\boldsymbol{R}_\epsilon(\boldsymbol{p},\Omega t)&=\boldsymbol{R}(\boldsymbol{p})+\epsilon\boldsymbol{S}(\boldsymbol{p},\Omega t)+\mathcal{O}(\epsilon^2).
	\end{aligned}
\end{equation}  

Plugging~\eqref{eq:red-reps} into~\eqref{eq:full-first}, we obtain the invariant equation as
\begin{equation}
	\begin{aligned}
		\boldsymbol{B}\partial_{\boldsymbol{p}} \boldsymbol{W}_\epsilon \boldsymbol{R}_\epsilon 
	+ \partial_{\Omega t} \boldsymbol{W}_\epsilon \Omega 
	= A \boldsymbol{W}_\epsilon 
	+ \boldsymbol{F}(\boldsymbol{W}_\epsilon) 
	+ \epsilon \, \boldsymbol{F}^{\mathrm{ext}} \cos \Omega t.
	\label{eq:invar-balance}
	\end{aligned}
\end{equation}
The autonomous SSM map $\boldsymbol{z} = \boldsymbol{W}(\boldsymbol{p})$ and its associated reduced dynamics $\dot{\boldsymbol{p}} = \boldsymbol{R}(\boldsymbol{p})$  can be obtained by solving the invariant equation~\eqref{eq:invar-balance} at $\epsilon = 0$, using a Taylor expansion in $\boldsymbol{p}$ truncated at cubic order, namely
\begin{equation}
	\begin{aligned}
		\boldsymbol{W}(\boldsymbol{p}) = \sum_{1 \leq m+n \leq 3} \boldsymbol{W}_{mn}p^m \bar{p}^n, \\
		\boldsymbol{R}(\boldsymbol{p}) =  \sum_{1 \leq m+n \leq 3} \boldsymbol{R}_{mn}p^m \bar{p}^n.
		\label{eq:Taylor Expansion W and R}
	\end{aligned}
\end{equation}
Cohomological equations can be derived, and the expansion coefficients can be obtained following a normal-form-style parameterization. More details can be found at~\cite{li_explicit_2024}.

The non-autonomous part of the periodic SSM mapping $\boldsymbol{X}(\boldsymbol{p}, \Omega t)$ can be approximated by the leading-order terms as
\begin{equation}
	\begin{aligned}
		\boldsymbol{X}(\boldsymbol{p},\Omega t) = \boldsymbol{x}_0 e^{\mathrm{i} \Omega t} + \bar{\boldsymbol{x}}_0 e^{- \mathrm{i} \Omega t}
	\end{aligned}.
\end{equation}
Accordingly, the non-autonomous part of the reduced dynamics truncated at leading order can be written as
\begin{equation}
    \boldsymbol{S}(\boldsymbol{p},\Omega t)= \boldsymbol{s}_0^+ e^{\mathrm{i} \Omega t} + \boldsymbol{s}_0^- e^{- \mathrm{i} \Omega t}.
\end{equation}
One can again use the invariance equation to solve for these unknown coefficients. Interested readers can refer to~\cite{li_explicit_2024} for more details.

We express the reduced coordinates in polar form as
 \begin{equation}
 	\label{eq:transform}
 	p=\rho e^{\mathrm{i}(\theta+\Omega t)}
 \end{equation}
 and plug it into the reduced dynamics in~\eqref{eq:red-reps}. Then the third-order reduced dynamics can be expressed as~\cite{li_explicit_2024}
\begin{equation}
\begin{aligned}
 	\dot{\rho} &= \mathrm{Re}(\lambda) \rho + \mathrm{Re}(\gamma) \rho^3 + \epsilon ( \mathrm{Re}(\tilde{f}) \cos \theta + \mathrm{Im}(\tilde{f}) \sin \theta), \\
	\dot{\theta} &= \mathrm{Im}(\lambda) - \Omega + \mathrm{Im}(\gamma) \rho^2 + \epsilon ( \mathrm{Im}(\tilde{f}) \cos \theta - \mathrm{Re}(\tilde{f}) \sin \theta)/\rho,
\end{aligned}
\label{eq:reduc-nonauto}
\end{equation}
where $\lambda$ is the eigenvalue of master subspace, i.e.,~\textcolor{blue}{$\lambda = \lambda_1=\textcolor{blue}{-\xi\,\omega_1 + i\,\sqrt{1-\xi^2}\omega_1}$, as seen in~\eqref{eq:lambda1}}; $\gamma$ is the third-order backbone coefficient; $\tilde{f}$ is a modal force. Explicit expressions for $\gamma$ and $\tilde{f}$ are given as
\begin{equation}
		\gamma = -\boldsymbol{\psi}^{\mathrm{T}}\boldsymbol{f}_{21}, \quad 
		\tilde{f} = 0.5\boldsymbol{\psi}^\mathrm{T}\boldsymbol{f}^{\mathrm{ext}},
	\label{eq:compute_gammma_force}
\end{equation}
where $\boldsymbol{f}_{21}$ is associated with cubic nonlinearity, given in the Eq.(19) of~\cite{li_explicit_2024}. It is noted that a fixed point of the ROM~\eqref{eq:reduc-nonauto} corresponds to a forced periodic orbit of the full system~\eqref{eq:eom-second-full}, and the forced response curve (FRC) of the full system can be analytically extracted via the SSM-based ROM~\cite{li_nonlinear_2022,li_explicit_2024}.

\textcolor{blue}{
\begin{remark}
    The reduced-order model in Eq.~\eqref{eq:reduc-nonauto} is based on a third-order SSM truncation and is therefore valid only within a range of forcing amplitudes for which the third-order truncation already converges. In practice, one should check the convergence of the optimal FRC with increasing truncation orders to ensure the reliability of the SSM prediction. We will apply this checking procedure in all examples.
\end{remark}
}
\subsection{Optimization problem statement}

\begin{figure}[!ht]
\centering
\includegraphics[width=0.45\textwidth]{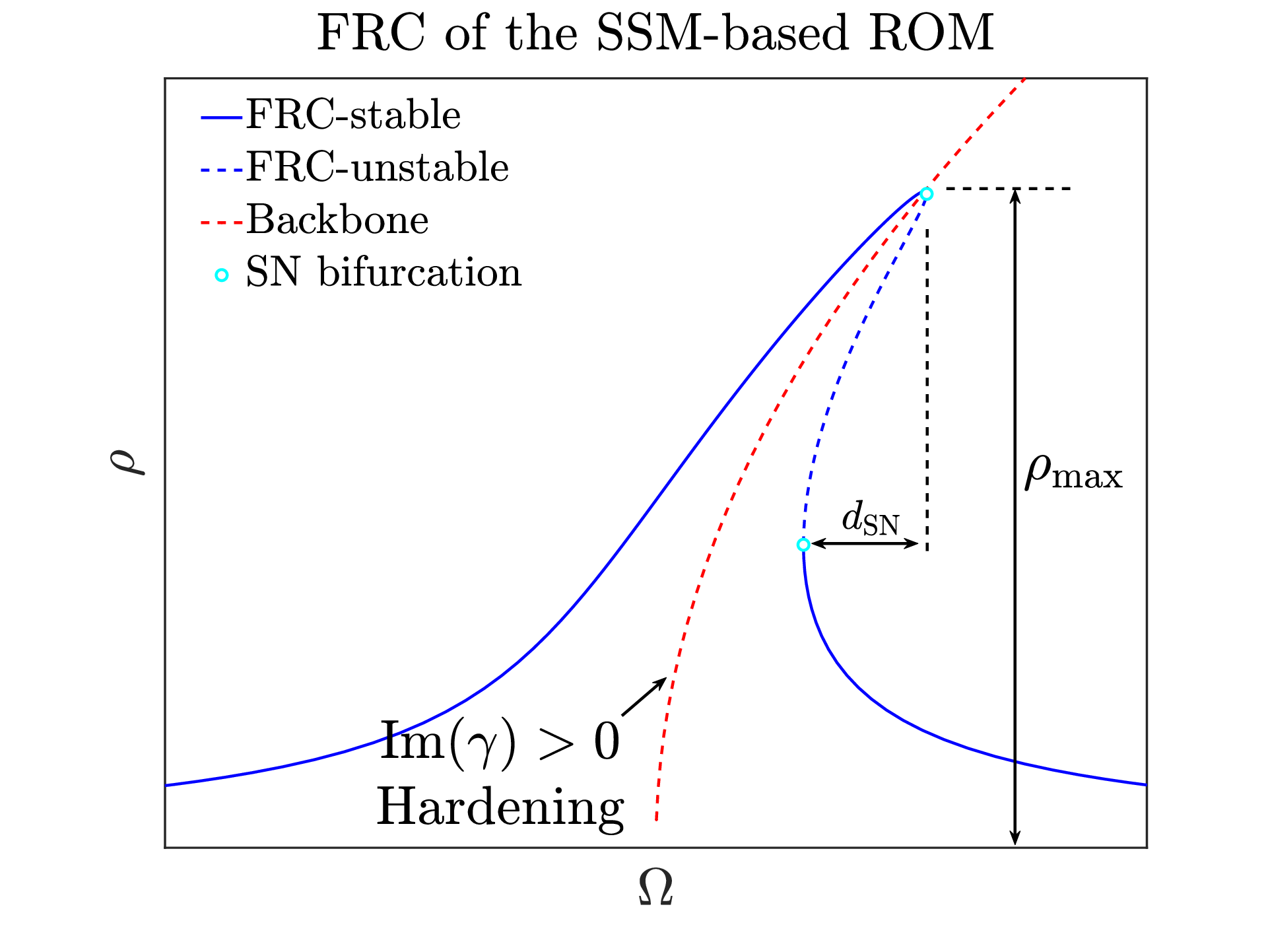}
\caption{\textcolor{blue}{FRC and backbone curve of the SSM-based ROM in Eq.~\eqref{eq:reduce-FRC}. The blue solid line represents the stable segment of the FRC, the blue dashed line represents the unstable segment of the FRC, and the circle on the FRC marks the saddle-node (SN) bifurcation. The associated backbone curve is represented by the red dashed line.}}
\label{fig:FRC-ROM-objf}
\end{figure}

In this paper, we consider a few optimization problems associated to the tuning of the FRC. As we will see, these optimization problems enable the tuning of the peak and the hardening/softening behavior of FRC, and also the SN bifurcations on the FRC \textcolor{blue}{(cf.~Fig.~\ref{fig:FRC-ROM-objf})}.

\subsubsection{Minimize the peak of FRC with target backbone}
\label{ssec:min-FRC}

By using SSM reduction, we transform the optimization of the nonlinear finite element system~\eqref{eq:eom-second-full} into the optimization of a reduced system~\eqref{eq:reduc-nonauto}. In this subsection, we will present the formulation for minimizing the peak of FRC in reduced coordinates.

Since the fixed point of the reduced dynamics~\eqref{eq:reduc-nonauto} corresponds to a periodic orbit of the full system~\cite{li_nonlinear_2022,li_explicit_2024}, setting $(\dot{\rho}, \dot{\theta}) = \boldsymbol{0}$ and eliminating $\theta$, we obtain~\cite{li_explicit_2024}
\begin{equation}
	\left(\mathrm{Re}(\lambda)\rho + \mathrm{Re}(\gamma)\rho^3 \right)^2
	+ \left(\mathrm{Im}(\lambda) - \Omega + \mathrm{Im}(\gamma)\rho^2 \right)^2 \rho^2
	= \epsilon^2 |\tilde{f}|^2.	
\label{eq:reduce-FRC}
\end{equation}
This algebraic equation gives the FRC in reduced coordinates $\rho(\Omega)$~\textcolor{blue}{as shown in Fig.~\ref{fig:FRC-ROM-objf}. This FRC} can be further mapped to the FRC of periodic orbits of the full system via the periodic SSM map $\boldsymbol{W}_\epsilon(\boldsymbol{p},\Omega t)$ in~\eqref{eq:red-reps}. 

Furthermore, the peak point $(\rho_{\rm{max}}, \Omega_{\rm{max}})$ on the reduced FRC $\rho(\Omega)$ corresponds to the peak of FRC in physical coordinates~\cite{li_explicit_2024,breunung_explicit_2018}. Let $d\rho/d\Omega = 0$, the peak point in reduced coordinates can be expressed as~\cite{li_explicit_2024}
\begin{align}
	&\Omega_{\max} = \mathrm{Im}(\lambda) + \mathrm{Im}(\gamma)\, \rho_{\max}^2, \label{eq:backbone} \\
	&\mathrm{Re}(\lambda)\, \rho_{\max} + \mathrm{Re}(\gamma)\, \rho_{\max}^3 
	= \epsilon |\tilde{f}|\, \mathrm{sign}(\mathrm{Re}(\lambda)). \label{eq:compute_rho_max}
\end{align}
Equation~\eqref{eq:backbone} provides the analytical expression of the backbone curve in reduced coordinates. Specifically, when $\mathrm{Im}(\gamma) > 0$, we obtain $\Omega_{\rm{max}} > \mathrm{Im}(\lambda)$, corresponding to a hardening behavior~\textcolor{blue}{as shown in Fig.~\ref{fig:FRC-ROM-objf}}; when $\mathrm{Im}(\gamma) < 0$, we obtain $\Omega_{\rm{max}} < \mathrm{Im}(\lambda)$, corresponding to a softening behavior. The corresponding amplitude $\rho_{\rm{max}}$ can be computed by solving Equation~\eqref{eq:compute_rho_max}. 

Now, we aim to minimize the peak of FRC by setting $\rho_{\rm{max}}$ as the objective function. Meanwhile, we add a constraint to $\gamma$ such that the backbone curve is also tuned. The corresponding optimization problem is formulated as
\begin{equation}
\begin{aligned}
	&\underset{\boldsymbol{\mu}}{\min} \quad c_{\mathrm{nl}} = \rho_{\max}  \\
	&\text{s.t.:} \\
	&\quad \left(\frac{\mathrm{Im}(\gamma)}{\gamma_{\text{target}}} -1\right)^2  \leq \varepsilon^2\\
	&\quad (\omega_Y/\omega_{Y,\text{target}} - 1)^2 \leq \varepsilon^2 \\
	&\quad \omega_X \geq \omega_{X,\text{target}} \\
	&\quad A \leq A_{\max}. \\
\end{aligned}
\label{eq:opt-rho_max}
\end{equation}
Here, the constraint on $\mathrm{Im}(\gamma)$ is imposed to control the hardening/softening behavior. 
\textcolor{blue}{The constraint on $\omega_Y$ (i.e., the first natural frequency $\omega_1$ associated to the vibration mode along \emph{vertical} direction) is introduced to maintain sufficient stiffness in the $Y$ direction throughout the optimization process, yielding well-connected optimal topology layout (cf. Remark~\ref{remark3:constraint_omX_OMY}), while the constraint on $\omega_X$ (i.e., the second natural frequency $\omega_2$ associated to the vibration mode along \emph{horizontal} direction) is imposed to prevent the occurrence of internal resonance between the first and second modes.} 
Specifically, the condition $\omega_{X,\text{target}} > 3\omega_{Y,\text{target}}$ must be satisfied, since the highest-order nonlinear term in the system is cubic. 
Further, the structural area fraction $A$ is subject to an upper bound.

We note that the optimization problem~\eqref{eq:opt-rho_max} provides a concurrent design of the FRC, namely, the peak of the FRC and the softening or hardening of the associated backbone curve. To illustrate the effectiveness of the formulation~\eqref{eq:opt-rho_max}, we also consider two reference optimization problems: an optimization problem on minimizing the peak of FRC without tuning the backbone, and an optimization problem solely on tuning the backbone.

For the first reference problem, we consider a linear optimization formulation for periodic response~\cite{dou_optimization_2015,niu_objective_2018}, namely
\begin{equation}
	\begin{aligned}
		&\underset{\boldsymbol{\mu}}{\min} \quad c_\mathrm{lin} = |\boldsymbol{U}|^{\mathrm{T}}\boldsymbol{L}|\boldsymbol{U}|  \\
		&\text{s.t.:} \\
		&\quad (\omega_Y/\omega_{Y,\text{target}} - 1)^2 \leq \varepsilon^2 \\
		&\quad \omega_X \geq \omega_{X,\text{target}} \\
		&\quad A \leq A_{\max}, \\
	\end{aligned}
	\label{eq:opt-linear}
\end{equation}
where $\boldsymbol{U}$ denotes the linear displacement vector obtained from $(\boldsymbol{K} + \mathrm{i} \Omega \boldsymbol{C} - \Omega^2 \boldsymbol{M}) \boldsymbol{U} = \boldsymbol{f}^{\text{ext}}$, and the symbol $|\cdot|$ represents taking the absolute value of each component of the vector. Here, $\boldsymbol{L}$ is a diagonal selection matrix with ones at DOFs corresponding to the node, line, or domain of interest~\cite{niu_objective_2018}. To maintain consistency, \eqref{eq:opt-linear} has the same constraints as~\eqref{eq:opt-rho_max}, except for the constraint on $\mathrm{Im}(\gamma)$, since linear optimization cannot capture or control the hardening/softening behavior. We will give an example to compare the optimization result of formulations~\eqref{eq:opt-rho_max} and~\eqref{eq:opt-linear} in Sec.~\ref{ssec: example1}.

\begin{remark}
    We consider the linear reference problem~\eqref{eq:opt-linear} instead of a nonlinear one similar to~\eqref{eq:opt-rho_max} but with the first constraint regarding $\gamma$ deleted for two reasons. Firstly, we would like to compare the results of linear and nonlinear optimizations. Secondly, numerical experiments show that the nonlinear one without constraint in $\gamma$ yields similar results with the linear reference problem in the domain of convergence for $\epsilon$. Discrepancies can be observed if $\epsilon$ is large, but the cubic truncation for SSM reduction~\eqref{eq:reduc-nonauto} is not sufficient in such cases.
\end{remark}

As for the second reference problem, we consider the following optimization problem in which a target value of $\mathrm{Im}(\gamma)$ is included in the objective function:
\begin{equation}
\begin{aligned}
	&\underset{\boldsymbol{\mu}}{\min} \quad c_{\gamma} = \left(\frac{\mathrm{Im}(\gamma)}{\gamma_{\text{target}}} -1\right)^2  \\
	&\text{s.t.:} \\
	&\quad (\omega_Y/\omega_{Y,\text{target}} - 1)^2 \leq \varepsilon^2 \\
	&\quad \omega_X \geq \omega_{X,\text{target}} \\
	&\quad A \leq A_{\max}. \\
\end{aligned}
\label{eq:opt-gamma}
\end{equation}
It is clear from the objective function that $c_\gamma = 0$ is achieved when $\mathrm{Im}(\gamma) = \gamma_{\text{target}}$. The constraints in the formulation~\eqref{eq:opt-gamma} are consistent with those in~\eqref{eq:opt-rho_max}, ensuring a fair comparison between the two formulations. We will give an example to compare the result of formulations~\ref{eq:opt-rho_max} and~\ref{eq:opt-gamma} in Sec.~\ref{ssec:example2}.

\begin{remark}
    We consider the reference problem~\eqref{eq:opt-gamma} to tune the backbone curve. Since forcing is not involved in the problem, it has no effect on the peak of FRC. The regulation of softening/hardening behavior through the manipulation of backbone curve has been demonstrated in~\cite{pozzi_topology_2025,pozzi2025adjointsensitivitiesoptimizationnonlinear}. The formulation~\eqref{eq:opt-gamma} is different from that of~\cite{pozzi_topology_2025} in two perspectives: 1) we manipulate \emph{damped} backbone curves~\cite{breunung_explicit_2018} instead of \emph{undamped} backbone curves as in~\cite{pozzi_topology_2025}; and 2) we set target value of $\gamma$ in the objective function instead of a constraint as in~\cite{pozzi_topology_2025}. In addition, our formulation~\eqref{eq:opt-gamma} tunes the backbone curve in reduced coordinates while the backbone curve is tailored by defining target points in terms of response frequency and physical amplitude in~\cite{pozzi2025adjointsensitivitiesoptimizationnonlinear}. 
\end{remark}

\subsubsection{Control of SN bifurcations on FRC}
\label{ssec:2.5 Control SN}

As discussed in~\cite{li_nonlinear_2022,liang_bifurcation_2025}, the SN bifurcations of the fixed points of the reduced dynamics~\eqref{eq:reduc-nonauto} correspond to those of the periodic orbits of the full system~\eqref{eq:eom-second-full}. This correspondence allows us to control the occurrence of SN bifurcations in reduced coordinates. In this subsection, we derive the expression of the SN bifurcation points in the reduced coordinates and eliminate their occurrence \textcolor{blue}{by controlling the frequency separation between the two SN bifurcation points,  namely, the $d_\mathrm{SN}$ shown in Fig.~\ref{fig:FRC-ROM-objf}}. In the final step, we formulate an optimization problem to suppress the occurrence of SN bifurcations.

We first need to find the fixed point of SN bifurcation $(\rho_{\mathrm{SN}}, \Omega_{\mathrm{SN}})$ in reduced coordinates. The Jacobian of the reduced dynamics~\eqref{eq:reduc-nonauto} can be expressed as~\cite{li_explicit_2024}
\begin{equation}
\boldsymbol{A}(\rho,\Omega) = \begin{pmatrix}
	\mathrm{Re}(\lambda) + 3\mathrm{Re}(\gamma) \rho^2 & -\left(\mathrm{Im}(\lambda) - \Omega + \mathrm{Im}(\gamma) \rho^2\right) \rho \\
	2\mathrm{Im}(\gamma) \rho + \left(\mathrm{Im}(\lambda) - \Omega + \mathrm{Im}(\gamma) \rho^2 \right)/\rho & \mathrm{Re}(\lambda) + \mathrm{Re}(\gamma) \rho^2
\end{pmatrix}.
\end{equation}
At the SN bifurcation of a fixed point, the Jacobian $\boldsymbol{A}$ has a simple zero eigenvalue~\cite{kuznetsov_elements_2023}. Thus, the determinant of $\boldsymbol{A}$ is 0 at SN bifurcation, namely
\begin{equation}
\begin{aligned}
\left(\mathrm{Re}(\lambda) + 3\mathrm{Re}(\gamma) \rho^2\right)\left(\mathrm{Re}(\lambda) + \mathrm{Re}(\gamma) \rho^2\right) + \left(\mathrm{Im}(\lambda) - \Omega + \mathrm{Im}(\gamma) \rho^2\right) \\ \left(2\mathrm{Im}(\gamma) \rho^2 + \left(\mathrm{Im}(\lambda) - \Omega + \mathrm{Im}(\gamma) \rho^2 \right)\right) =0.
\end{aligned}
\label{eq:SN-deteq0}
\end{equation}
By combining~\eqref{eq:reduce-FRC} and~\eqref{eq:SN-deteq0}, and eliminating $\Omega$, we derive the following algebraic equation for the response amplitude $\rho_{\mathrm{SN}}$ in the reduced coordinates
\begin{equation}
\begin{aligned}
	\left( 4\mathrm{Re}(\gamma)^4 + 4\mathrm{Im}(\gamma)^2 \mathrm{Re}(\gamma)^2 \right) \rho_{\mathrm{SN}}^{12}
	+ 8\mathrm{Re}(\lambda)\mathrm{Re}(\gamma) \left( \mathrm{Re}(\gamma)^2 + \mathrm{Im}(\gamma)^2 \right) \rho_{\mathrm{SN}}^{10} \\
	+ 4\mathrm{Re}(\lambda)^2 \left( \mathrm{Re}(\gamma)^2 + \mathrm{Im}(\gamma)^2 \right) \rho_{\mathrm{SN}}^8
	+ 4\epsilon^2 |\tilde{f}|^2 \left( \mathrm{Re}(\gamma)^2 - \mathrm{Im}(\gamma)^2 \right) \rho_{\mathrm{SN}}^6 \\
	+ 4\epsilon^2 |\tilde{f}|^2 \mathrm{Re}(\lambda)\mathrm{Re}(\gamma) \rho_{\mathrm{SN}}^4
	+ \epsilon^4 |\tilde{f}|^4 = 0. 
\end{aligned}
\label{eq:rho_SN}
\end{equation}
Substituting the value of $\rho_{\mathrm{SN}}$ into~\eqref{eq:reduce-FRC}, we obtain the corresponding frequency $\Omega_{\mathrm{SN}}$ as:
\begin{equation}
\begin{aligned}
	\Omega_{\mathrm{SN}} &= \mathrm{Im}(\lambda) + \mathrm{Im}(\gamma) \rho_{\mathrm{SN}}^2 + \mathrm{sign}(\mathrm{Im}(\gamma)) k_{\mathrm{SN}},
\end{aligned}
\label{eq:omega_SN}
\end{equation}
where
\begin{equation}
\begin{aligned}
	k_{\mathrm{SN}} = \sqrt{{\epsilon^2 |\tilde{f}|^2}/{\rho_{\mathrm{SN}}^2} - \left( \mathrm{Re}(\lambda(\mu)) + \mathrm{Re}(\gamma(\mu)) \rho_{\mathrm{SN}}^2 \right)^2}.
\end{aligned}
\end{equation} 
Note that there are usually two SN bifurcation points on the FRC, \textcolor{blue}{each corresponding to an excitation frequency at which an SN bifurcation occurs in the forced response of the system. By controlling the frequency separation between these two SN points, the SN bifurcation behavior can be effectively tailored, thereby regulating the associated jump phenomena in frequency sweeps. Next, we aim to suppress the occurrence of these two SN bifurcation points by controlling the magnitude of $d_{\text{SN}}:=|\Omega_{\mathrm{SN1}} - \Omega_{\mathrm{SN2}}|$ }.

Many MEMS sensors (e.g., accelerometers, gyroscopes) are designed to operate within the linear regime to avoid hysteresis and excess noise caused by nonlinear effects~\cite{tiwari_using_2019}. To ensure such behavior, we aim to control the system's bifurcation structure while enhancing its performance. Specifically, we formulate an optimization problem that maximizes the response amplitude $\rho_{\max}$, subject to a constraint on the parameter \textcolor{blue}{$d_{\text{SN}}$} to limit the distance between the two SN bifurcation points. The resulting optimization problem is given as 
\textcolor{blue}{\begin{equation}
	\begin{aligned}
		&\underset{\boldsymbol{\mu}}{\max} \quad c_{\mathrm{SN}} = \rho_{\max}  \\
		&\text{s.t.:} \\
		&\quad d_{\text{SN}}=|\Omega_{\text{SN1}}-\Omega_{\text{SN2}}| \leq d_\text{target} \\
        &\quad (\omega_Y/\omega_{Y,\text{target}} - 1)^2 \leq \varepsilon^2\\
		&\quad \omega_X \geq \omega_{X,\text{target}} \\
		&\quad (A/A_\text{target} - 1)^2 \leq \varepsilon^2, \\
	\end{aligned}
	\label{eq:opt-SNbifur}
\end{equation}}
where the constant \textcolor{blue}{$d_{\text{target}}$} regulates the distance between the two SN bifurcation points. \textcolor{blue}{As $d_{\text{target}}$ approaches zero, these two points coalesce, and the jump phenomenon is suppressed.} \textcolor{blue}{The roles of the constraints on $\omega_Y$ and $\omega_X$ are discussed in Remark~\ref{remark3:constraint_omX_OMY}}. A two-sided constraint is also imposed on the structural area to improve the stability of the optimization process.


\textcolor{blue}{\begin{remark}
We have consistently added the constraint $(\omega_Y/\omega_{Y,\text{target}} - 1)^2 \leq \varepsilon^2$ in the topology optimization problem formulations. This constraint enforces sufficient stiffness in the $Y$ direction throughout the optimization process, which is essential for obtaining a well-connected optimal topology layout. Indeed, we observe that the optimization iteration in~\eqref{eq:opt-gamma} and~\eqref{eq:opt-SNbifur} has difficulty converging to a well-defined optimal layout if this constraint is removed. We further add a constraint on $\omega_X$ to prevent the internal resonance between the master mode and other modes. 
\label{remark3:constraint_omX_OMY}
\end{remark}}

\section{Sensitivity analysis}
\label{section3}

In the previous section, we formulated several optimization problems. Since the Method of Moving Asymptotes (MMA) is used to solve it, sensitivity analysis with respect to the design variables is essential for updating the design. In this section, we derive the sensitivity expressions for the key quantities $\omega_j$, $\lambda$, $\gamma$, $\tilde{f}$, $\rho_{\max}$, and \textcolor{blue}{$\Omega_{\text{SN}}$}, which are critical for the optimization process.

The sensitivity of the eigenfrequencies $\omega_j$ and  eigenvalues $\lambda$ has already been derived in Appendix.C of~\cite{li_explicit_2024}, namely 
\begin{equation}
	\begin{aligned}
	\omega_j' = \frac{\boldsymbol{\phi}^{\text{T}}(\boldsymbol{K}' - \omega_j^2 \boldsymbol{M}')\boldsymbol{\phi}}{2\omega_j}, \\
	\lambda' = -\frac{\xi' \omega \lambda + (\xi \lambda + \omega) \omega'}{\lambda + \xi \omega}.
	\end{aligned}
\end{equation}
where $\omega$ is the eigenfrequency of the master subspace, i.e., $\omega_1$. Here and throughout this paper, the apex $'$ denotes the derivative with respect to one of the system parameters contained in the vector $\boldsymbol{\mu}$ introduced in~\eqref{eq:eom-second-full}.

Following~\cite{li_explicit_2024}, the variation of $\gamma$ can be obtained by the adjoint method, which can be expressed as 
\begin{equation}
	\begin{aligned}
		\delta \gamma = 
		& - \boldsymbol{\lambda}_{21}^{\mathrm{T}} \Big( 
		\partial_{\boldsymbol{\mu}} \boldsymbol{f}_2(\boldsymbol{\mu}, \boldsymbol{\phi}, \boldsymbol{W}_{20}^{(1)}) 
		+ \partial_{\boldsymbol{\mu}} \boldsymbol{f}_2(\boldsymbol{\mu}, \boldsymbol{W}_{20}^{(1)}, \boldsymbol{\phi}) \\
		&  + \partial_{\boldsymbol{\mu}} \boldsymbol{f}_2(\boldsymbol{\mu}, \boldsymbol{\phi}, \boldsymbol{W}_{11}^{(1)}) 
		+ \partial_{\boldsymbol{\mu}} \boldsymbol{f}_2(\boldsymbol{\mu}, \boldsymbol{W}_{11}^{(1)}, \boldsymbol{\phi}) \delta \boldsymbol{\mu} \Big) \\
		& - 3 \boldsymbol{\lambda}_{21}^{\mathrm{T}} \partial_{\boldsymbol{\mu}} \boldsymbol{f}_3(\boldsymbol{\mu}, \boldsymbol{\phi}, \boldsymbol{\phi}, \boldsymbol{\phi})  \delta \boldsymbol{\mu}\\
		& + \boldsymbol{\lambda}_{20}^{\mathrm{T}} \Big( (4\lambda^2 + 2\lambda \alpha)\delta \boldsymbol{M} + (2\lambda \beta + 1)\delta \boldsymbol{K} \Big) \boldsymbol{W}_{20}^{(1)} \\
		& + \boldsymbol{\lambda}_{11}^{\mathrm{T}} \Big( (4 [\mathrm{Re}(\lambda)]^2 + 2\mathrm{Re}(\lambda) \alpha) \delta \boldsymbol{M} 
		+ (2\mathrm{Re}(\lambda)\beta + 1)\delta \boldsymbol{K} \Big) \boldsymbol{W}_{11}^{(1)} \\
		& + (\boldsymbol{\lambda}_{20} + \boldsymbol{\lambda}_{11})^{\mathrm{T}} \partial_{\boldsymbol{\mu}} \boldsymbol{f}_2(\boldsymbol{\mu}, \boldsymbol{\phi}, \boldsymbol{\phi}) \delta \boldsymbol{\mu} \\
		& + \boldsymbol{\lambda}_{\boldsymbol{\phi}}^{\mathrm{T}} \Big( \delta \boldsymbol{K} - \omega^2 \delta \boldsymbol{M} \Big) \boldsymbol{\phi} 
		+ \lambda_{\mathrm{norm}} \boldsymbol{\phi}^{\mathrm{T}} \delta \boldsymbol{M} \boldsymbol{\phi}.
	\end{aligned}
\end{equation}
In the above equation, the damping coefficients $\alpha$ and $\beta$ are fixed, while the damping ratio $\xi$ varies throughout the optimization process. This differs from the formulation in~\cite{li_explicit_2024}, where the damping ratio remains constant. We follow the tensorial approach in~\cite{pozzi_topology_2025} to ensure efficient implementation.

The variation of $\tilde{f}$ can also be obtained using the adjoint method. As shown in Appendix~\ref{appA: sen_f}, the sensitivity of $\tilde{f}$ is given by
\begin{equation}
	\begin{aligned}
		\delta \tilde{f} = \boldsymbol{\eta}_{\boldsymbol{\phi}}^{\mathrm{T}}
		\bigl(\delta\boldsymbol{K}\boldsymbol{\phi} - \omega^2 \delta \boldsymbol{M}\boldsymbol{\phi}\bigr)
		+ \eta_{\mathrm{norm}} \boldsymbol{\phi}^{\mathrm{T}} \delta \boldsymbol{M} \boldsymbol{\phi}
	\end{aligned},
\end{equation}
where $\boldsymbol{\eta}_{\boldsymbol{\phi}}$ and $\eta_{\mathrm{norm}}$ are obtained from~\eqref{eq:solve_lambda}.
The sensitivity of $\rho_{\max}$ can be obtained by taking derivative of~\eqref{eq:compute_rho_max}, resulting in
\begin{equation}
	\begin{aligned}
		\rho_{\max}' = \frac{\epsilon|\tilde{f}|' \mathrm{sign}(\mathrm{Re}(\lambda)) - \mathrm{Re}(\lambda')\rho_{\max} - \mathrm{Re}(\mathrm{\gamma'})\rho_{\max}^3}{\mathrm{Re}(\lambda) + 3\mathrm{Re}(\gamma)\rho_{\max}^2}
	\end{aligned}.
\end{equation}
\textcolor{blue}{Similarly, by taking derivative of~\eqref{eq:reduce-FRC}, the sensitivity of $\Omega_{\text{SN}}$ can be expressed as
 \begin{equation}
	\begin{aligned}
\Omega_{\text{SN}}'=c_3 - \frac{c_4}{c_5}
	\end{aligned}
\end{equation} 
where
\begin{equation}
	\begin{aligned}
		c_3 = 
		&\mathrm{Im}(\lambda)' + \mathrm{Im}(\gamma)'\rho_\text{SN}^2 + 2\mathrm{Im}(\gamma)\rho_\text{SN}\rho_\text{SN}', \\
		c_4 = 
		& 2\epsilon^2|\tilde{f}||\tilde{f}|' - 2(\mathrm{Re}(\lambda)\rho_\text{SN} + \mathrm{Re}(\gamma)\rho_\text{SN}^3) \\
		&(\mathrm{Re}(\lambda)'\rho_\text{SN} + \mathrm{Re}(\lambda)\rho_\text{SN}' + \mathrm{Re}(\gamma)'\rho_\text{SN}^3+3\mathrm{Re}(\gamma)\rho_\text{SN}^2\rho_\text{SN}') \\
		&-2(\mathrm{Im}(\lambda) - \Omega_{\mathrm{SN}} + \mathrm{Im}(\gamma)\rho_\text{SN}^2)^2\rho_\text{SN}\rho_\text{SN}', \\
		c_5 = 
		&2(\mathrm{Im}(\lambda)-\Omega_{\mathrm{SN}}+\mathrm{Im}(\gamma)\rho_{\text{SN}}^2)\rho_{\text{SN}}^2,
	\end{aligned}
    \label{eq:sens_omega_SN}
\end{equation}
and $\rho_\mathrm{SN}'$ is given by Eq.~\eqref{eq:rhosn-prime} in Appendix~\ref{appA: sen_b}. One can further easily compute $d_{\text{SN}}'$ via the sensitivity of $\Omega_{\text{SN}}'$.}

\textcolor{blue}{We conclude this section with an intuitive interpretation of how the sensitivities of $\rho_{\max}$, $\gamma$, and $d_{\text{SN}}$ guide the design evolution. As seen in Fig.~\ref{fig:FRC-ROM-objf}, the parameter $\rho_{\max}$ corresponds to the peak amplitude of the FRC, and reducing $\rho_{\max}$ is generally associated with an increase in the effective stiffness of the structure. The parameter $\gamma$ characterizes the hardening or softening behavior of the nonlinear response. In particular, increasing $\mathrm{Im}(\gamma)$ weakens softening effects or enhances hardening behavior. Finally, the parameter $d_{\text{SN}}$ is related to the frequency range with the coexistence of stable and unstable periodic orbits on the FRC. A larger magnitude of $d_{\text{SN}}$ corresponds to a larger interval of unstable periodic orbits, leading to a more pronounced jump phenomenon in forward frequency sweep.}
  
\section{SIMP Model for Topology Optimization}

In this work, the Solid Isotropic Material with Penalization (SIMP) method~\cite{bendsoe_material_1999} is employed as the foundation for density-based topology optimization. In the SIMP-based framework, the material properties of the structure are interpolated element-wise based on the physical density variables $\hat{\mu}_e$. These densities are used to assemble the global matrices and nonlinear force vectors. Specifically, the global mass and stiffness matrices, as well as the quadratic and cubic stiffness tensors, are computed as follows~\cite{pozzi_topology_2025}:
\begin{equation}
\begin{aligned}
    \boldsymbol{M} &= \bigcup_{e=1}^{N_{\mathrm{el}}} \hat{\mu}_e \boldsymbol{M}_e,  \quad 
    \boldsymbol{K} = \bigcup_{e=1}^{N_{\mathrm{el}}} \hat{\mu}_e \boldsymbol{K}_e, \\
    F_2 &= \bigcup_{e=1}^{N_{\mathrm{el}}} \hat{\mu}_e F_{2e},  \quad
    F_3 = \bigcup_{e=1}^{N_{\mathrm{el}}} \hat{\mu}_e F_{3e}, 
\end{aligned}    
\end{equation}
where the symbol $\bigcup$ denotes the standard finite element assembly process, and $N_{\mathrm{el}}$ is the total number of finite elements. The quadratic tensor $F_2$ and cubic tensor $F_3$ are used to compute the nonlinear internal force via Einstein notation, e.g.,
\begin{equation}
\begin{aligned}
   f_2^i(\boldsymbol{a},\boldsymbol{b})=F_2^{ijk}a^jb^k, \quad f_3^i(\boldsymbol{a},\boldsymbol{b},\boldsymbol{c})=F_3^{ijkl}a^jb^kc^l,
\end{aligned}    
\end{equation}
where the vectors $\boldsymbol{a}$, $\boldsymbol{b}$, and $\boldsymbol{c}$ are the inputs of the nonlinear force. The physical densities $\hat{\mu}_e$ are obtained by applying filtering and projection operations to the design variables $\mu_e$, as described below.

To ensure mesh-independent and well-posed topology optimization results, the design variable $\mu \in [0, 1]$ is first smoothed using a density filter~\cite{wang_projection_2011}, defined as
\begin{equation}
    \tilde{\mu}_e = \frac{\sum_{j \in \mathcal{N}_e} w_{j,e} \mu_j}{\sum_{j \in \mathcal{N}_e} w_{j,e}}, 
\end{equation}
where $\tilde{\mu}_e$ is the filtered density of element $e$, $\mathcal{N}_e$ is the set of neighboring elements within a radius $R$, and $w_{j,e} = R - \lvert \mathbf{x}_j - \mathbf{x}_e \rvert$ is the weight based on centroid distance. Unless otherwise specified, the filter radius is set to $R = 4$.

Next, a Heaviside-type projection~\cite{wang_projection_2011} is applied to obtain the projected density $\bar{\mu}_e$, which sharpens the transition between solid and void regions:
\begin{equation}
    \bar{\mu}_e = \frac{\tanh(\sigma \eta) + \tanh(\sigma (\tilde{\mu}_e - \eta))}{\tanh(\sigma \eta) + \tanh(\sigma (1 - \eta))}, 
\end{equation}
where $\sigma$ is the projection steepness parameter and $\eta \in (0, 1)$ is the threshold. In this work, we initially set $\sigma = 10$ and $\eta = 0.5$, and increase $\sigma$ progressively during the optimization, unless otherwise specified.

Finally, the physical density $\hat{\mu}_e$ used to interpolate material properties is computed via the SIMP scheme:
\begin{equation}
    \hat{\mu}_e = \hat{\mu}_0 + (1 - \hat{\mu}_0)\bar{\mu}_e^p, 
\end{equation}
where $\hat{\mu}_0 \ll 1$ is a small lower bound to avoid singular stiffness in void regions, and $p$ is the penalization exponent to promote 0-1 designs. In this work, the value of $\hat{\mu}_0$ is set to $10^{-6}$, while $p$ is initially set to 1 and gradually increased during the optimization process.

\textcolor{blue}{
\begin{remark}
    The material distribution is initialized according to initial layouts (cf. Figs.~\ref{fig:beam_linear}, \ref{fig:initial-MBB}, and \ref{fig:initial-microbeam}). Specifically, the design variables of the elements in the non-design region (black area) are fixed to $\mu_e = 1$, whereas the design variables of the elements in the designable region (gray area) are uniformly initialized to the \emph{prescribed area fraction upper-bound} $A_{\max}$, i.e., $\mu_e = A_{\max}$, such that the area fraction constraint is active.
\end{remark}}

\section{Numerical examples}
\label{section4}

In this section, three numerical examples are presented to show the result of minimizing the peak of FRC, controlling hardening/softening behavior, and suppressing the occurrence of SN bifurcations. In all the examples, the 2-dimensional structures are discretized using 4-node, square elements. Following~\cite{pozzi_topology_2025}, material properties are specified with the density $\rho = 2330 \times 10^{-6}  \ \rm{ng/um^3}$, Young's modulus $E=148 \times 10^9 \ \rm{Pa}$, and Poisson's ratio $\nu = 0.23$. The plane stress approximation is used, with a thickness of 24 $\rm{\mu m}$.

The matrix assembly of the finite element model and the solution of the optimization problems are carried out using the \texttt{YetAnotherFEcode} package~\cite{Jain2022YetAnotherFEcode}. The FRCs of the optimized layouts are computed using \texttt{SSMTool}~\cite{Jain2023SSMTool}, an open-source software package for the computation of invariant manifolds and their reduced dynamics. In the Sec.~\ref{ssec:reduc_on_SSM} and~\ref{ssec:min-FRC}, we employed a third-order reduced-order model~\eqref{eq:reduc-nonauto} to obtain the expression of FRC in the reduced coordinates~\eqref{eq:reduce-FRC}. To assess the reliability of the optimized result, we further compute the FRC using \texttt{SSMTool} at higher-order truncations for the optimized structure. If the FRC at higher-order truncations agrees well with that at the third-order truncation, it indicates that the predicted FRC has converged and is hence reliable.

\subsection{Minimizing the peak of FRC: linear vs. nonlinear optimizations}
\label{ssec: example1}

\begin{figure}[!ht]
\centering
\includegraphics[width=0.7\textwidth]{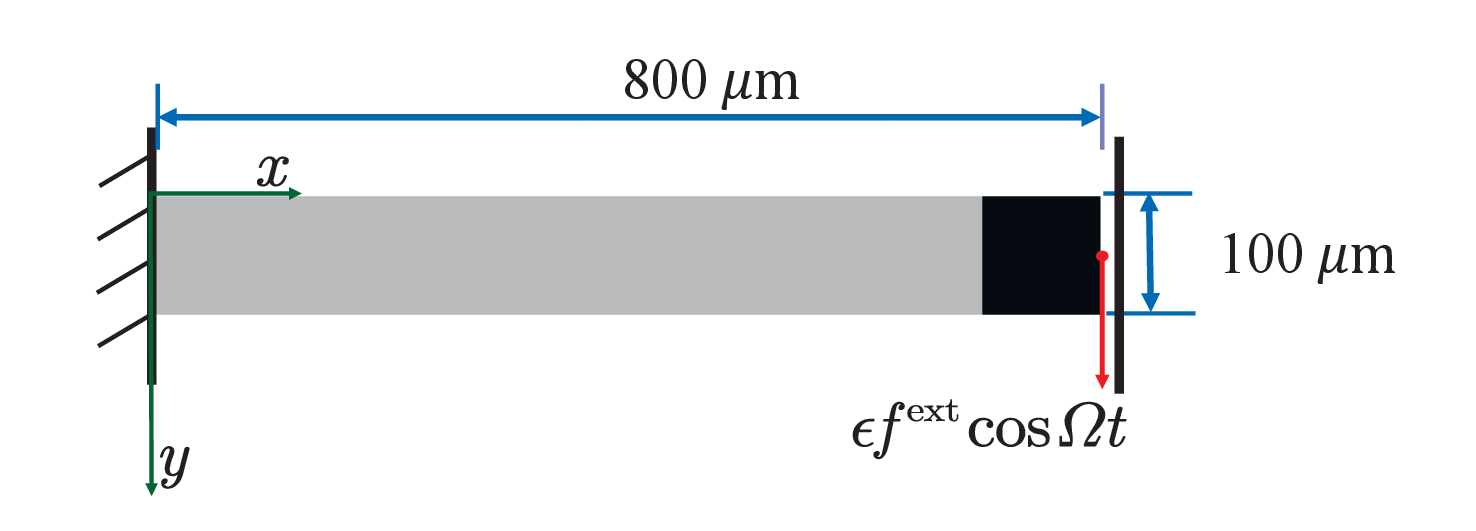}
\caption{Initial layout of the half MBB beam considered in the example of Sec.~\ref{ssec: example1}. The total domain is 800~$\mathrm{\mu m}$ long and 100~$\mathrm{\mu m}$ high, and includes a non-design region of 100~$\mathrm{\mu m}$ in length. The gray area represents the designable region, while the black area indicates the fixed non-design region.}
\label{fig:beam_linear}
\end{figure}

We consider a Messerschmitt-Bölkow-Blohm (MBB) beam with a fixed segment located at its center. Due to structural symmetry, the optimization is performed only on the left half of the beam~\cite{pozzi_topology_2025}, as illustrated in Fig.~\ref {fig:beam_linear}. The half MBB beam measures 800~$\mathrm{\mu m}$ in length and 100~$\mathrm{\mu m}$ in height, including a fixed region with a length of 100~$\mathrm{\mu m}$. The left end of the beam is fully fixed, and the right end is constrained in the $x$-direction. A periodic excitation is applied at the midpoint of the right edge.

A finite element mesh of $160 \times 20$ elements is employed, resulting in a system with $n = 6699$ degrees of freedom in~\eqref{eq:eom-second-full}. The damping ratio of the initial layout is set to $\xi$ = 0.1\%, which yields the damping constants $\alpha = 2.99$ and $\beta = 1.72 \times 10^{-7}$. During the optimization process, the forcing amplitude is given as $f^{\mathrm{ext}} = 5 \times 10^{9} \ \mathrm{ng \cdot \mu m/ms^2}$. Unless otherwise specified, the parameter $\epsilon$ is set to $0.01$ throughout the computation. The target values for other constraints are set as follows: $\omega_{Y,\text{target}} = 300 ~\mathrm{kHz}$, $\omega_{X,\text{target}} = 1200 ~\mathrm{kHz}$, and $A_{\max} = 50\%$. Based on these parameter settings, we compare the results obtained from the nonlinear optimization formulation~\eqref{eq:opt-rho_max} with those from the linear optimization formulation~\eqref{eq:opt-linear}.

\begin{figure}[!ht]
\centering
\includegraphics[width=0.45\textwidth]{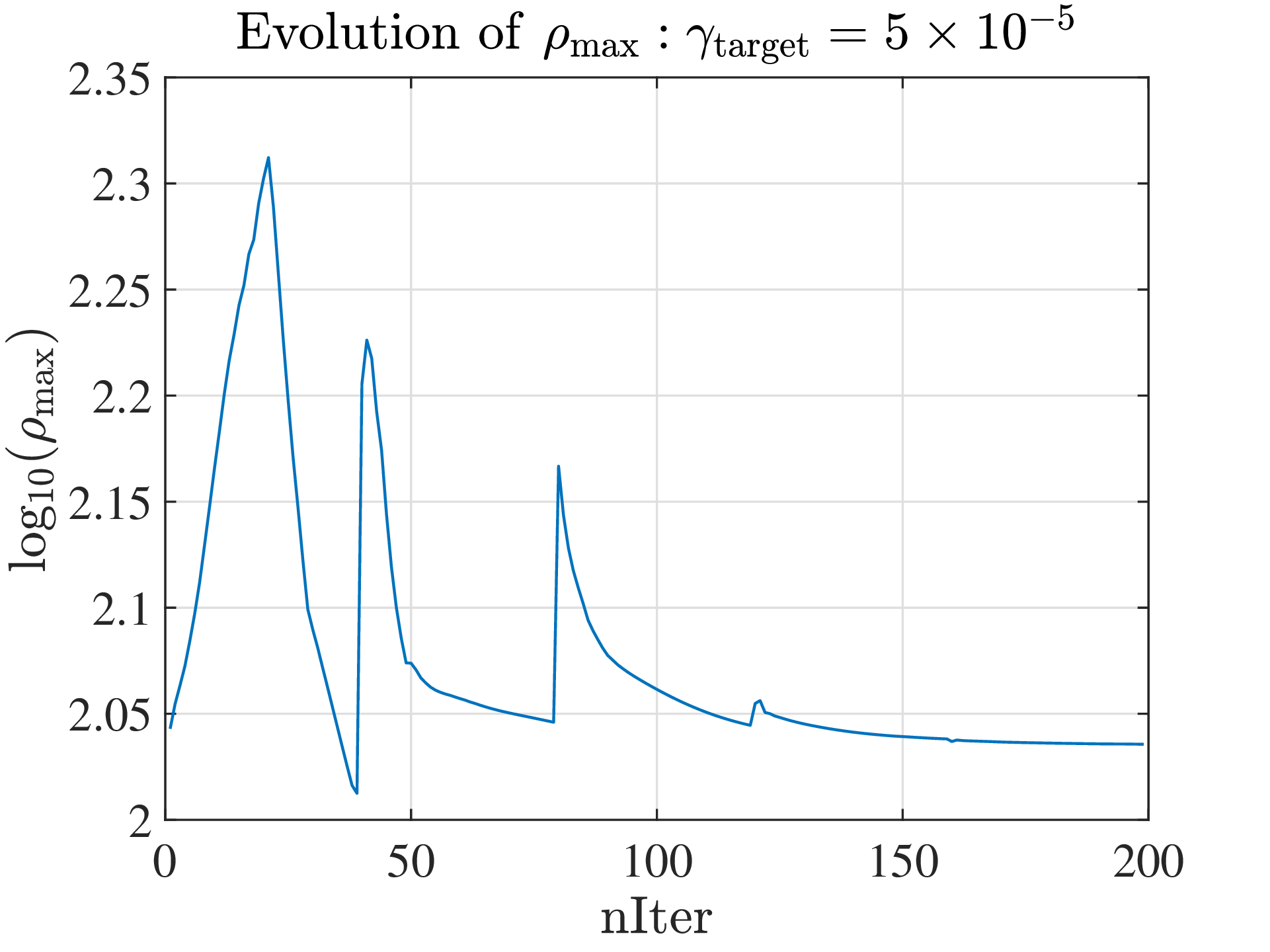}
\includegraphics[width=0.45\textwidth]{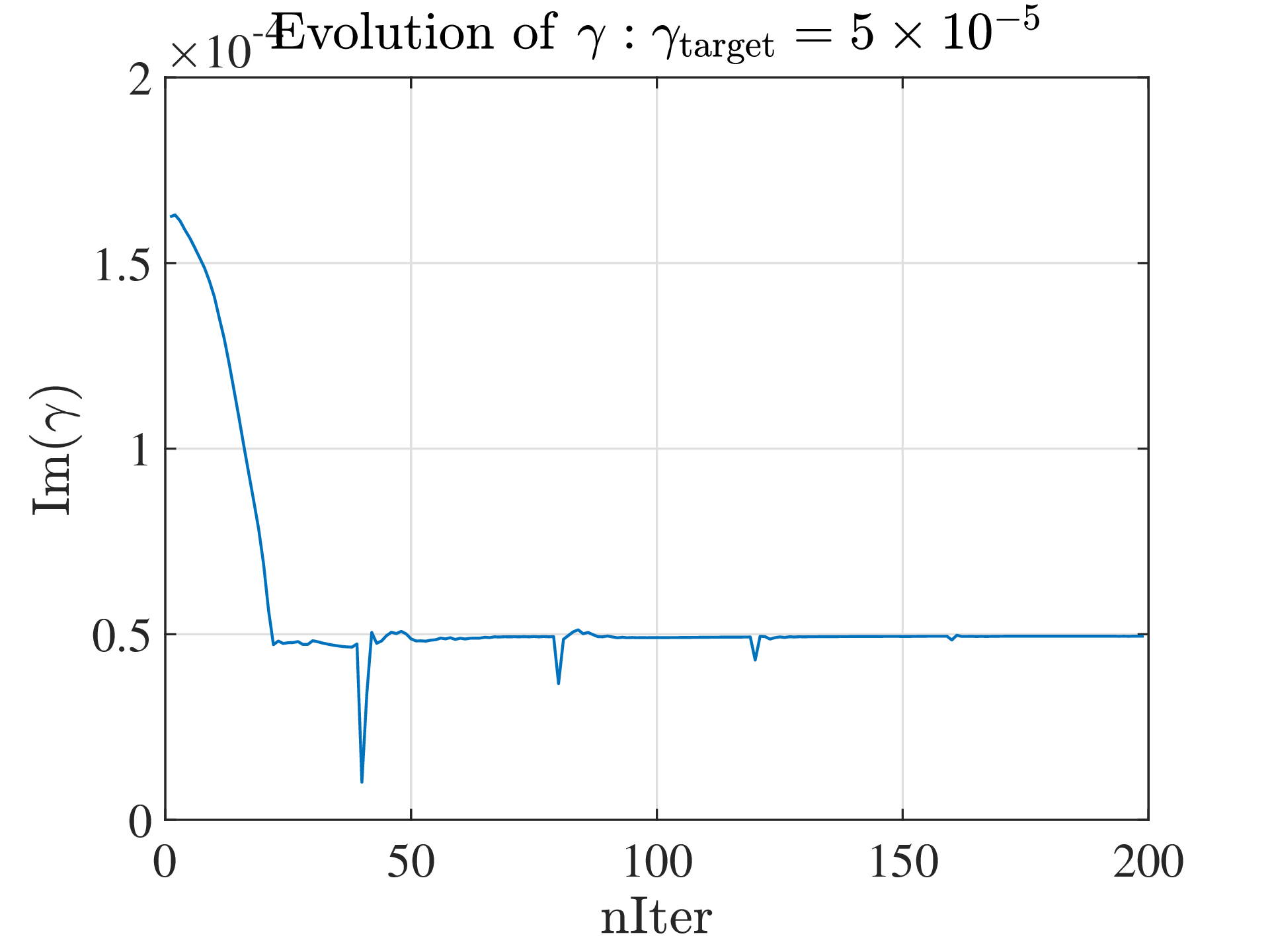}
\caption{Iteration history of the objective $\rho_{\max}$ (left panel) and the constraint value $\mathrm{Im}(\gamma)$ (right panel) for the case of $\gamma_{\mathrm{target}} = 5 \times 10^{-5}$.}
\label{fig:iteration_gamma00005}
\end{figure}

\begin{figure}[!ht]
 \centering
 \includegraphics[width=0.5\textwidth]{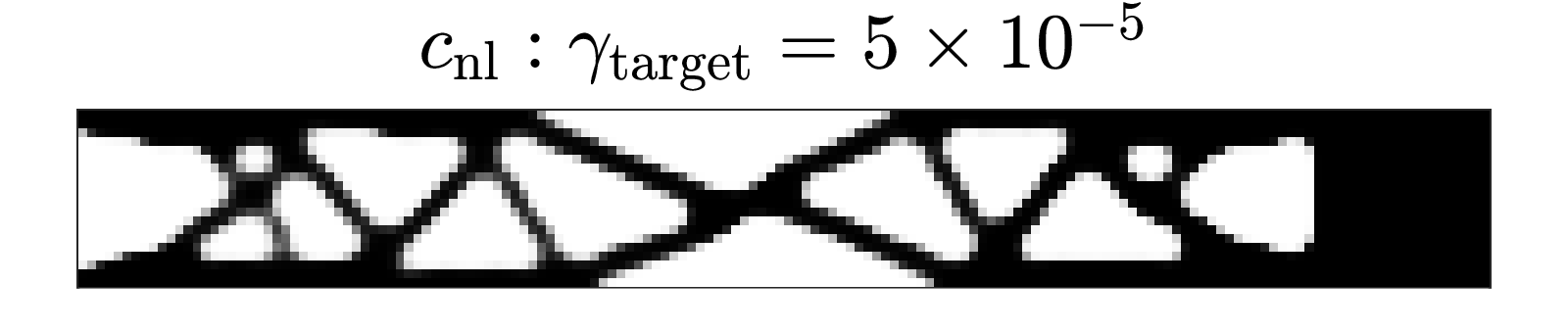}\\
 \includegraphics[width=0.5\textwidth]{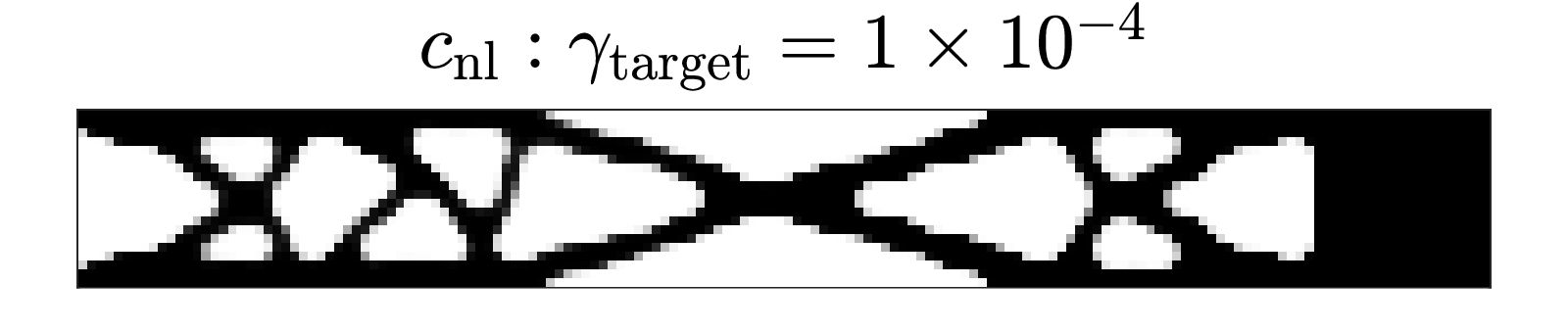}\\
 \includegraphics[width=0.5\textwidth]{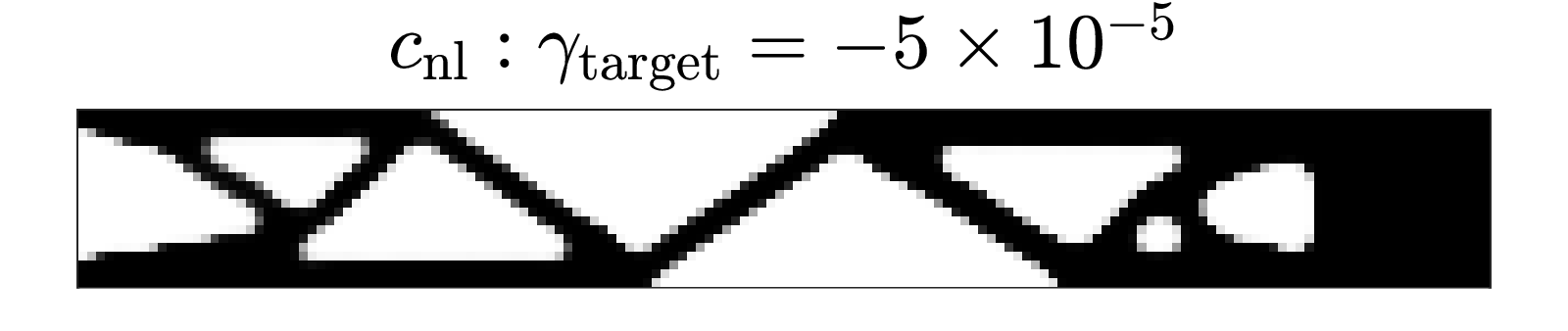}\\
 \includegraphics[width=0.5\textwidth]{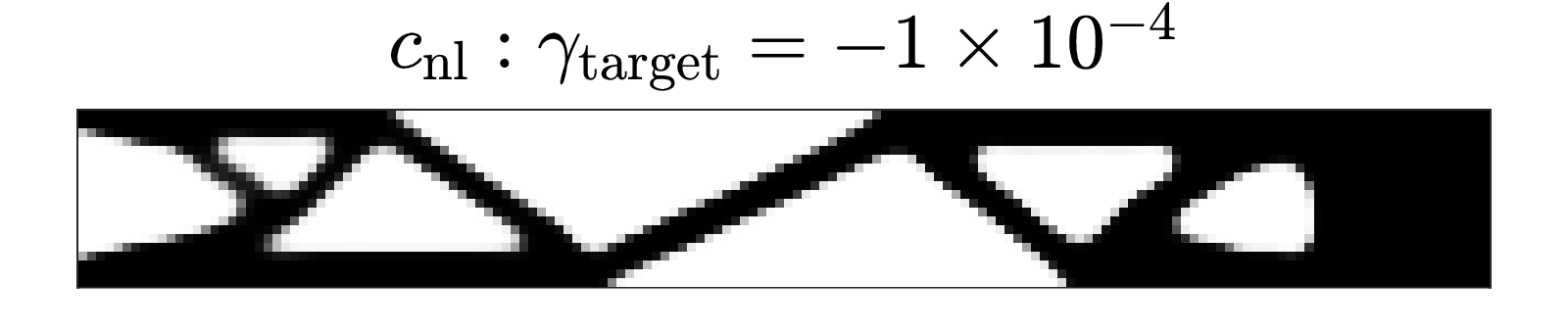}\\
 \includegraphics[width=0.5\textwidth]{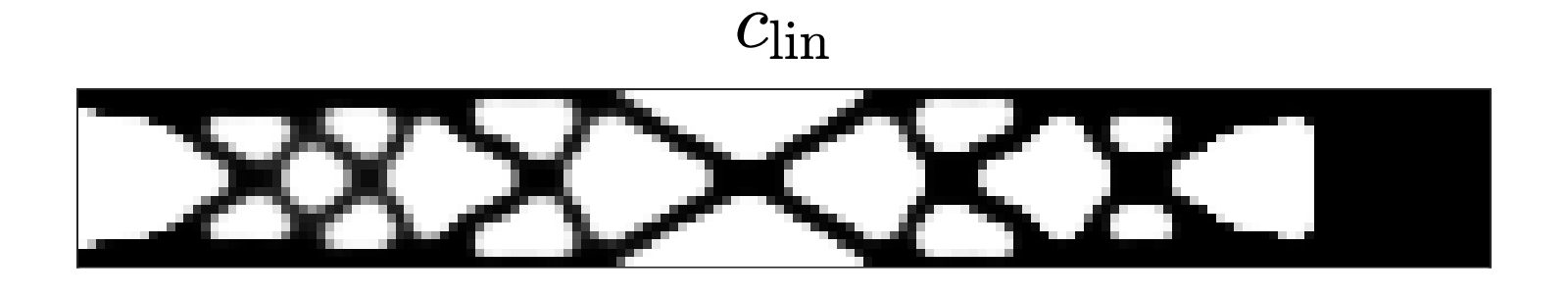}\\
 \caption{Optimal layouts obtained from the nonlinear optimization formulation~\eqref{eq:opt-rho_max} and the linear optimization formulation~\eqref{eq:opt-linear}. The first four panels correspond to nonlinear optimal layouts with different values of $\gamma_{\mathrm{target}}$. The last panel shows the layout obtained from the linear optimization formulation.}
 \label{fig:layouts_linear_nonlinear}
 \end{figure}

With $\gamma_{\mathrm{target}} = 5 \times 10^{-5}$, the iterative process for the nonlinear optimization formulation~\eqref{eq:opt-rho_max} is provided in Fig.~\ref{fig:iteration_gamma00005}. The objective $\rho_{\max}$ exhibits an initial increase, followed by a rapid decrease until convergence. Meanwhile, the constraint on $\mathrm{Im}(\gamma)$ is violated initially, and the value of $\mathrm{Im}(\gamma)$ gradually converges and ultimately reaches the target $\mathrm{Im}(\gamma) = \gamma_{\mathrm{target}}$. The abrupt changes observed in the objective $\rho_{\max}$ and $\mathrm{Im}(\gamma)$ during the iteration are caused by the scheduled updates of the penalization exponent $p$ and the projection parameter \textcolor{blue}{$\sigma$} in the SIMP algorithm. \textcolor{blue}{Specifically, $\sigma$ is initialized at $\sigma_0 = 10$ and multiplied by $1.2$ every 40 iterations until it reaches a maximum value of $\sigma_{\max} = 50$. Meanwhile, the parameter $p$ is initialized at $p_0 = 1$ and increased by one every 40 iterations, with an upper bound of $p_{\max} = 8$.} The obtained optimal layout is given in the upper panel of Fig.~\ref{fig:layouts_linear_nonlinear}. The computational performance of this iterative process is summarized in Table~\ref{tab:comp_perf_ex1}.~\textcolor{blue}{Here and throughout this study, the stopping criterion for the optimization of $c_{\text{nl}}$ is defined by the simultaneous satisfaction of all constraints and the convergence of the objective function, i.e., $\max_{j\in\{i-3,i-2,i-1\}}\ |{c_{\text{nl}}^{(i)}-c_{\text{nl}}^{(j)}}|/{c_{\text{nl}}^{(i)}}<10^{-3}$. We check the convergence by a comparison with several previous iterations to ensure a steady convergence. In addition, the efficiency listed in all tables of this article is defined as the total optimization time divided by the number of optimization iterations, i.e., the average computational time per iteration, which includes the time to compute the SSM reduction, the objective function and also its sensitivity.}

\begin{table}[H]
\centering
\caption{Computational performance of the iterative process in Fig.~\ref{fig:iteration_gamma00005}. All computations in this paper were performed on a Windows desktop equipped with Intel Core i9-12900K CPU (3.20~GHz) and 64.0~GB RAM.}
\label{tab:comp_perf_ex1}
\begin{tabular}{cccc}
\hline
\textbf{Mesh size} & \textbf{Total time} & \textbf{Iterations} & \textbf{Efficiency} \\
\hline
$160 \times 20$ & 44 mins & 199 & 13.3 s/step \\
\hline
\end{tabular}
\end{table}

We solve the same nonlinear optimization formulation~\eqref{eq:opt-rho_max} but with varied $\gamma_{\mathrm{target}}$. The optimal layouts obtained for different values of $\gamma_{\mathrm{target}}$, as well as from the linear optimization formulation~\eqref{eq:opt-linear}, are shown in Fig.~\ref{fig:layouts_linear_nonlinear}. As $\gamma_{\mathrm{target}}$ varies, the optimal layouts from the nonlinear optimization exhibit significant changes. Moreover, all layouts obtained from the nonlinear optimization differ noticeably from the linear layout.
 
\begin{figure}[!ht]
\centering
\includegraphics[width=0.45\textwidth]{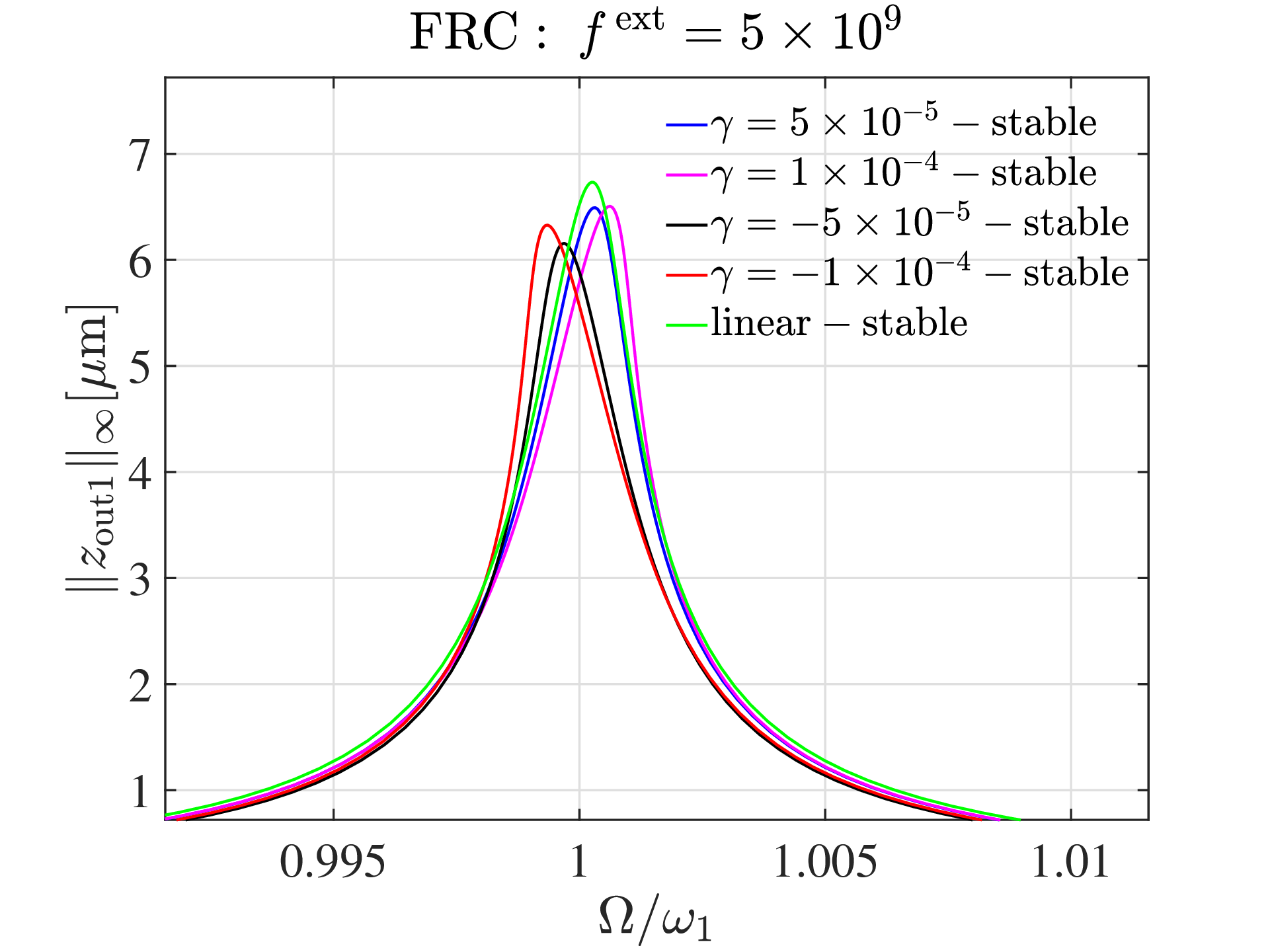}
\includegraphics[width=0.45\textwidth]{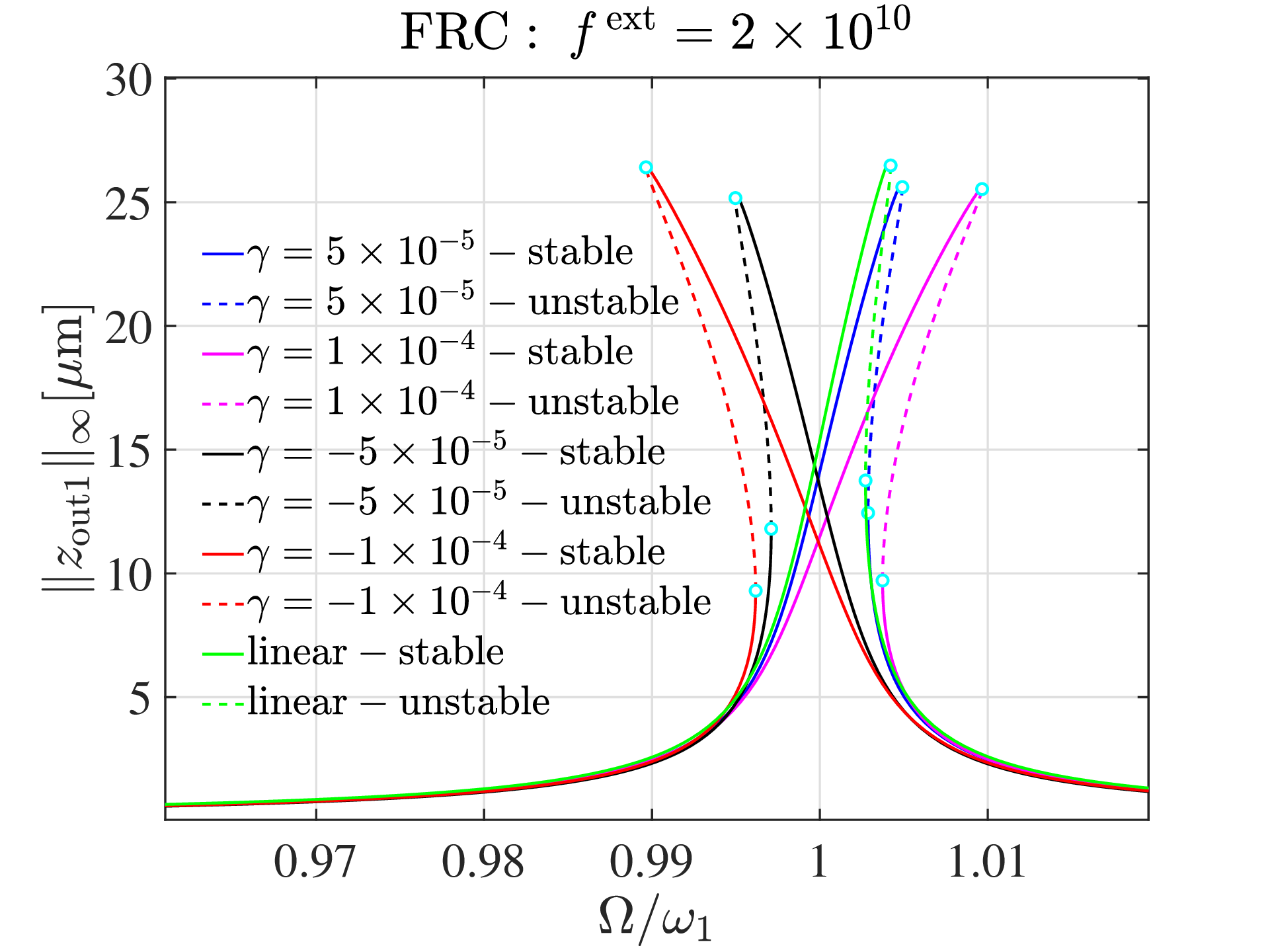}
\caption{FRCs of the optimal layouts shown in Fig.~\ref{fig:layouts_linear_nonlinear} under two levels of forcing amplitude: $f^{\mathrm{ext}} = 5 \times 10^{9}$ (left panel) and $2 \times 10^{10}$ (right panel) $\mathrm{ng \cdot \mu m/ms^2}$.}
\label{fig:FRCs_linear_nonlinear}
\end{figure}

We now compare the FRCs of the optimal layouts shown in Fig.~\ref{fig:layouts_linear_nonlinear}. The vertical axis $\Vert z_{\rm{out1}} \Vert_\infty$ represents the vibration amplitude of the node located at 
($x$,$y$)=(800, 50) in this example. As shown in the left panel of Fig.~\ref{fig:FRCs_linear_nonlinear}, the FRCs corresponding to the nonlinear optimization exhibit distinct hardening or softening behaviors depending on the value of $\gamma_{\mathrm{target}}$. To make these nonlinear characteristics more evident, we increase the forcing amplitude to $f^{\mathrm{ext}} = 2 \times 10^{10} \ \mathrm{ng \cdot \mu m/ms^2}$ and recompute the FRCs for the optimized structures. The results, presented in the right panel of Fig.~\ref{fig:FRCs_linear_nonlinear}, show that the FRC behavior gradually shifts from softening to hardening as $\gamma_{\mathrm{target}}$ increases. Thus, while both linear and nonlinear optimization approaches succeed in controlling the FRC amplitude, only the nonlinear optimization is capable of adjusting the hardening/softening behavior. A posteriori computation shows that the structure obtained from the linear optimization yields $\mathrm{Im}(\gamma) = 3.89 \times 10^{-5}$. This value is close to $5\times10^{-5}$, which explains the FRC for the $\gamma_{\mathrm{target}}=5\times10^{-5}$ case looks similar to the linear one, as seen in Fig.~\ref{fig:FRCs_linear_nonlinear}. 

Since third-order SSM reduction is employed to compute the FRCs of the optimal layouts, it is necessary to assess the convergence of the SSM-based model reduction. The convergence analysis is presented in Appendix~\ref{appA: SSM_order_test_ex1}, which shows that the FRC obtained using third-order SSM reduction agrees well with that obtained using seventh-order SSM reduction under the forcing amplitude $f^{\mathrm{ext}} = 2 \times 10^{10} \ \mathrm{ng \cdot \mu m/ms^2}$.

\textcolor{blue}{To assess the robustness of the optimized structures, we vary both the forcing amplitude and the material stiffness and examine their effects on the resulting optimal layouts and dynamic responses. The corresponding optimization results are presented in Appendix~\ref{appA2: robustness_ex1}, which show that, within the range where the third-order SSM remains convergent, the optimized layouts and their dynamic responses show good robustness to these variations.}

\textcolor{blue}{Our optimization framework assumes no internal resonance. To justify this assumption, the evolution of the first three eigenfrequencies during the iterative process for the case of $\gamma_{\mathrm{target}} = 5 \times 10^{-5}$, is presented in Appendix~\ref{appA3: test_mode_collisions}. The results show that the natural frequencies are well separated and neither modal collisions nor internal resonances occur. As a result, we can track the target master mode in optimization easily by choosing the mode with lowest natural frequency. As detailed in Appendix~\ref{appA3: test_mode_collisions}, the initial and optimized structures exhibit
similar linear modal shapes for the master mode, which further confirms the consistency of master mode during optimization process.}

\textcolor{blue}{The optimization results presented in this subsection show that the nonlinear optimization formulation in Eq.~\eqref{eq:opt-rho_max} achieves a minimization of the FRC peak comparable to that obtained with the linear optimization formulation in Eq.~\eqref{eq:opt-linear}. This suggests that the SSM-based framework provides an effective approach for reducing the peak response of the FRC. More importantly, in contrast to the linear optimization formulation, the nonlinear framework allows the hardening or softening characteristics of the system to be actively tailored, highlighting an additional advantage of the SSM-based approach beyond peak-response reduction.}

\subsection{Tailoring hardening/softening behavior: concurrent design vs. backbone design only}
\label{ssec:example2}

\begin{figure}[!ht]
\centering
\includegraphics[width=0.5\textwidth]{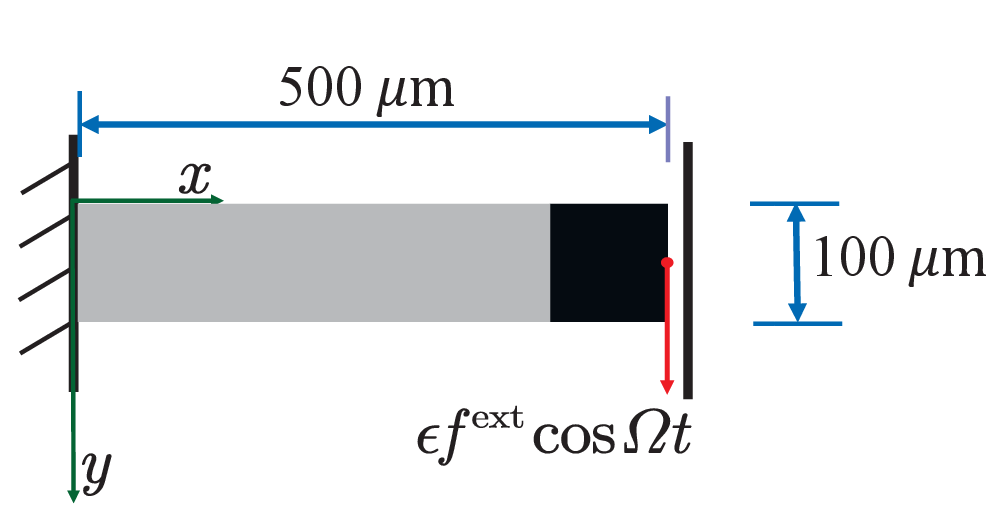}
\caption{Initial layout of the half MBB beam considered in the example of Sec.~\ref{ssec:example2}. The total domain is 500~$\mathrm{\mu m}$ long and 100~$\mathrm{\mu m}$ high, and includes a non-design region of 100~$\mathrm{\mu m}$ in length. The gray area represents the designable region, while the black area indicates the fixed non-design region.}
\label{fig:initial-MBB}
\end{figure}

Tailoring the hardening/softening behavior of the backbone curve in an MBB beam has been discussed in~\cite{pozzi_topology_2025}. In this subsection, we aim to minimize the response amplitude while simultaneously tailoring the softening–hardening behavior, namely, the concurrent design~\eqref{eq:opt-rho_max}. We further compare this concurrent design with the design given by the reference problem~\eqref{eq:opt-gamma} that tunes the backbone only. As shown in Fig.~\ref{fig:initial-MBB}, the design domain of the beam has a length of 500~$\mathrm{\mu m}$ and a height of 100~$\mathrm{\mu m}$. A fixed region, occupying 20\% of the total beam length, is located at the center and serves as a proof mass. The left end of the beam is fully fixed, while the right end is constrained in the $x$-direction. A periodic excitation is applied at the midpoint of the right edge.

The FE model uses a mesh of $100 \times 20$ elements, leading to a system with $n$= 4179 DOF in~\eqref{eq:eom-second-full}. The damping ratio of initial layout $\xi$ = 0.1\%, which leads to the values of damping constants $\alpha = 6.90$ and $\beta = 7.41 \times 10^{-8}$. During the optimization iteration process, the forcing amplitude is given as $ f^{\mathrm{ext}}= 5 \times 10^{9} \ \mathrm{ng \cdot \mu m/ms^2}$. In the following computation, we set $\epsilon = 0.01$ unless otherwise stated. The projection parameter is fixed at $\sigma = 10$ throughout the optimization, \textcolor{blue}{and the penalization parameter $p$ is initialized at $p_0 = 1$ and increased by one every 40 iterations, up to a maximum value of $p_{\max} = 8$.} The filter radius is set to $R = 3$. In this example, no constraint is imposed on $\omega_X$ since $\omega_X > 3\omega_Y$ is naturally satisfied throughout the optimization iterations. Furthermore, the two-sided constraint on $\gamma$ is replaced by a one-sided constraint. Specifically, $\gamma > \gamma_{\mathrm{target}}$ is enforced when $\gamma_{\mathrm{target}} > 0$, and $\gamma < \gamma_{\mathrm{target}}$ is enforced when $\gamma_{\mathrm{target}} < 0$. The target values for other constraints are set as follows: $\omega_{Y,\text{target}} = 600 ~\mathrm{kHz}$, and $A_{\max} = 50\%$. Based on these parameter settings, we compare the optimization results obtained from formulations~\eqref{eq:opt-rho_max} and~\eqref{eq:opt-gamma} for $\gamma_{\mathrm{target}} = \pm 1 \times 10^{-3}$.

\subsubsection{Hardening optimization}

\begin{figure}[!ht]
	\centering
	\includegraphics[width=0.4\textwidth]{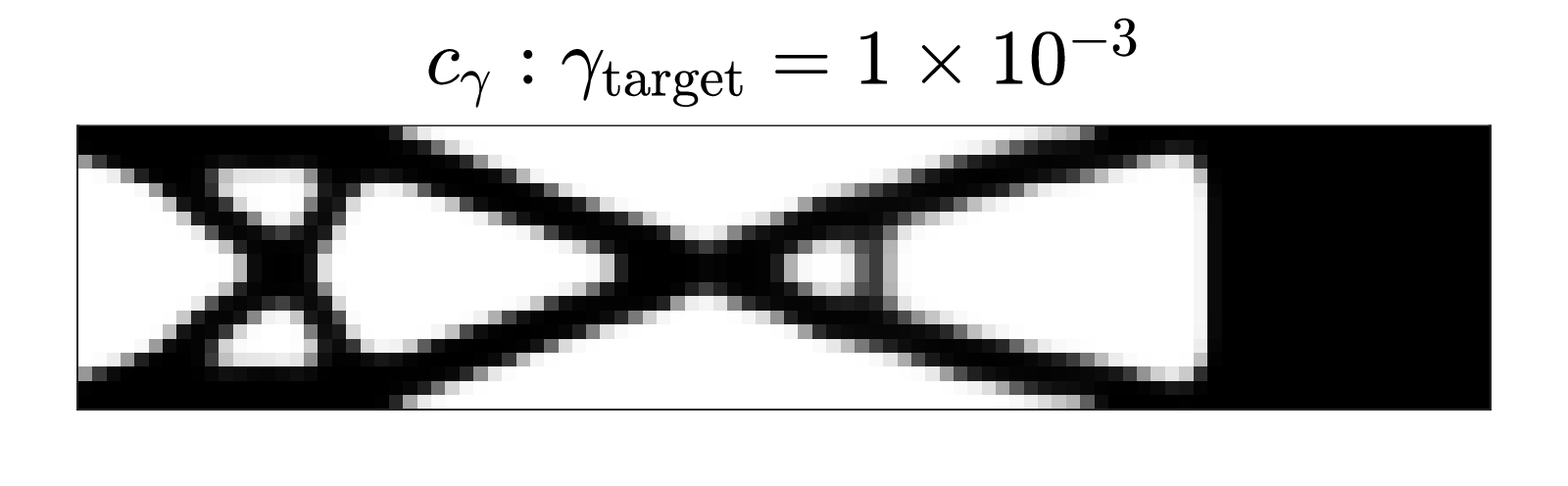}
	\includegraphics[width=0.4\textwidth]{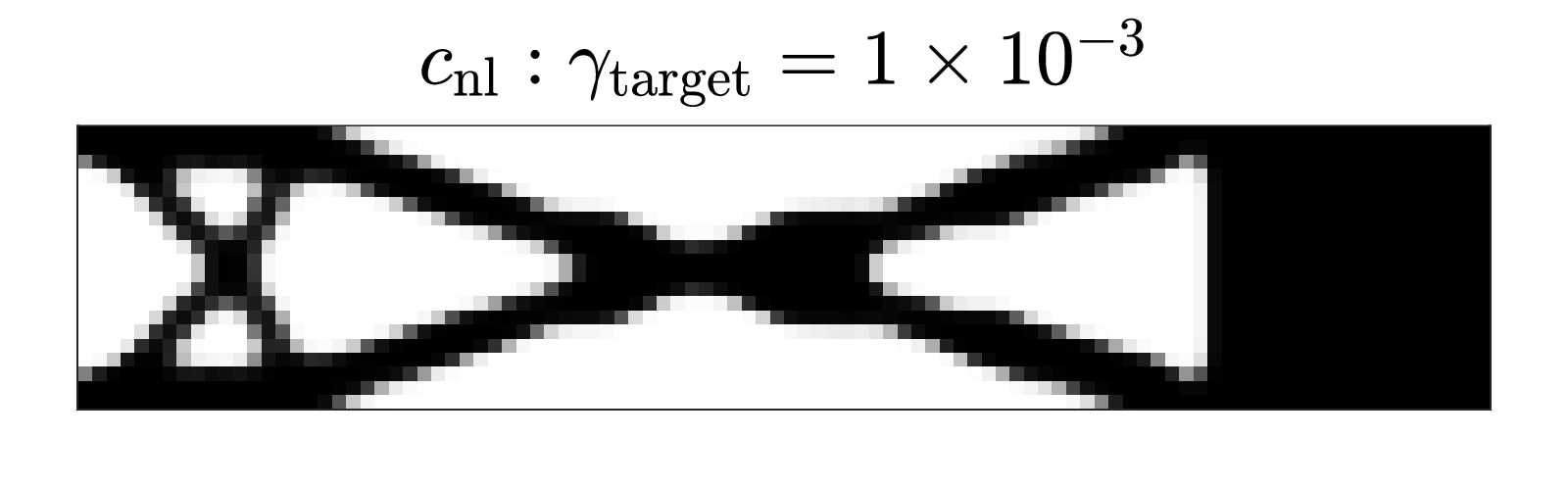} \\
	\includegraphics[width=0.3\textwidth]{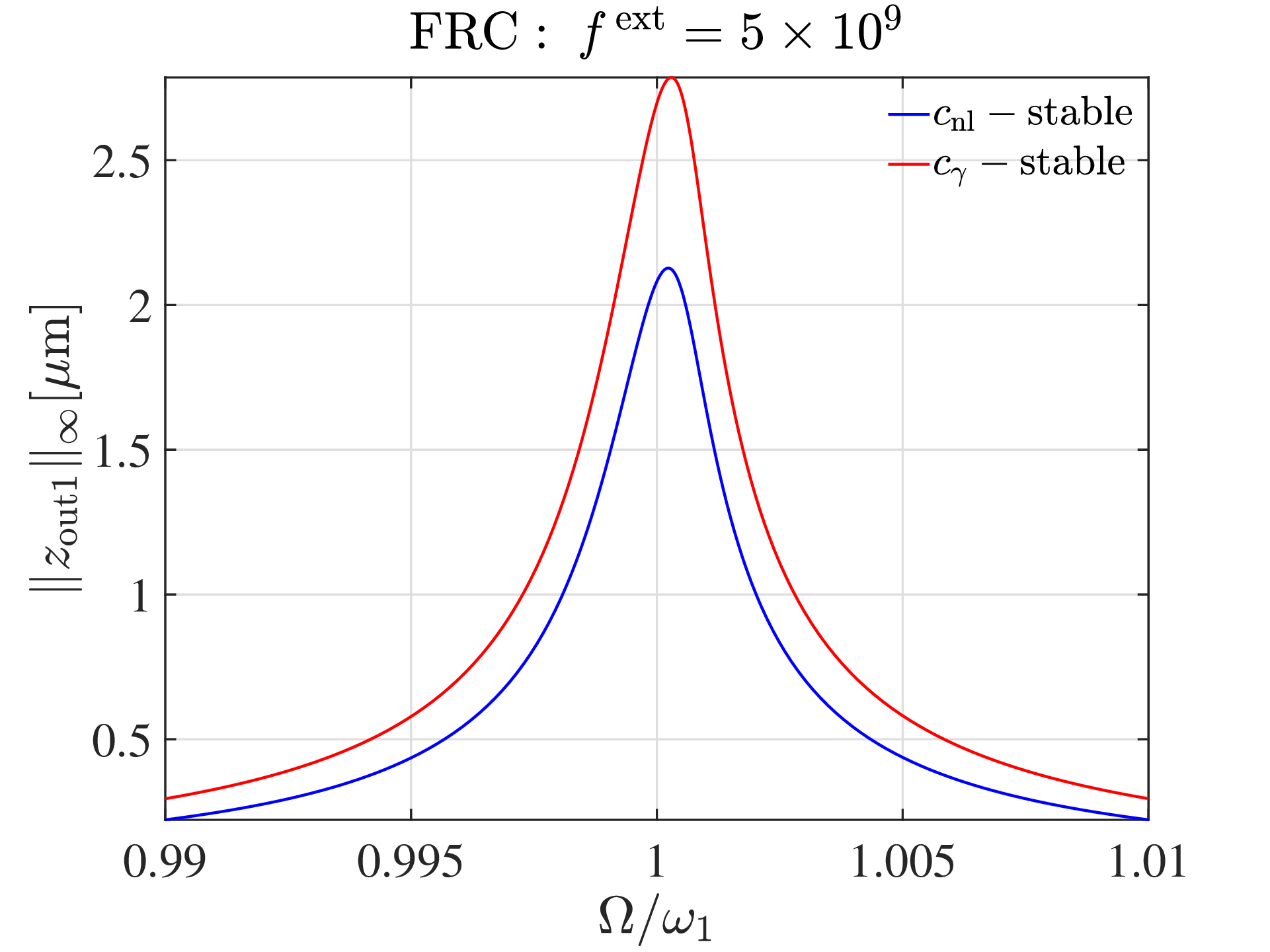} 
	\includegraphics[width=0.3\textwidth]{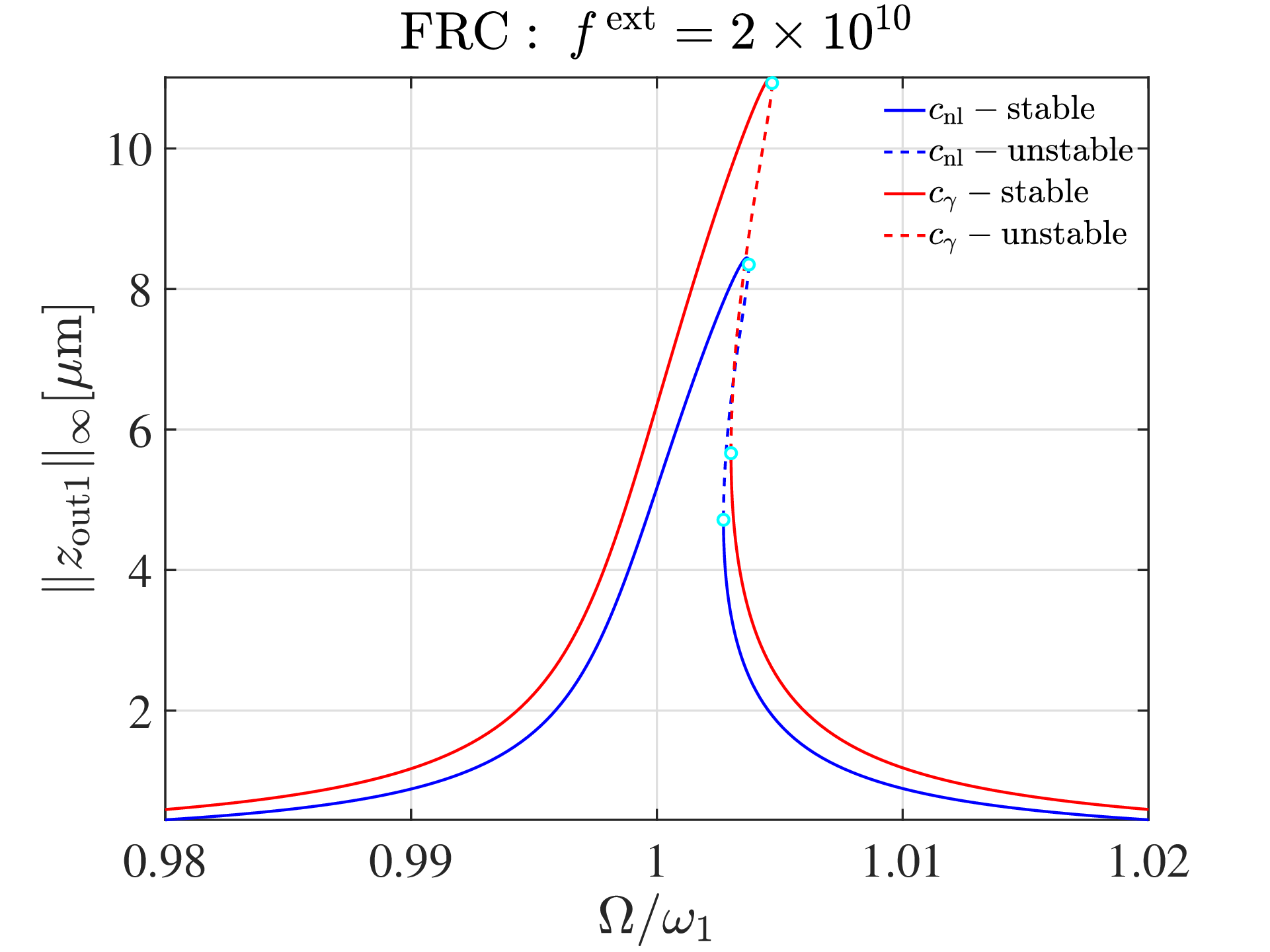} 
	\includegraphics[width=0.3\textwidth]{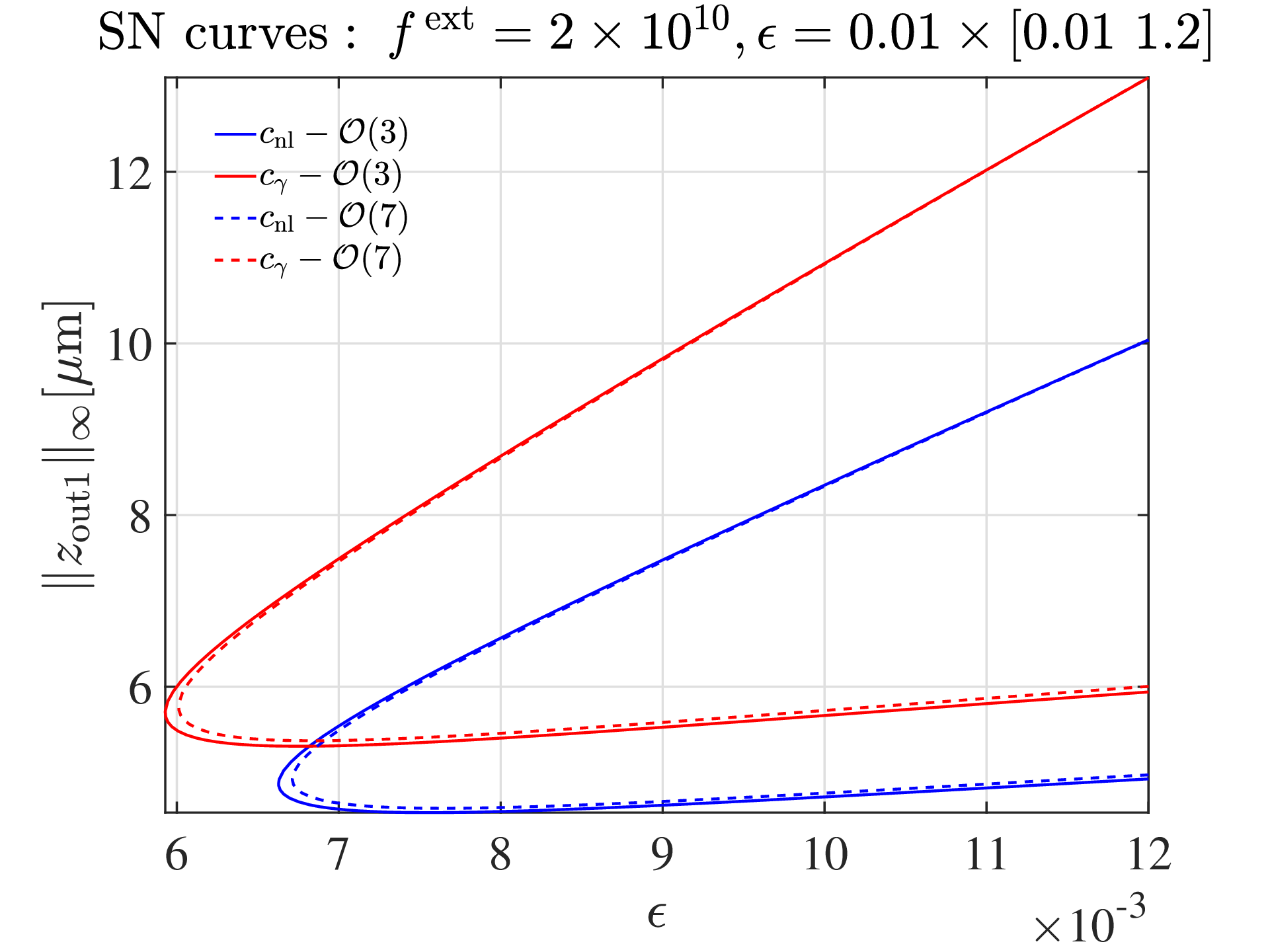} 
	\caption{Optimal results for $\gamma_{\mathrm{target}} = 1 \times 10^{-3}$. Top row: optimal layouts obtained from formulations~\eqref{eq:opt-gamma} (left panel) and~\eqref{eq:opt-rho_max} (right panel). 
Bottom row: FRCs of the optimal layouts under two levels of forcing amplitude: $f^{\mathrm{ext}} = 5 \times 10^{9}$ (left panel) and $2 \times 10^{10}$ (middle panel) $\mathrm{ng \cdot \mu m/ms^2}$, and SN bifurcation curves (right panel) computed using third-order and seventh-order SSM reductions.}
\label{fig:optHarden-gamma0001}
\end{figure}

At $\gamma_{\mathrm{target}} = 1 \times 10^{-3}$, the optimal layouts for $c_{\mathrm{nl}}$ in~\eqref{eq:opt-rho_max} and $c_{\gamma}$ in~\eqref{eq:opt-gamma} are shown in Fig.~\ref{fig:optHarden-gamma0001}. The optimal layouts of $c_{\mathrm{nl}}$ and $c_{\gamma}$ are both symmetrical structures with respect to the axis $y = 50$. Although both designs exhibit global symmetry, noticeable local differences can be observed between the two layouts. These differences reflect the influence of the chosen optimization objective on the final layout.

We then compare the FRCs of two optimal structures shown in Fig.~\ref{fig:optHarden-gamma0001} using \texttt{SSMTool}. The lower-left panel of Fig.~\ref{fig:optHarden-gamma0001} presents the FRCs computed under a forcing amplitude of $f^{\mathrm{ext}} = 5 \times 10^{9}$, where the vertical axis $\Vert z_{\rm{out1}} \Vert_\infty$ denotes the vibration amplitude at node ($x$, $y$) = (500, 50). At this excitation level, the FRC peak of  $c_\mathrm{nl}$ is lower than that of $c_{\gamma}$. Moreover, the hardening behavior of the FRCs is not prominent, as the forcing is too weak to activate significant nonlinear effects in the optimized layouts. Therefore, we increase the forcing amplitude to $f^{\mathrm{ext}} = 2 \times 10^{10}$ and recompute the FRCs for both designs. The resulting curves are shown in the lower-middle panel of Fig.~\ref{fig:optHarden-gamma0001}, where the FRC peak of $c_\mathrm{nl}$ remains lower than that of $c_{\gamma}$. Additionally, as the forcing amplitude increases, two SN bifurcation points are observed on the FRCs.

Since one of the SN bifurcation points lies close to the peak of the FRC, the variation of the FRC peak with respect to the forcing amplitude $\epsilon$ can be studied via the SN bifurcation curves. As seen in the lower-right panel of Fig.~\ref{fig:optHarden-gamma0001}, we allow for the variations in the forcing amplitude with $\epsilon = [0.01, 1.2]\epsilon_o$ with $\epsilon_o= 0.01$ to investigate how the SN bifurcations evolve with varying $\epsilon$. We can see that the variation amplitude of SN bifurcations of optimal layout $c_\mathrm{nl}$ is lower than that of $c_{\gamma}$. Consequently, as the forcing amplitude increases, the peak of FRC of $c_\mathrm{nl}$ remains lower than that of $c_{\gamma}$. To verify the convergence of the reduced-order model, we recomputed the SN bifurcation curves using an SSM reduction truncated at $\mathcal{O}(7)$. As shown in the same panel, the results confirm the convergence.  Finally, the computational performance for $\gamma_{\mathrm{target}} = 1 \times 10^{-3}$ is summarized in Table~\ref{tab:comp_perf_ex2_gamma1e-3}.~\textcolor{blue}{Here the stopping criterion for the optimization of $c_{\text{nl}}$ is the same as that of Sec 5.1. As for the optimization problem with objective $c_\gamma$, since the objective function $c_\gamma \approx 0$ during optimization iterations, it is not suitable for judgment to optimize convergence. Thus, we take satisfaction of all constraints and the convergence of the design variable as the judgment, i.e., $\max|{\bar{\mathbf{\mu}}^{(i)}-\bar{\mathbf{\mu}}^{(i-1)}}|<10^{-2}$. }

\begin{table}[H]
\centering
\caption{Computational performance of the case of $\gamma_{\mathrm{target}} = 1 \times 10^{-3}$ in Fig.~\ref{fig:optHarden-gamma0001}.}
\label{tab:comp_perf_ex2_gamma1e-3}
\begin{tabular}{ccccc}
\hline
\textbf{} & \textbf{Mesh size} & \textbf{Total time} & \textbf{Iterations} & \textbf{Efficiency} \\
\hline
 $c_\gamma$ & $100 \times 20$ & 35 mins & 250 & 8.4 s/step \\
\hline
 $c_{\text{nl}}$ & $100 \times 20$ & 21 mins & 154 & 8.0 s/step \\
 \hline
\end{tabular}
\end{table}

\subsubsection{Softening Optimization}

\begin{figure}[!ht]
	\centering
	\includegraphics[width=0.4\textwidth]{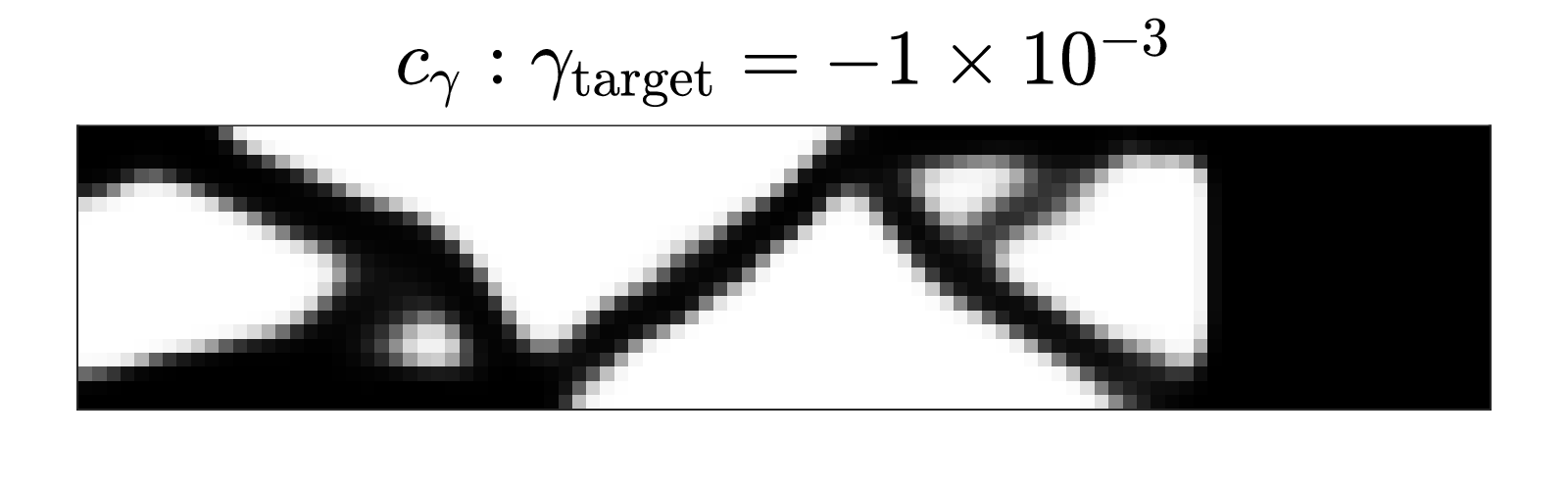}
	\includegraphics[width=0.4\textwidth]{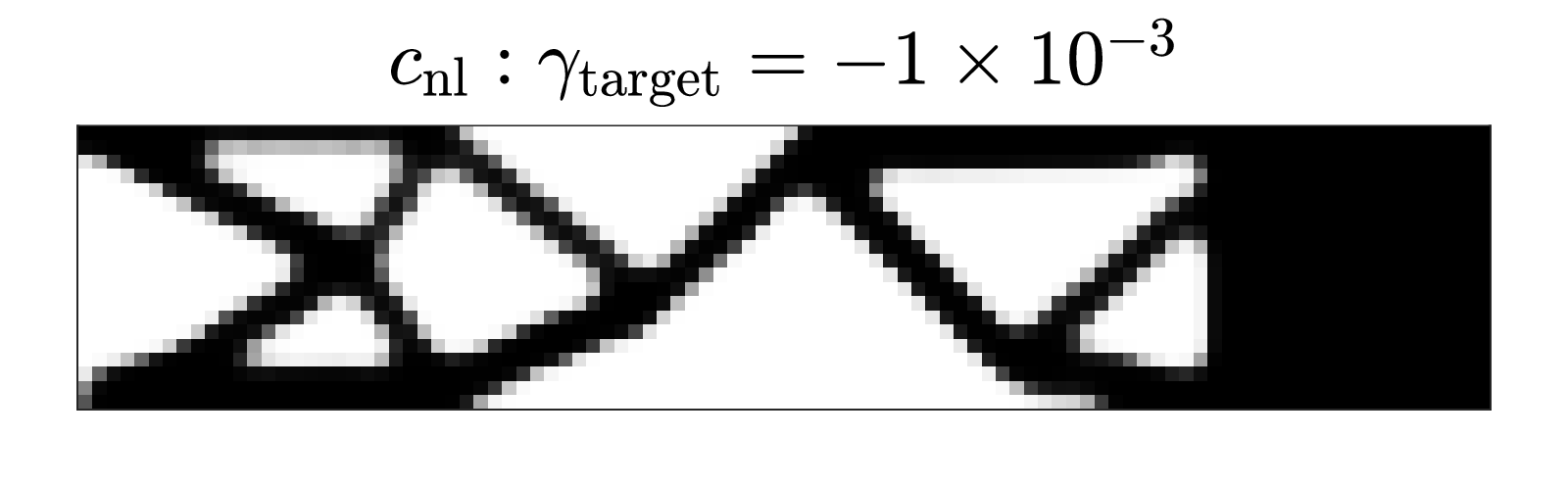} \\
	\includegraphics[width=0.3\textwidth]{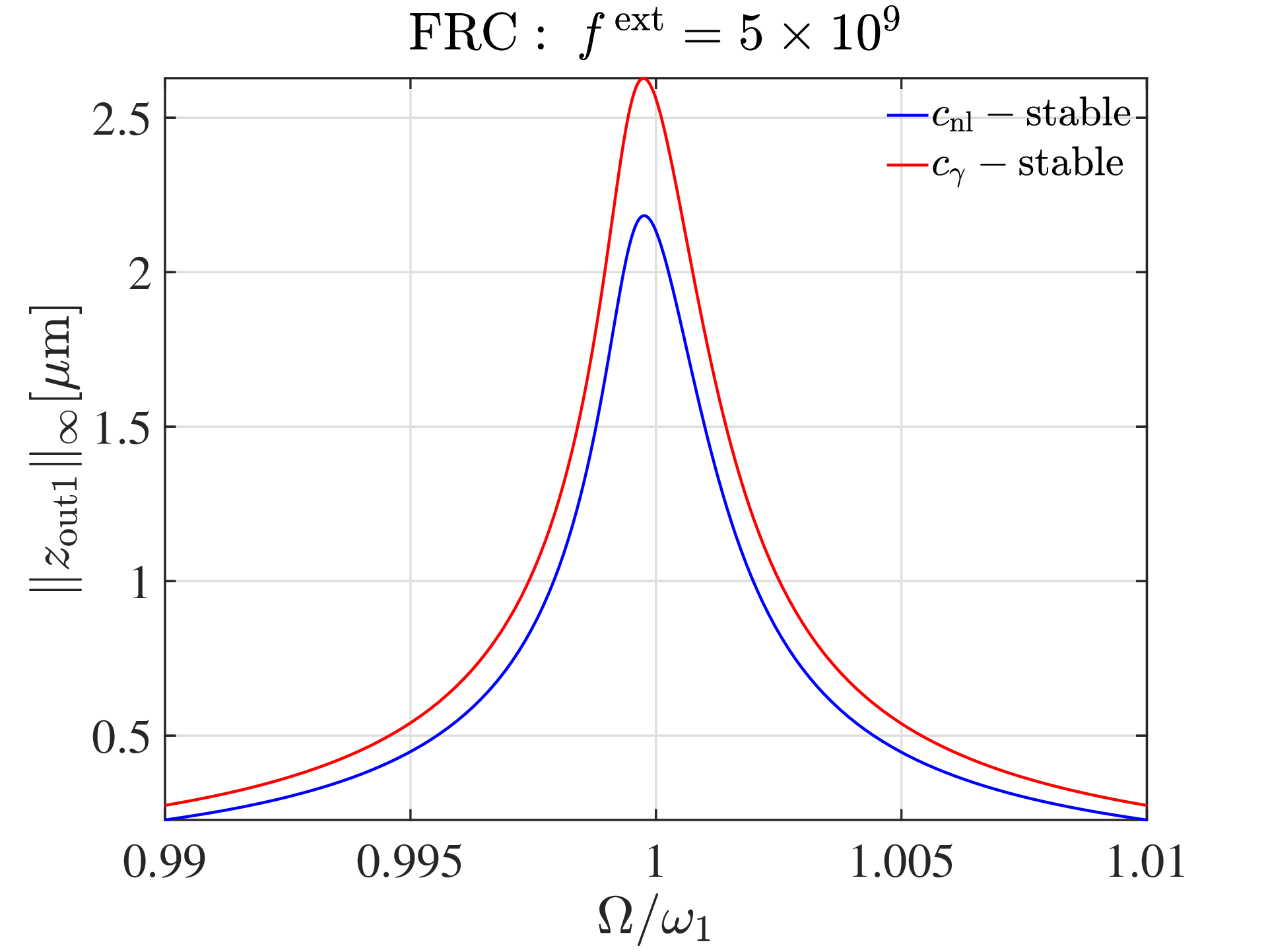} 
	\includegraphics[width=0.3\textwidth]{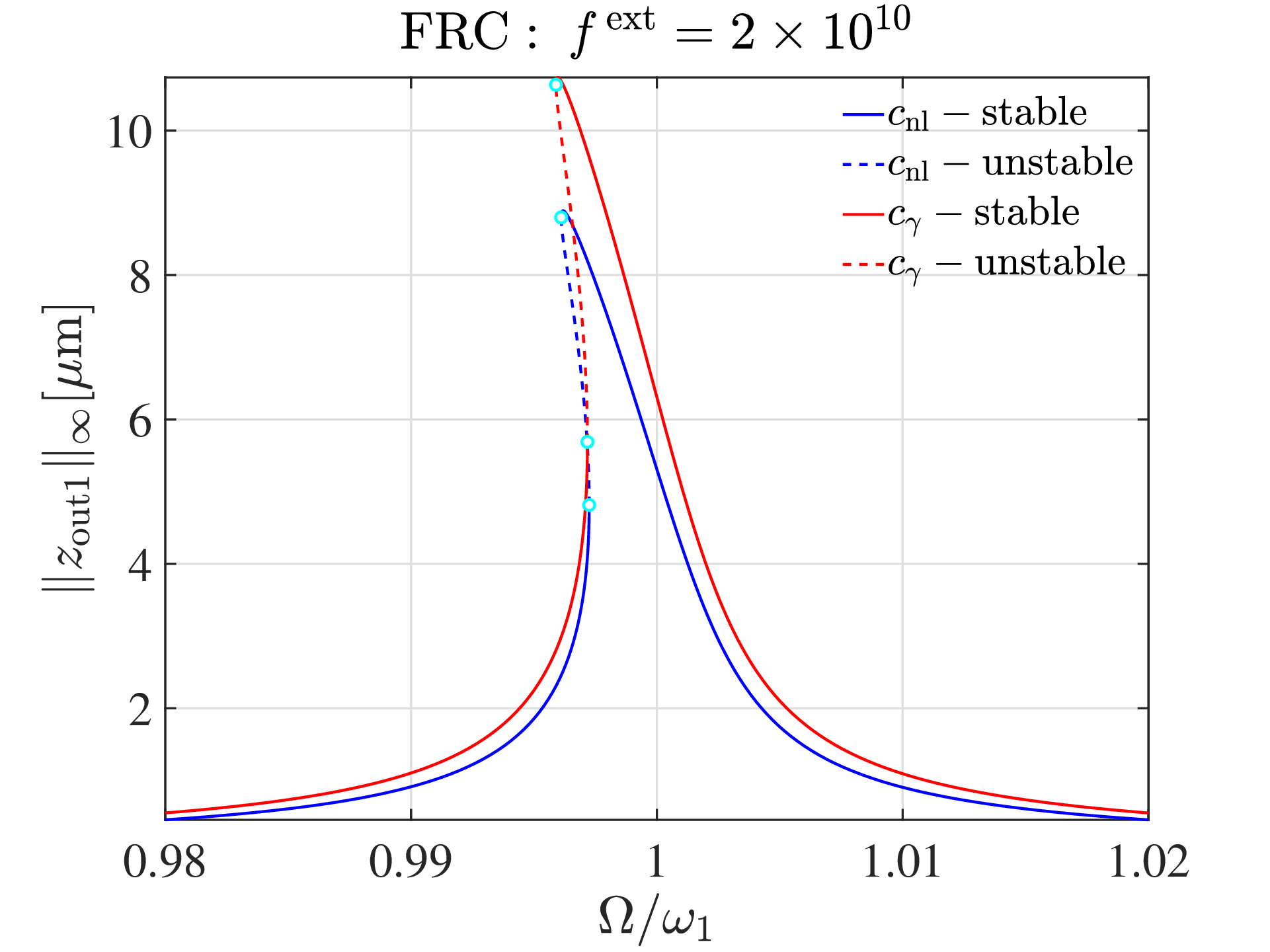} 
	\includegraphics[width=0.3\textwidth]{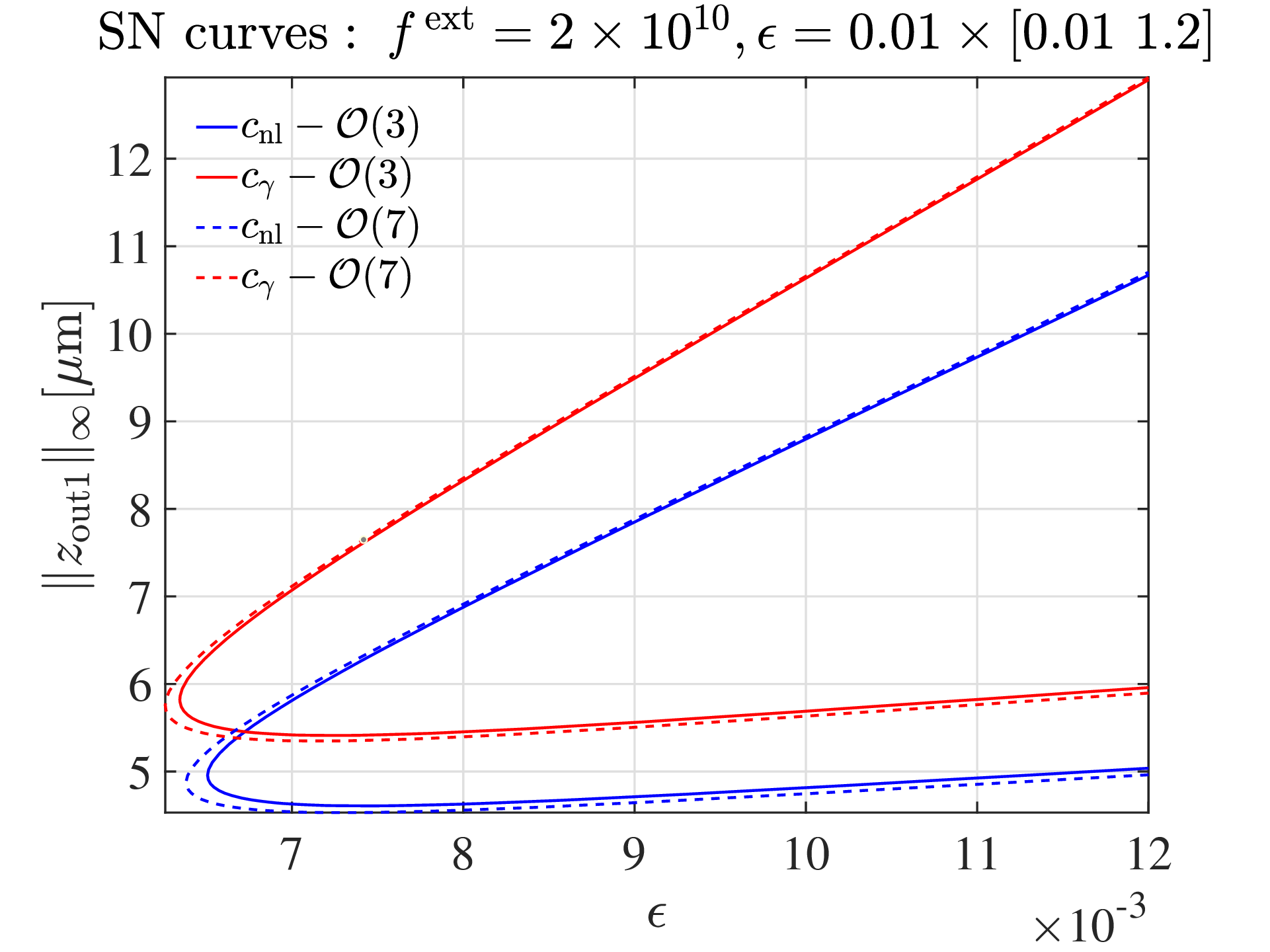} 
	\caption{Optimal results for $\gamma_{\mathrm{target}} = -1 \times 10^{-3}$. Top row: optimal layouts obtained from formulations~\eqref{eq:opt-gamma} (left panel) and~\eqref{eq:opt-rho_max} (right panel). 
Bottom row: FRCs of the optimal layouts under two levels of forcing amplitude: $f^{\mathrm{ext}} = 5 \times 10^{9}$ (left panel) and $2 \times 10^{10}$ (middle panel) $\mathrm{ng \cdot \mu m/ms^2}$, and SN bifurcation curves (right panel) computed using third-order and seventh-order SSM reductions.}
\label{fig:optSoften-gamma_0001}
\end{figure}

Now we consider the case of $\gamma_{\mathrm{target}}<0$, which corresponds to softening behavior of FRCs. At $\gamma_{\mathrm{target}} = -1 \times 10^{-3}$, the optimal layouts of the optimization problems~\eqref{eq:opt-rho_max} and~\eqref{eq:opt-gamma} are shown in Fig.~\ref{fig:optSoften-gamma_0001}. Unlike the case of hardening optimization, the optimal layouts of $c_{\gamma}$ and $c_\mathrm{nl}$ are both asymmetrical structures. Further, the optimal layout of $c_\mathrm{nl}$ is different from that of $c_{\gamma}$. 

We then compare the FRCs of these two optimal layouts. The FRCs under forcing amplitude $ f^{\mathrm{ext}}= 5 \times 10^{9}$ are shown in the lower-left panel of Fig~\ref{fig:optSoften-gamma_0001}, from which we can see that the peak of FRC of $c_\mathrm{nl}$ is lower than that of $c_{\gamma}$. To observe the softening behavior more clearly, we increase the forcing amplitude to $f^{\mathrm{ext}} = 2 \times 10^{10}$. As seen in the lower-middle panel of Fig~\ref{fig:optSoften-gamma_0001}, the peak of FRC of $c_\mathrm{nl}$ is still lower than that of $c_{\gamma}$. Also, the softening behavior can be observed clearly for both optimal layouts. Further, two SN bifurcation points emerge as the forcing amplitude increases.

Similar to the case of hardening optimization, we perform parameter continuation of the SN bifurcation points with varying forcing amplitude $\epsilon = [0.01,1.2]\epsilon_o$. The projections of the obtained SN bifurcation curves are shown in the lower-right panel of Fig~\ref{fig:optSoften-gamma_0001}, from which we can see that the vibration amplitude of SN bifurcation points of $c_\mathrm{nl}$ is lower than that of $c_{\gamma}$. Therefore, the optimized structure $c_\mathrm{nl}$ demonstrates consistently enhanced resistance to external disturbances compared to $c_{\gamma}$ under various forcing amplitudes. Finally, we examine the computational performance for the case of $\gamma_{\mathrm{target}} = -1 \times 10^{-3}$, as summarized in Table~\ref{tab:comp_perf_ex2}.~\textcolor{blue}{The stopping criteria for $c_\gamma$ and $c_{\text{nl}}$ here are the same as that of Table~\ref{tab:comp_perf_ex2_gamma1e-3}}.

\begin{table}[H]
\centering
\caption{Computational performance of the case of $\gamma_{\mathrm{target}} = -1 \times 10^{-3}$ in Fig.~\ref{fig:optSoften-gamma_0001}.}
\label{tab:comp_perf_ex2}
\begin{tabular}{ccccc}
\hline
\textbf{} & \textbf{Mesh size} & \textbf{Total time} & \textbf{Iterations} & \textbf{Efficiency} \\
\hline
 $c_\gamma$ & $100 \times 20$ & 29 mins & 203 & 8.5 s/step \\
\hline
 $c_{\text{nl}}$ & $100 \times 20$ & 20 mins & 154 & 7.9 s/step \\
 \hline
\end{tabular}
\end{table}

\textcolor{blue}{Similar to Sec.~\ref{ssec: example1}, we find that modes are well separated such that neither modal collisions nor internal resonances occur in this example. Detailed results and analysis are shown in Appendix~\ref{appD3: test_mode_collisions_ex2}. In addition, results in the appendix show that the initial and optimized structures exhibit similar linear modal shapes for the first mode, which again indicates that the consistency of master mode is ensured in the optimization process.}

\textcolor{blue}{The optimization results presented in this subsection show that the optimal layouts consistently exhibit hardening behavior when $\gamma_{\text{target}} > 0$ and softening behavior when $\gamma_{\text{target}} < 0$. This indicates that the proposed optimization framework can steer the nonlinear dynamic response of the structure toward a hardening/softening behavior. Further, a comparison with the reference formulation in Eq.~\eqref{eq:opt-gamma} shows that Eq.~\eqref{eq:opt-rho_max} achieves concurrent minimization of the peak amplitude of FRC and the regulation of backbone curve.}

\subsection{Tuning saddle-node (SN) bifurcations on FRC}
\label{ssec: control_SN}

\begin{figure}[!ht]
\centering
\includegraphics[width=0.5\textwidth]{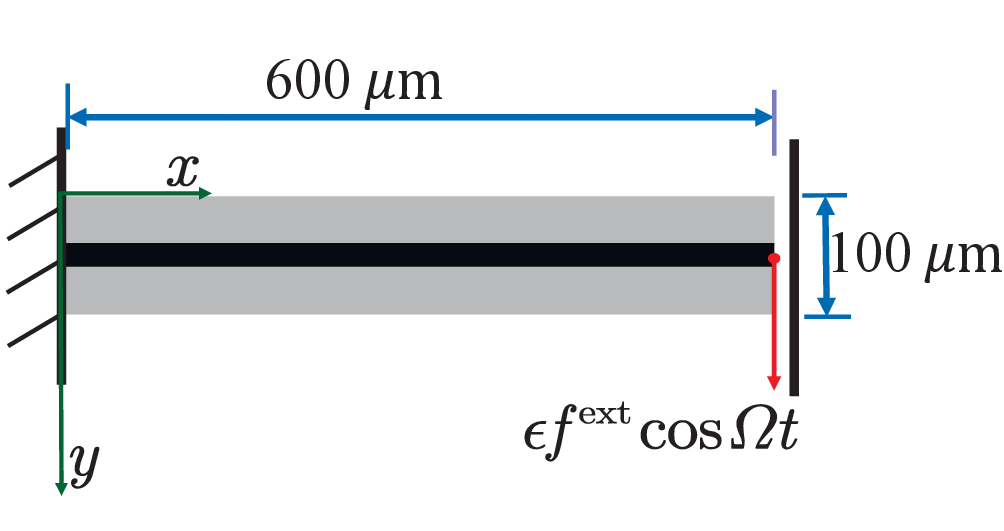}
\caption{Initial layout of the half microbeam considered in the example of Sec.~\ref{ssec: control_SN}. The total domain is 600~$\mathrm{\mu m}$ long and 100~$\mathrm{\mu m}$ high, and includes a non-design region of \textcolor{blue}{20~$\mathrm{\mu m}$ in height located at the center of the domain.} The gray area represents the designable region, while the black area indicates the fixed non-design region.}
\label{fig:initial-microbeam}
\end{figure}

The final example is a microbeam in a mass sensor~\cite{sun_new_2025}. The goal is to maximize the vibration amplitude of the microbeam to improve its measurement accuracy. Meanwhile, to prevent excessive amplitude fluctuations, we control the SN bifurcation points on the FRC to maximize the region without bistability. As seen in Fig.~\ref{fig:initial-microbeam}, the design domain of the beam has a length of 600~$\mathrm{\mu m}$ and a height of 100~$\mathrm{\mu m}$. A fixed region in the middle accounts for 20\% of the design domain.  The left end of the beam is fixed, while the right end is constrained in the $x$-direction. A periodic excitation is applied at the center of the right edge.

Similar to the case of the MBB beam, a mesh of $120 \times 20$ elements is used in the FE model, leading to a system with $n$= 5019 DOF in~\eqref{eq:eom-second-full}. The damping ratio of initial layout $\xi$ = 0.1\%, which leads to the values of damping constants $\alpha = 1.97$ and $\beta = 2.77 \times 10^{-7}$. During the optimization iteration process, the forcing amplitude is given as $ f^{\mathrm{ext}}= 2 \times 10^{8} \ \mathrm{ng \cdot \mu m/ms^2}$. In the following computation, we set $\epsilon = 0.01$. \textcolor{blue}{The projection parameter is initialized at $\sigma_0=10$ and multiplied by 1.5 every 60 iterations until it reaches a maximum value of $\sigma_{\max}=50$, while the parameter $p$ is initialized at $p_0=1$ and increased by one every 60 iterations, up to a maximum value of $p_{\max}=8$. The target values for constraints in formulation~\eqref{eq:opt-SNbifur} are set as $\omega_{Y,\text{target}} = 100~\mathrm{kHz}$, $\omega_{X,\text{target}} = 500~\mathrm{kHz}$, and $A_{\mathrm{target}} = 40\%$.} Based on these parameter settings, we present the optimization result of optimization formulation~\eqref{eq:opt-SNbifur} with different values of \textcolor{blue}{$d_{\mathrm{target}}$}. 

\begin{figure}[!ht]
\centering
\includegraphics[width=0.3\textwidth]{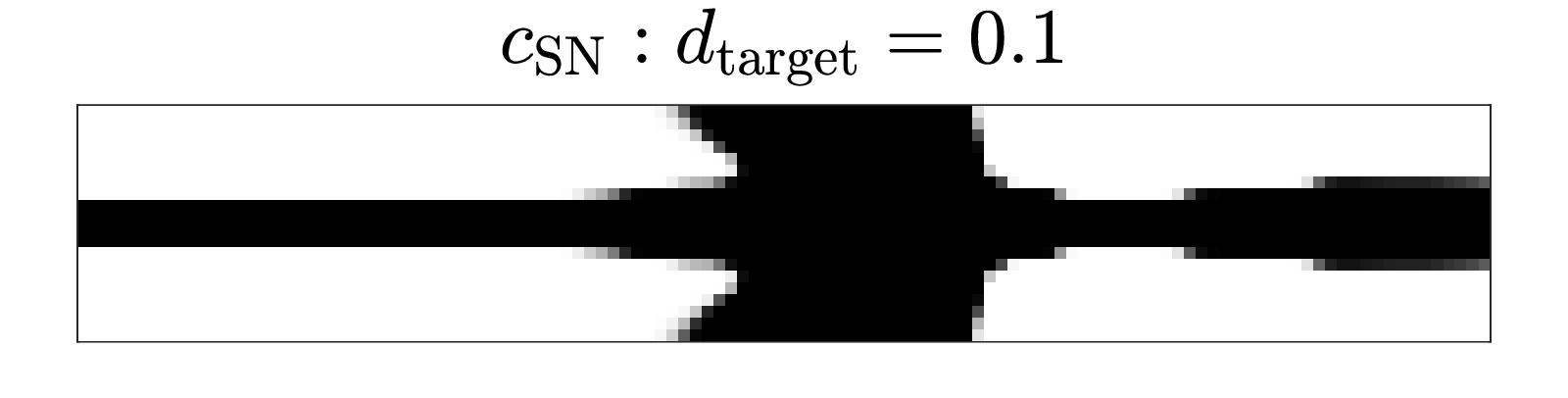}
\includegraphics[width=0.3\textwidth]{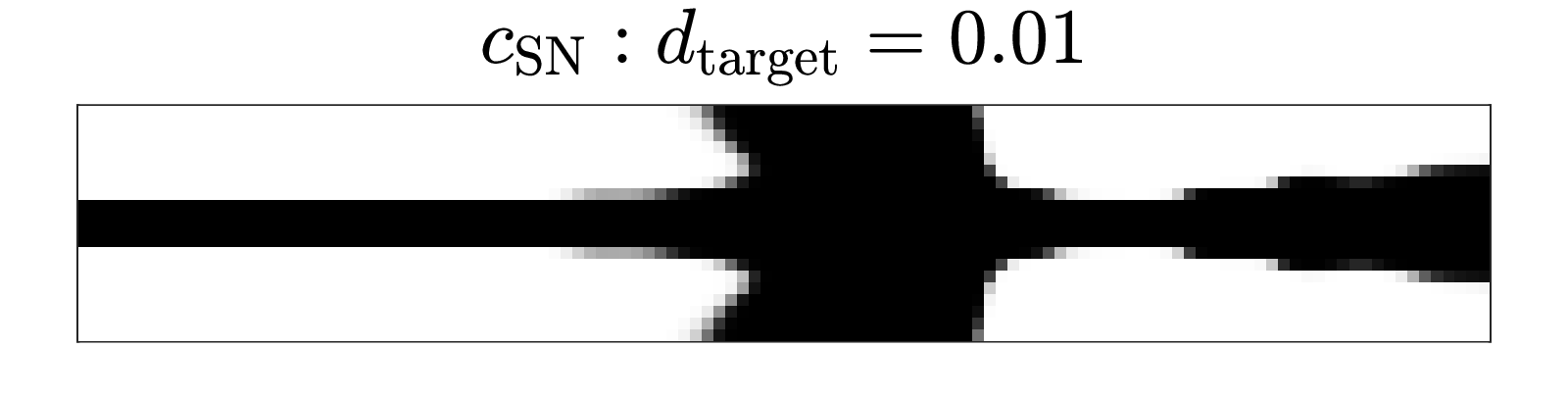} 
\includegraphics[width=0.3\textwidth]{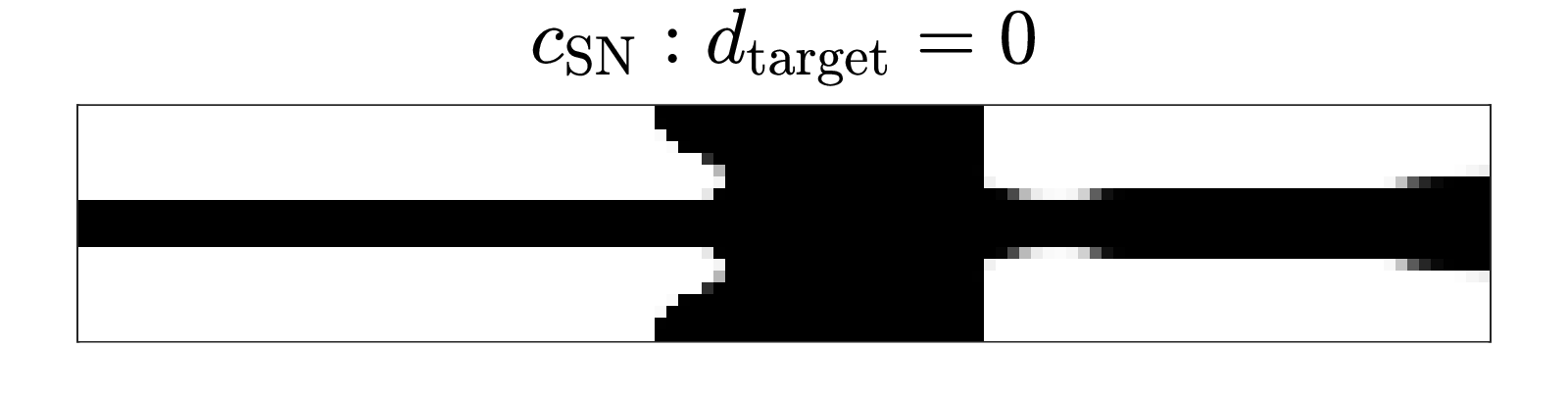} \\
\includegraphics[width=0.3\textwidth]{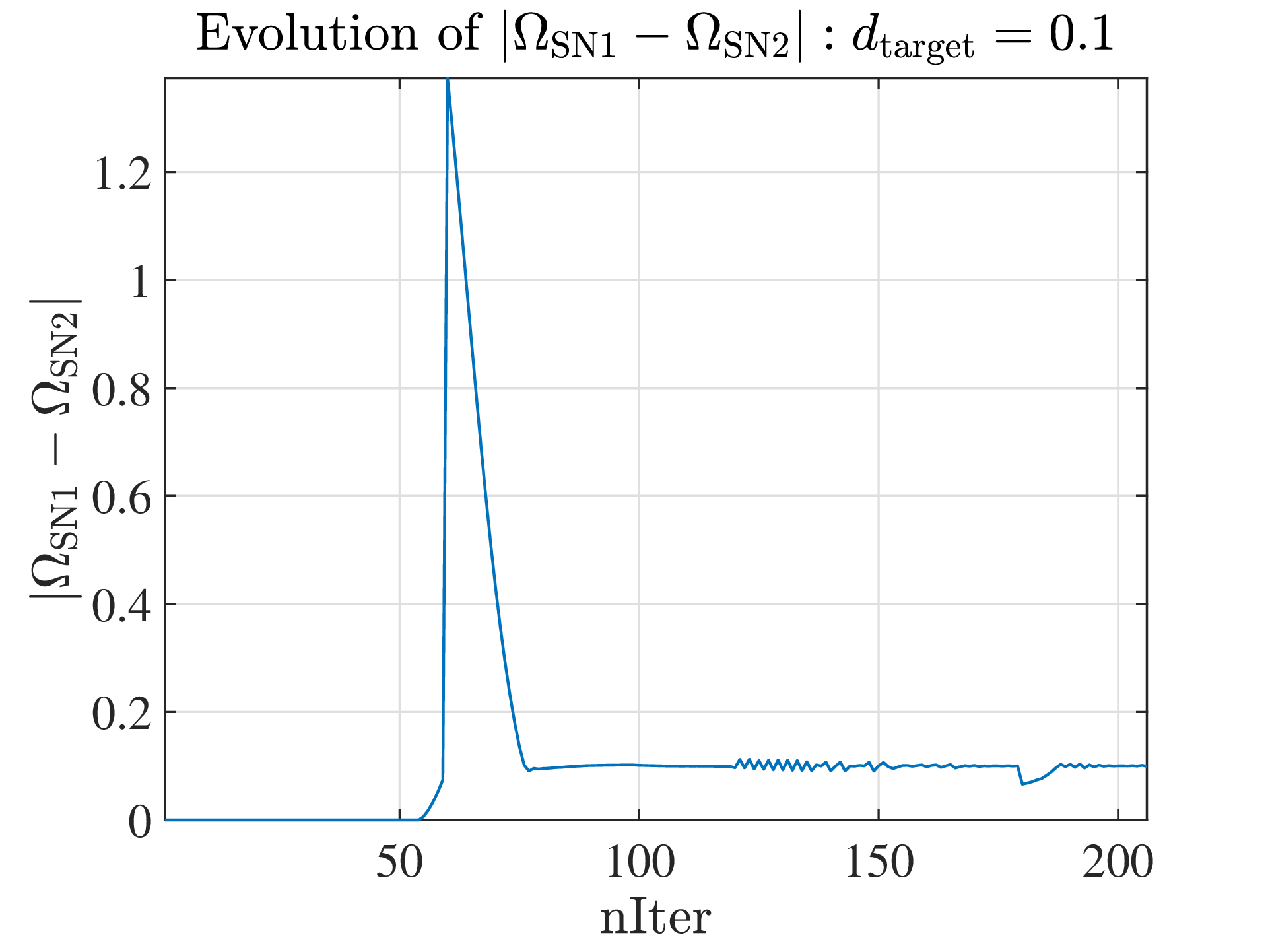} 
\includegraphics[width=0.3\textwidth]{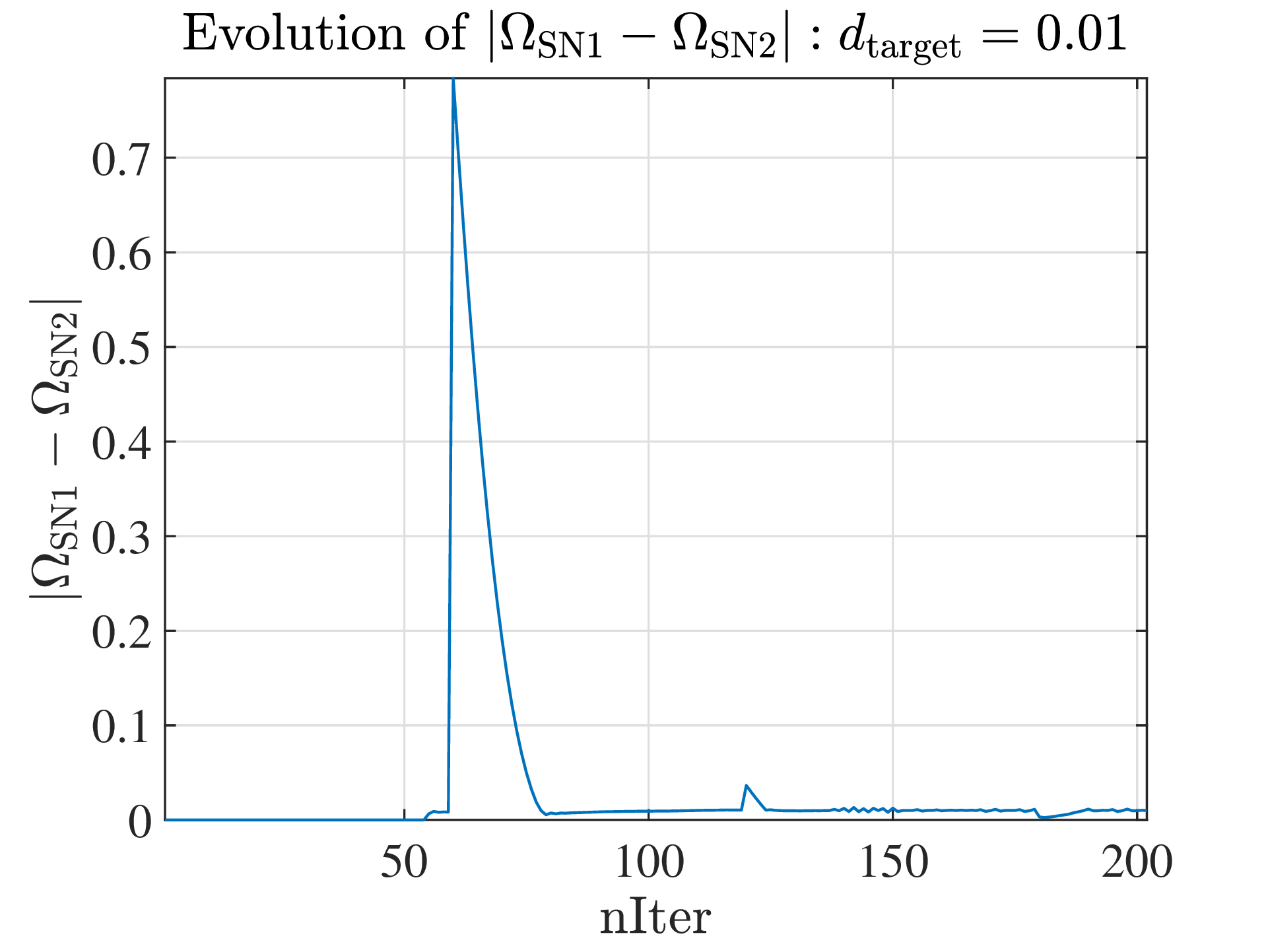} 
\includegraphics[width=0.3\textwidth]{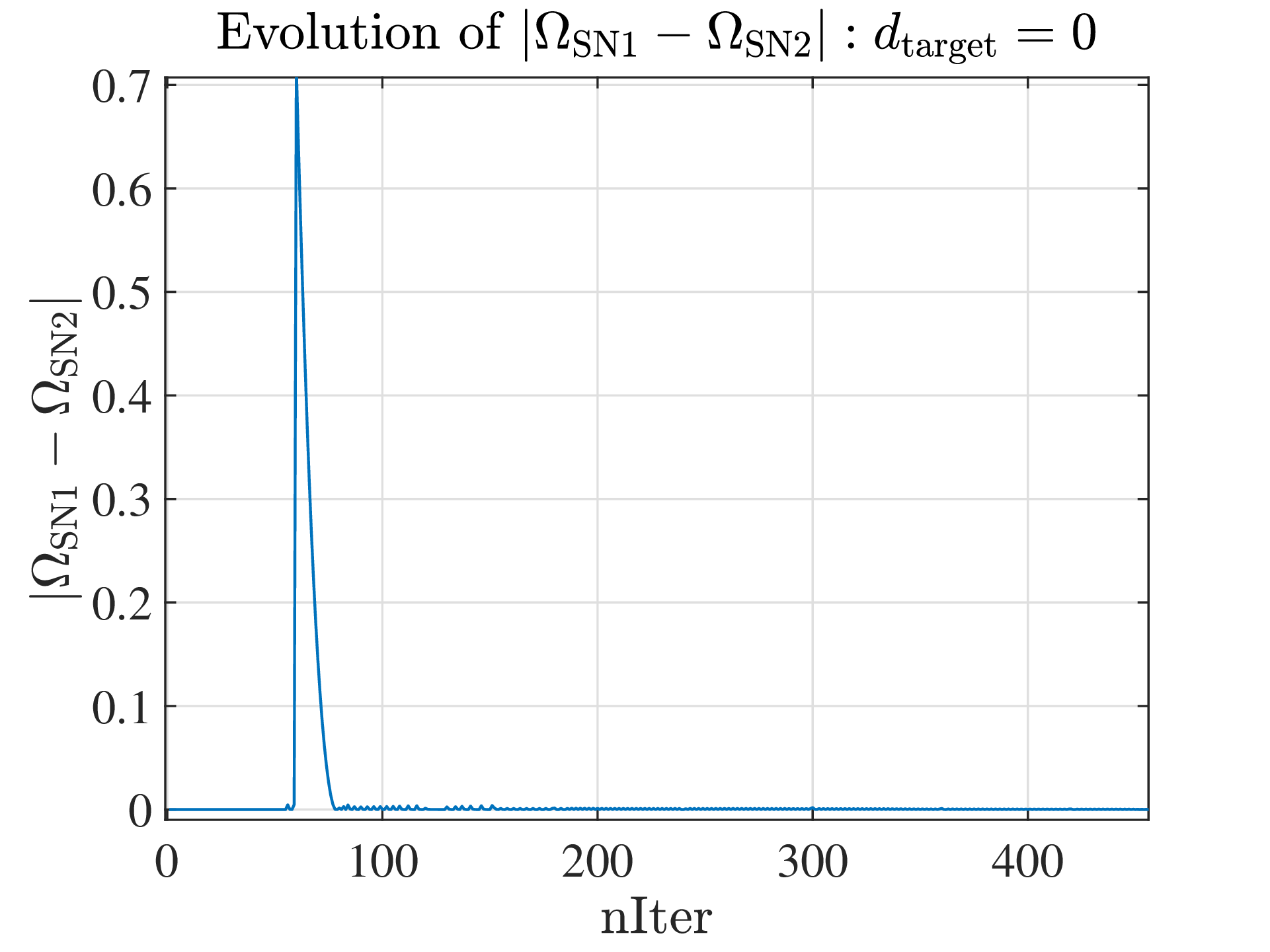} \\
\includegraphics[width=0.3\textwidth]{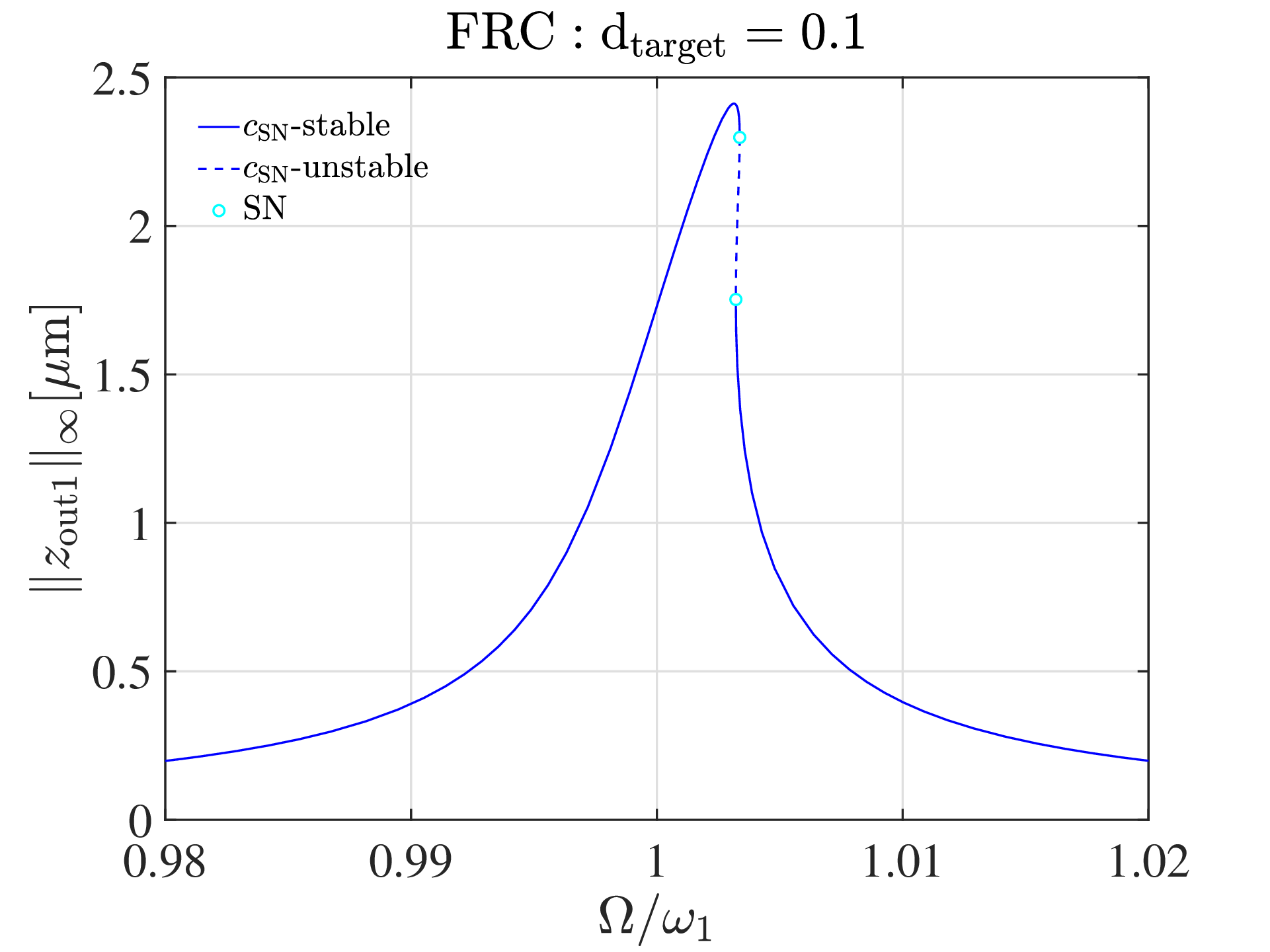} 
\includegraphics[width=0.3\textwidth]{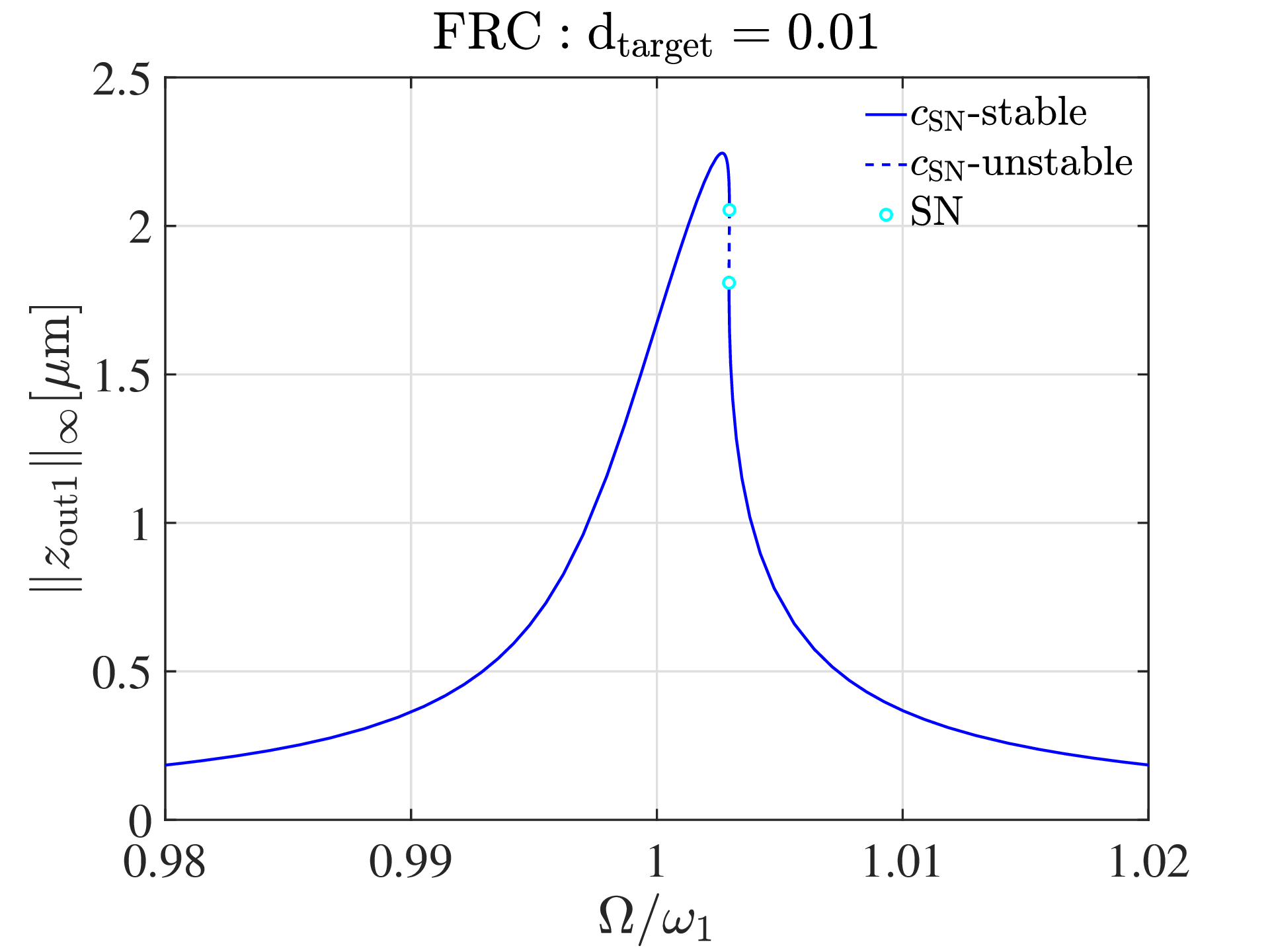} 
\includegraphics[width=0.3\textwidth]{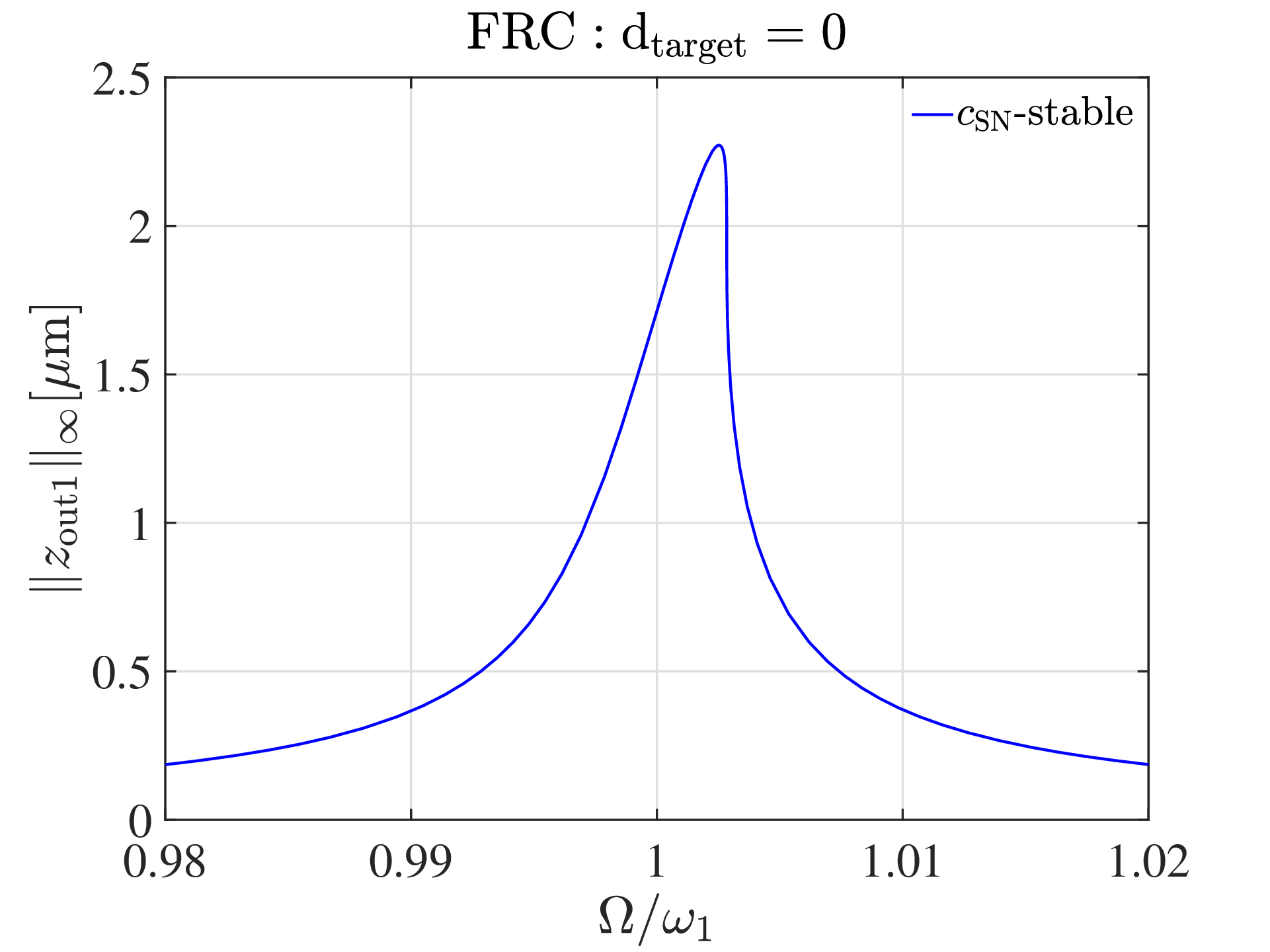} 
\caption{Optimization results of formulation~\eqref{eq:opt-SNbifur} with different values of \textcolor{blue}{$d_{\mathrm{target}}$: 0.1 (left column), 0.01 (middle column), and 0 (right column)}. Top row: Optimal layouts. Middle row: Evolution of \textcolor{blue}{$|\Omega_{\text{SN1}}-\Omega_{\text{SN2}}|$} during the optimization iterations. Bottom row: FRCs of the optimal layouts computed using third-order SSM reduction.}
\label{fig:optSN-layoutFRC}
\end{figure}

We consider three cases \textcolor{blue}{$d_{\mathrm{target}} \in \{0.1,0.01,0\}$}. The obtained results are shown in Fig.~\ref{fig:optSN-layoutFRC}. \textcolor{blue}{From the first row, we observe that the optimal layouts evolve with decreasing $d_{\mathrm{target}}$, exhibiting noticeable changes in local shape features}. The second row shows the variation of \textcolor{blue}{$|\Omega_{\text{SN1}}-\Omega_{\text{SN2}}|$} during the iterative process. At the beginning of the iteration, the value of \textcolor{blue}{$|\Omega_{\text{SN1}}-\Omega_{\text{SN2}}|$} is 0, which indicates that the SN bifurcation point has not yet occurred. As the number of iteration steps increases, the vibration amplitude of the system becomes larger, the SN bifurcation points emerge, the value of \textcolor{blue}{$|\Omega_{\text{SN1}}-\Omega_{\text{SN2}}|$} is greater than 0, and it finally tends to the value of \textcolor{blue}{$d_{\mathrm{target}}$}. Further, when \textcolor{blue}{$d_{\mathrm{target}} = 0$}, more iterations are required for the convergence of \textcolor{blue}{$|\Omega_{\text{SN1}} - \Omega_{\text{SN2}}|$}. \textcolor{blue}{The abrupt change of $|\Omega_{\text{SN1}} - \Omega_{\text{SN2}}|$ observed during the iteration process is due to the update of the penalization exponent $p$ and the projection parameter $\sigma$}. Finally, we consider the FRCs of the three optimal layouts, as shown in the third row of Fig.~\ref{fig:optSN-layoutFRC}. In these plots, the vertical axis $\Vert z_{\rm{out1}} \Vert_\infty$ denotes the vibration amplitude at the node located at $(x, y) = (600, 50)$. We observe that two SN bifurcation points get closer as the value of \textcolor{blue}{$d_{\mathrm{target}}$} decreases.

\begin{figure}[!ht]
\centering
\includegraphics[width=0.4\textwidth]{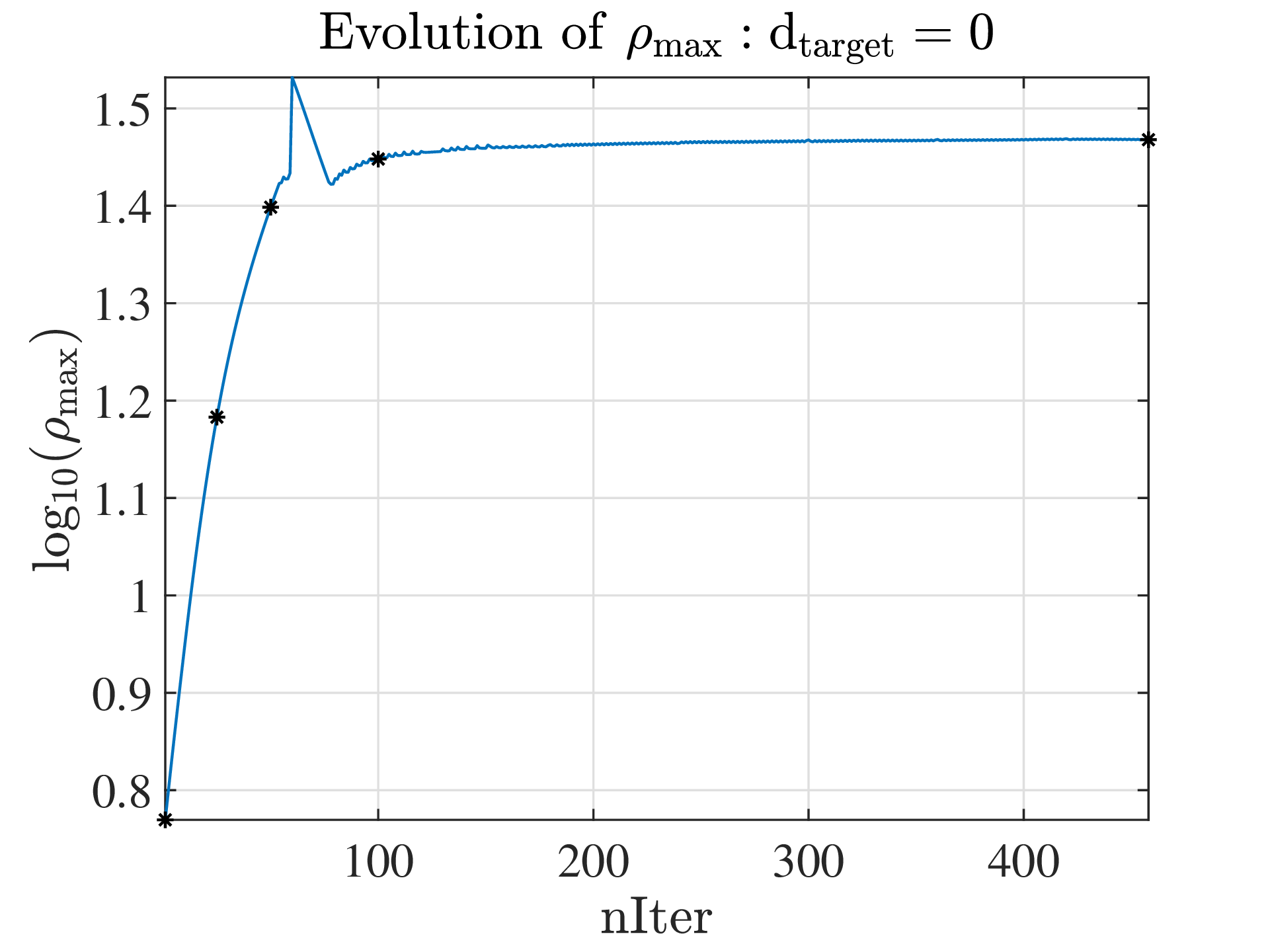}
\includegraphics[width=0.4\textwidth]{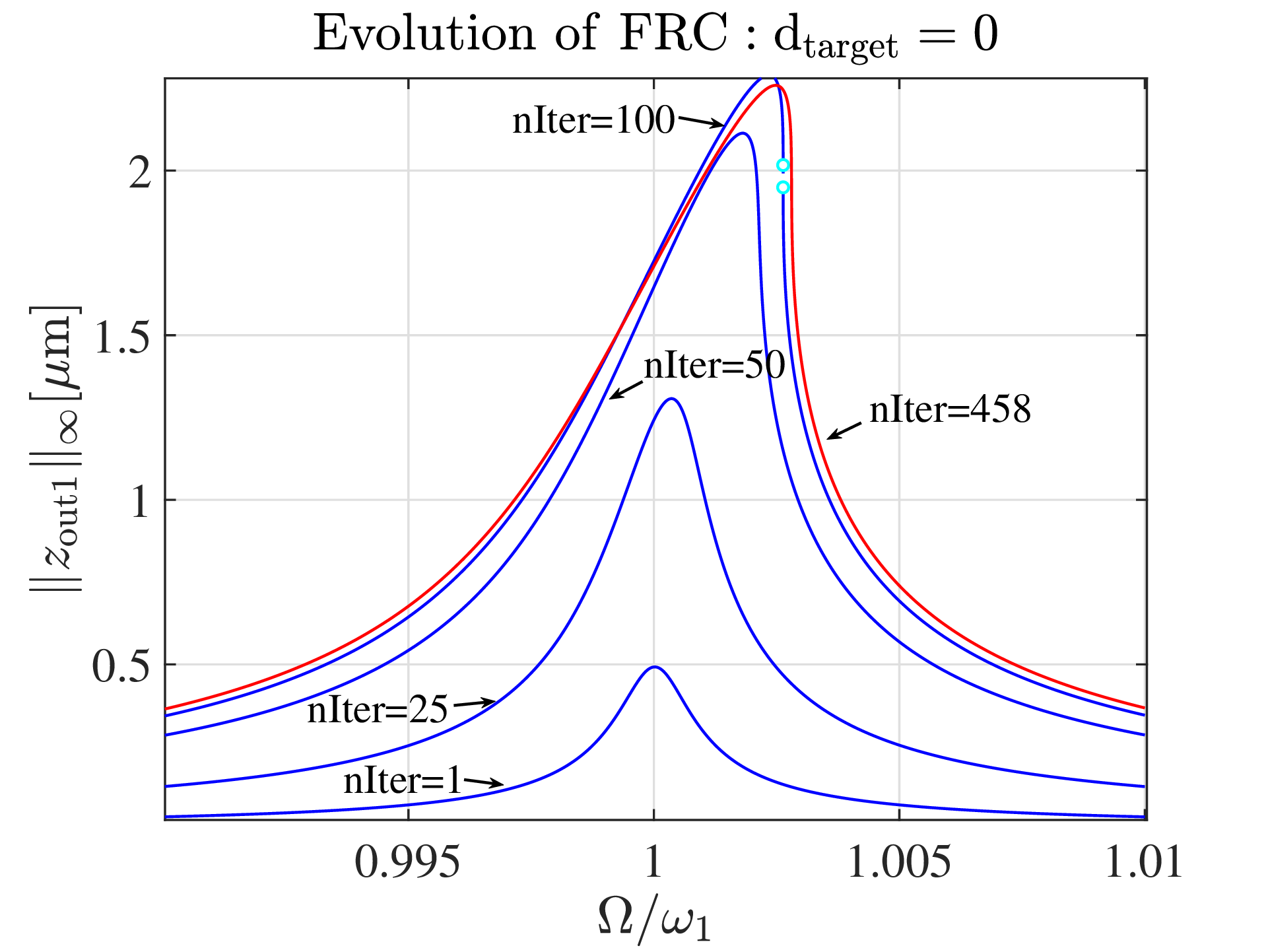} 
\caption{Evolution of the optimization process for \textcolor{blue}{$d_{\mathrm{target}}$ = 0}. Left panel: convergence history of the objective function $\rho_{\max}$. Right panel: corresponding evolution of the FRCs over iterations.}
\label{fig:optSN-b005}
\end{figure}

Now we present a close look at the iterative process for the case of \textcolor{blue}{$d_{\mathrm{target}}$ = 0}. The evolution of the objective function $\rho_{\max}$ in optimization formulation~\eqref{eq:opt-SNbifur} is shown in the left panel of Fig.~\ref{fig:optSN-b005}. It can be observed that the value of the objective function increases during the initial iterations and gradually stabilizes as the optimization progresses. The corresponding evolution of the FRCs is shown in the right panel of Fig.\ref{fig:optSN-b005}, where the peak amplitude of the FRCs also increases with the number of iterations. The computational performance of the above iterative process is summarized in Table~\ref{tab:comp_perf_ex3}.~\textcolor{blue}{Here, the stopping criterion for the optimization of $c_{\text{SN}}$ is similar to that of $c_{\text{nl}}$. Specifically, the stopping criterion is defined by the simultaneous satisfaction of all constraints and the steady convergence of the objective function $c_{\text{SN}}$.}

\begin{table}[H]
\centering
\caption{Computational performance of the iterative process in Fig.~\ref{fig:optSN-b005}}
\label{tab:comp_perf_ex3}
\begin{tabular}{cccc}
\hline
\textbf{Mesh size} & \textbf{Total time} & \textbf{Iterations} & \textbf{Efficiency} \\
\hline
$120 \times 20$ & \textcolor{blue}{53 mins} & \textcolor{blue}{458} & \textcolor{blue}{7.0 s/step} \\
\hline
\end{tabular}
\end{table}

Similar to the previous two examples, it is necessary to examine the convergence of the SSM-based model reduction. In Appendix~\ref{appA: SSM_order_test_ex3}, we compute the FRC of the optimized structure shown in Fig.~\ref{fig:optSN-layoutFRC} using a seventh-order SSM-reduced model. The results indicate that the third-order SSM model has already achieved convergence.

\textcolor{blue}{To justify the assumption of no internal resonance, we present in Appendix~\ref{appE3: test_mode_collisions_ex3} the evolution of the first three eigenfrequencies throughout the iterative process for the case of $d_{\mathrm{target}} = 0$. From these results, we observe that modes are well separated and neither modal collisions nor internal resonance occur. In addition, the consistency of master mode during optimization is again validated in Appendix~\ref{appE3: test_mode_collisions_ex3}.}

\textcolor{blue}{The optimization results presented in this subsection indicate that decreasing the prescribed value of $d_{\mathrm{target}}$ in the optimization formulation leads to a reduced frequency distance between the two SN bifurcations on the FRC. This observation demonstrates the effectiveness of the SSM-based framework in regulating the length of segment of unstable periodic orbits on FRC.}

\section{Conclusion}
\label{section5}

In this paper, we have performed topology optimization to tune the forced response curves (FRCs) of high-dimensional nonlinear systems. By employing reduction on spectral submanifolds (SSMs), we have enabled efficient analysis and sensitivity computation of FRCs for large-scale finite element models. Several optimization objectives were considered, including minimizing the FRC peak, tailoring hardening/softening behavior, and tuning saddle-node (SN) bifurcations on FRC. 

To tune SN points on the FRC, we have derived an explicit expression for the coefficient governing the distance between two SN bifurcation points, along with its sensitivity, which allows for the effective control of jump phenomena in nonlinear MEMS devices. Numerical examples demonstrate the effectiveness of the proposed approach in designing structures with desired dynamic characteristics. This framework offers a practical approach for incorporating nonlinear dynamic behavior into topology optimization, with potential applications in devices such as MEMS sensors and energy harvesters. 

In future work, we aim to extend this framework to the control of internal resonances, which often arise in nonlinear systems and significantly influence complex dynamic responses. In particular, internal resonances may induce Hopf bifurcations, which lead to quasi-periodic motions on invariant tori~\cite{liang_bifurcation_2025,li_nonlinear_2022_part2}. We plan to explore how such modal interactions can be systematically controlled by manipulating reduced-order dynamics, including the regulation of Hopf bifurcations and the resulting quasi-periodic responses, to enable advanced control of multi-frequency behaviors in nonlinear structures. \textcolor{blue}{The proposed framework can also be extended to realistic MEMS devices by incorporating anchors, electrodes, and actuation gaps, implemented through non-design regions, fixed boundary conditions, and appropriate optimization constraints. We leave this extension to future study.}

\section*{Acknowledgements}
HL and ML acknowledge the financial support of the National Natural Science Foundation of China (No. 12302014) and State Key Laboratory of Structural Analysis, Optimization and CAE Software for Industrial Equipment (GZ24117).
MP and JM acknowledge the financial support of STMicroelectronics (award number 4000614871).

\section*{Data availability}
The data used to generate the numerical results included in this paper are available from the corresponding author on request.

\section*{Conflict of interest}
The authors declare that they have no conflict of interest.

\appendix

\section{The sensitivity of $\tilde{f}$}
\label{appA: sen_f}

We begin by listing all relevant constraints
\begin{equation}
	\begin{aligned}
		\tilde{f}-0.5\boldsymbol{\psi}^\mathrm{T}\boldsymbol{f}^{\mathrm{ext}}=0, \\
		\boldsymbol{K}\boldsymbol{\phi}=\omega^2\boldsymbol{M}\boldsymbol{\phi},
		\boldsymbol{\phi}^\mathrm{T}\boldsymbol{M}\boldsymbol{\phi}=1, \\
		\boldsymbol{\psi}=\kappa\boldsymbol{\phi}, \\
		\kappa \omega=\frac{-\mathrm{i}}{2\sqrt{1-\xi^2}}, \\
		2\xi\omega=\alpha+\beta\omega^2.	
	\end{aligned}
\end{equation}
Now we define a Lagrangian as
\begin{equation}
	\begin{aligned}
		L=&\tilde{f} + \eta_{\tilde{f}}(\tilde{f}-0.5\boldsymbol{\psi}^\mathrm{T}\boldsymbol{f}^{\mathrm{ext}})  
		+\eta_{\boldsymbol{\phi}}(\boldsymbol{K}\boldsymbol{\phi}-\omega^2\boldsymbol{M}\boldsymbol{\phi})+\eta_{\mathrm{norm}}(\boldsymbol{\phi}^\mathrm{T}\boldsymbol{M}\boldsymbol{\phi}-1)\\
		&+\eta_{\boldsymbol{\psi}}(\boldsymbol{\psi}-\kappa\boldsymbol{\phi})
		+\eta_{\kappa}\bigl(\kappa \omega+\tfrac{\mathrm{i}}{2\sqrt{1-\xi^2}}\bigr) 
		+\eta_{\xi}\bigl(2\xi\omega-(\alpha+\beta\omega^2)\bigr).	
	\end{aligned}
\end{equation}
The variation of the Lagrangian is given by
\begin{equation}
	\begin{aligned}
		\delta L =&
		\delta \eta_{\tilde{f}} (\tilde{f} - 0.5\boldsymbol{\psi}^{\mathrm{T}}\boldsymbol{f}^{\mathrm{ext}})
		+ \delta \boldsymbol{\eta}_{\boldsymbol{\phi}}^{\mathrm{T}}
		(\boldsymbol{K}\boldsymbol{\phi} - \omega^2\boldsymbol{M}\boldsymbol{\phi}) \\
		&+ \delta \eta_{\mathrm{norm}}
		(\boldsymbol{\phi}^{\mathrm{T}}\boldsymbol{M}\boldsymbol{\phi} - 1)
		+ \delta \boldsymbol{\eta}_{\boldsymbol{\psi}}^{\mathrm{T}}
		(\boldsymbol{\psi} - \kappa\boldsymbol{\phi}) \\
		&+ \delta \eta_{\kappa}
		\bigl(\kappa\omega + \tfrac{\mathrm{i}}{2\sqrt{1-\xi^2}}\bigr)
		+ \delta \eta_{\xi}
		\bigl(2\xi\omega - (\alpha + \beta\omega^2)\bigr) \\
		&+ \bigl(1 + \eta_{\tilde{f}}\bigr)\delta \tilde{f}
		+ \bigl(\boldsymbol{\eta}_{\boldsymbol{\psi}}^{\mathrm{T}}
		-0.5\eta_{\tilde{f}}{\boldsymbol{f}^{\mathrm{ext}}}^\mathrm{T}\bigr) \delta \boldsymbol{\psi} \\
		&+ \boldsymbol{\eta}_{\boldsymbol{\phi}}^{\mathrm{T}}
		\bigl(\delta\boldsymbol{K}\boldsymbol{\phi} - \omega^2 \delta \boldsymbol{M} \boldsymbol{\phi}\bigr) \\
		&+ \bigl(
		\boldsymbol{\eta}_{\boldsymbol{\phi}}^{\mathrm{T}}(\boldsymbol{K} - \omega^2 \boldsymbol{M})
		+ 2 \eta_{\mathrm{norm}} \boldsymbol{\phi}^{\mathrm{T}} \boldsymbol{M}
		- \kappa \boldsymbol{\eta}_{\boldsymbol{\psi}}^{\mathrm{T}}
		\bigr) \delta \boldsymbol{\phi} \\
		&+ \bigl(\eta_{\kappa} \kappa
		- 2\omega\boldsymbol{\eta}_{\boldsymbol{\phi}}^{\mathrm{T}}\boldsymbol{M}\boldsymbol{\phi}
		+ 2\xi\eta_{\xi}
		- 2\eta_{\xi} \beta \omega \bigr) \delta \omega \\
		&+ \eta_{\mathrm{norm}}\boldsymbol{\phi}^{\mathrm{T}}\delta \boldsymbol{M}\boldsymbol{\phi}
		+ \bigl(\eta_{\kappa}\omega
		- \boldsymbol{\eta}_{\boldsymbol{\psi}}^{\mathrm{T}}\boldsymbol{\phi}\bigr)\delta \kappa \\
		&+ \bigl(\eta_{\kappa}\tfrac{\xi\mathrm{i}}{2\sqrt{(1-\xi^2)^3}}
		+ 2\eta_{\xi}\omega \bigr)\delta \xi.
	\end{aligned}
\end{equation}
By collecting the coefficients of the independent variations, we obtain the following adjoint equations
\begin{equation}
	\begin{aligned}
		\delta \tilde{f}:&\quad 1 + \eta_{\tilde{f}} = 0,\\
		\delta \boldsymbol{\psi}:&\quad \boldsymbol{\eta}_{\boldsymbol{\psi}} - 0.5\eta_{\tilde{f}}{\boldsymbol{f}^\mathrm{ext}} = 0,\\
		\delta \boldsymbol{\phi}:&\quad \bigl(\boldsymbol{K} - \omega^2 \boldsymbol{M}\bigr)\boldsymbol{\eta}_{\boldsymbol{\phi}}
		+ 2\eta_{\mathrm{norm}}\boldsymbol{M}\boldsymbol{\phi}
		- \kappa \boldsymbol{\eta}_{\boldsymbol{\psi}} = 0,\\
		\delta \omega:&\quad \eta_\kappa\,\kappa
		- 2\omega\boldsymbol{\eta}_{\boldsymbol{\phi}}^{\mathrm{T}}\boldsymbol{M}\boldsymbol{\phi}
		+ 2\eta_\xi(\xi - \beta\omega) = 0,\\
		\delta \kappa:&\quad \eta_\kappa\,\omega - \boldsymbol{\eta}_{\boldsymbol{\psi}}^{\mathrm{T}}\boldsymbol{\phi} = 0,\\
		\delta \xi:&\quad \eta_{\kappa}\tfrac{\xi\mathrm{i}}{2\sqrt{(1-\xi^2)^3}}
		+ 2\eta_{\xi}\omega = 0.
	\end{aligned}
\end{equation}
Thus, we have $\eta_{\tilde{f}}=-1$, $\boldsymbol{\eta}_{\boldsymbol{\psi}}=0.5\eta_{\tilde{f}}\boldsymbol{f}^{\mathrm{ext}}$, $\eta_{\kappa}={\boldsymbol{\eta}_{\boldsymbol{\psi}}}^{\mathrm{T}}\boldsymbol{\phi}/\omega$, $\eta_{\xi}=-\frac{\eta_{\kappa}\xi\mathrm{i}}{4 \omega \sqrt{(1-\xi^2)^3}}$.
Let  $\boldsymbol{b}_{\mathrm{norm}}=\kappa \boldsymbol{\eta}_{\boldsymbol{\psi}}$, $b_{\boldsymbol{\phi}}=\eta_\kappa \kappa + 2 \eta_{\xi}(\xi-\beta \omega)$, $\boldsymbol{\eta}_{\boldsymbol{\phi}}$ and $\eta_{\mathrm{norm}}$ can be obtained by 
\begin{equation}
	\begin{aligned}
		\begin{pmatrix}
			2 \omega \boldsymbol{\phi}^{\mathrm{T}}\boldsymbol{M} & 0 \\
			\boldsymbol{K}-\omega^2 \boldsymbol{M} & 2 \boldsymbol{M} \boldsymbol{\phi}
		\end{pmatrix}
		\begin{pmatrix}
			\boldsymbol{\eta}_{\boldsymbol{\phi}} \\ \eta_{\mathrm{norm}}
		\end{pmatrix}
		=
		\begin{pmatrix}
			b_{\boldsymbol{\phi}} \\ \boldsymbol{b}_{\mathrm{norm}}
		\end{pmatrix}.
	\end{aligned}
	\label{eq:solve_lambda}
\end{equation}
Thus, we obtain the following expression for the sensitivity of $\tilde{f}$
\begin{equation}
	\begin{aligned}
		\delta \tilde{f} = \boldsymbol{\eta}_{\boldsymbol{\phi}}^{\mathrm{T}}
		\bigl(\delta\boldsymbol{K}\boldsymbol{\phi} - \omega^2 \delta \boldsymbol{M}\boldsymbol{\phi}\bigr)
		+ \eta_{\mathrm{norm}} \boldsymbol{\phi}^{\mathrm{T}} \delta \boldsymbol{M} \boldsymbol{\phi}.
	\end{aligned}
\end{equation}

\section{The sensitivity of \textcolor{blue}{$\Omega_{\text{SN}}$}}
\label{appA: sen_b}

\textcolor{blue}{We first compute the sensitivity of $\rho_{\mathrm{SN}}$ by taking derivative of~\eqref{eq:rho_SN}, namely
\begin{equation}
\label{eq:rhosn-prime}
	\begin{aligned}
		\rho_{\mathrm{SN}}' = \frac{c_1}{c_2},
	\end{aligned}
\end{equation}
where 
\begin{equation}
	\begin{aligned}
		c_1 = 
		& 8\mathrm{Im}(\gamma)\mathrm{Im}(\gamma)'\rho_\text{SN}^6(\epsilon^2|\tilde{f}|^2 - (\mathrm{Re}(\lambda) \rho_\text{SN} + \mathrm{Re}(\gamma) \rho_\text{SN}^3)^2) \\
		& + 4\mathrm{Im}(\gamma)^2\rho_\text{SN}^6(2\epsilon^2|\tilde{f}||\tilde{f}|' - 2(\mathrm{Re}(\lambda)\rho_\text{SN} + \mathrm{Re}(\gamma)\rho_\text{SN}^3)(\mathrm{Re}(\lambda)'\rho_\text{SN} + \mathrm{Re}(\gamma)'\rho_\text{SN}^3)) \\
		& - 2(\epsilon^2|\tilde{f}|^2 + 2 \mathrm{Re}(\lambda) \mathrm{Re}(\gamma) \rho_\text{SN}^4 + 2\mathrm{Re}(\gamma)^2 \rho_\text{SN}^6) \\
		& (2\epsilon^2 |\tilde{f}||\tilde{f}|' + 2\mathrm{Re}(\lambda)'\mathrm{Re}(\gamma)\rho_\text{SN}^4 + 2\mathrm{Re}(\lambda)\mathrm{Re}(\gamma)'\rho_\text{SN}^4 + 4\mathrm{Re}(\gamma)\mathrm{Re}(\gamma)'\rho_\text{SN}^6), \\
		c_2 =& 2(\epsilon^2|\tilde{f}|^2 + 2\mathrm{Re}(\lambda)\mathrm{Re}(\gamma)\rho_\text{SN}^4 + 2\mathrm{Re}(\gamma)^2\rho_\text{SN}^4 + 2\mathrm{Re}(\gamma)^2\rho_\text{SN}^6) \\
		&(8\mathrm{Re}(\lambda)\mathrm{Re}(\gamma)\rho_\text{SN}^3 + 12\mathrm{Re}(\gamma)^2\rho_\text{SN}^5) \\
		& + 24\mathrm{Im}(\gamma)^2\rho_\text{SN}^5((\mathrm{Re}(\lambda)\rho_\text{SN} + \mathrm{Re}(\gamma)\rho_\text{SN}^3)^2 - \epsilon^2|\tilde{f}|^2) \\
		& + 8\mathrm{Im}(\gamma)^2(\mathrm{Re}(\lambda)\rho_\text{SN} + \mathrm{Re}(\gamma)\rho_\text{SN}^3)(\mathrm{Re}(\lambda) +3\mathrm{Re}(\gamma)\rho_\text{SN}^2)\rho_\text{SN}^6. \\
	\end{aligned}
\end{equation}
Then, the sensitivity of $\Omega_{\rm{SN}}$ can be obtained by taking derivative of~\eqref{eq:reduce-FRC}, namely
\begin{equation}
	\begin{aligned}
		 \Omega_{\mathrm{SN}}' =  c_3 -\frac{c_4}{c_5},
	\end{aligned}
\end{equation}
where $c_3,c_4$ and $c_5$ have been defined in~\eqref{eq:sens_omega_SN}}.

\section{\textcolor{blue}{Supplementary materials for the example in Sec.~\ref{ssec: example1}}}

\subsection{The convergence validation of SSM-based reduction} 
\label{appA: SSM_order_test_ex1}

\begin{figure}[!ht]
\centering
\includegraphics[width=0.4\textwidth]{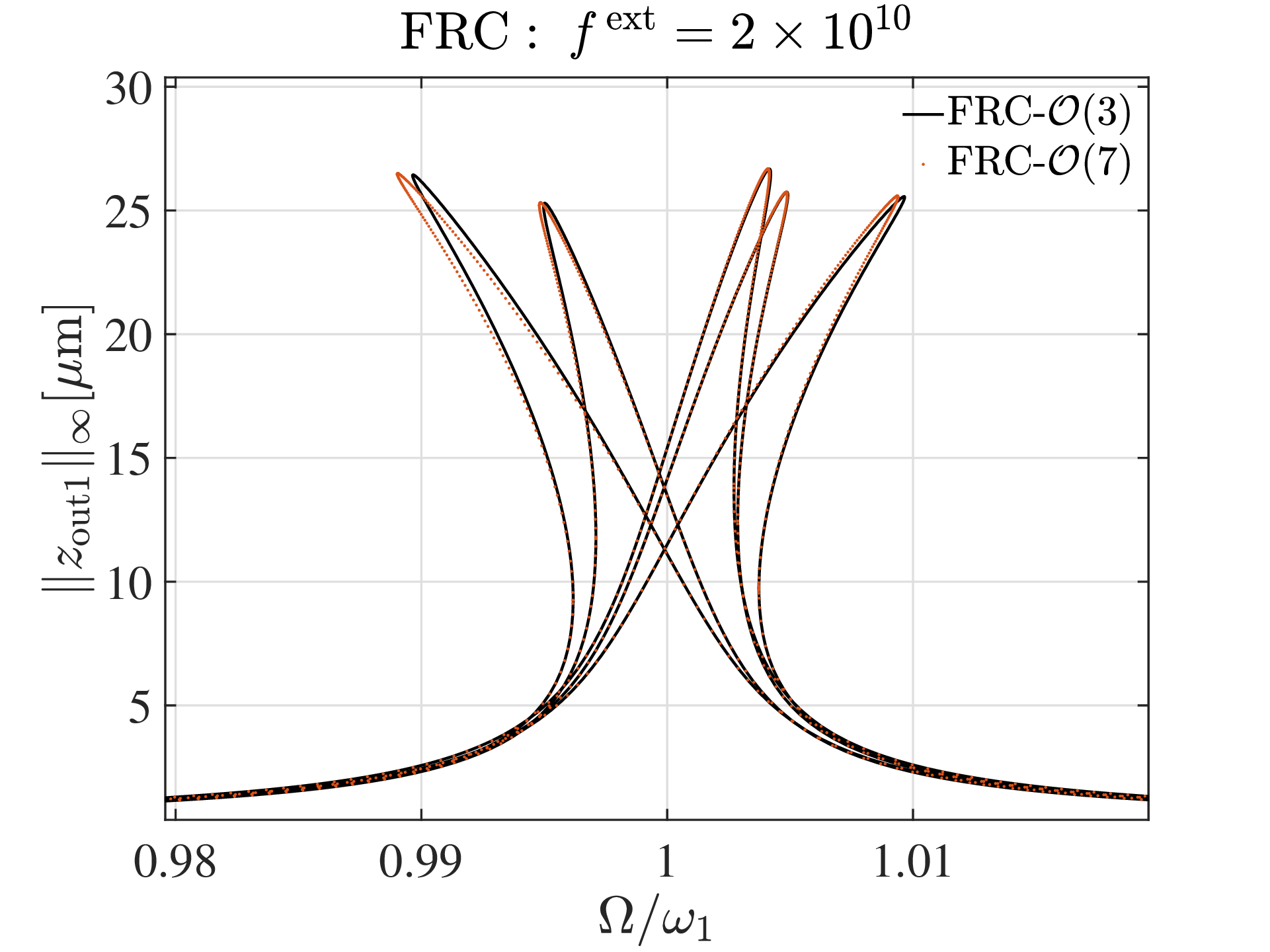} 
\caption{Comparison of the FRCs computed using third-order and seventh-order SSM reductions for the structures optimized by linear and nonlinear formulations, under forcing amplitude $f^{\mathrm{ext}} = 2 \times 10^{10} \ \mathrm{ng \cdot \mu m/ms^2}$.}
\label{fig:FRC_O3_fext2e10_O7}
\end{figure}

To assess the convergence of the SSM-based model reduction, we compute the FRCs of the optimized structures using seventh-order SSM reduction and compare them with those obtained from third-order SSM reduction in Fig.~\ref{fig:FRCs_linear_nonlinear}. As shown in Fig.~\ref{fig:FRC_O3_fext2e10_O7}, the FRCs for both cases exhibit excellent agreement between the two SSM orders under a forcing amplitude of $f^{\mathrm{ext}} = 2 \times 10^{10} \ \mathrm{ng \cdot \mu m/ms^2}$. This confirms that the third-order SSM reduction is sufficiently accurate for capturing the nonlinear dynamic behavior of the optimized structures in this study.

\subsection{\textcolor{blue}{Robustness of optimized structures}}
\label{appA2: robustness_ex1}

\subsubsection{\textcolor{blue}{Influence of forcing amplitude on optimization}} 
\label{appA2: changing_forcingAmplitude_ex1}

\textcolor{blue}{In this part, we vary the forcing amplitude to test the robustness of the optimized structures in Sec.~\ref{ssec: example1}. Specifically, we investigate how variations in $f^{\mathrm{ext}}$ affect the obtained optimal layouts and their dynamic responses. In Sec.~\ref{ssec: example1}, the forcing amplitude is set to $f^{\mathrm{ext}} = 5 \times 10^{9}\,\mathrm{ng \cdot \mu m/ms^2}$. Here, we fix $\gamma_{\mathrm{target}} = 5 \times 10^{-5}$ and increase the forcing amplitude to $f^{\mathrm{ext}} = 1 \times 10^{10}$ and $2 \times 10^{10}$, respectively.}

\begin{figure}[!ht]
 \centering
 \includegraphics[width=0.5\textwidth]{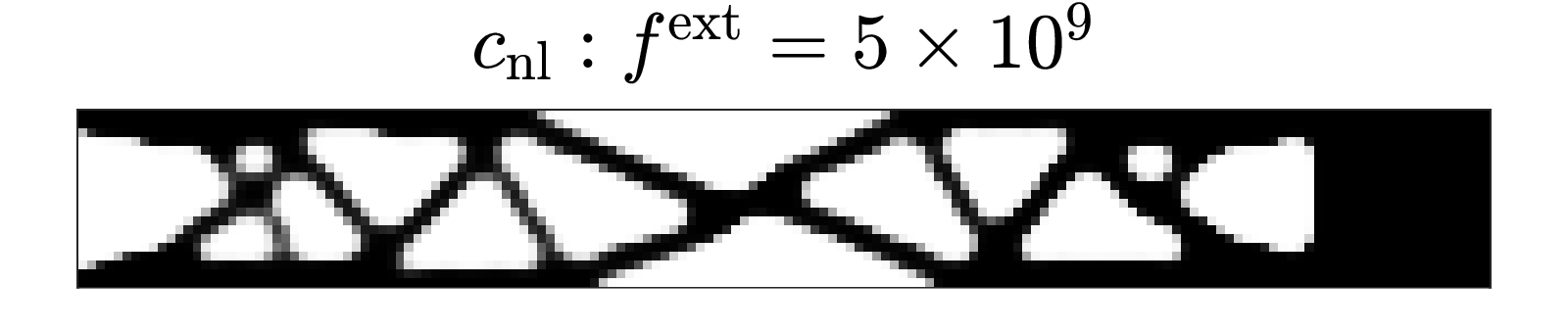}\\
 \includegraphics[width=0.5\textwidth]{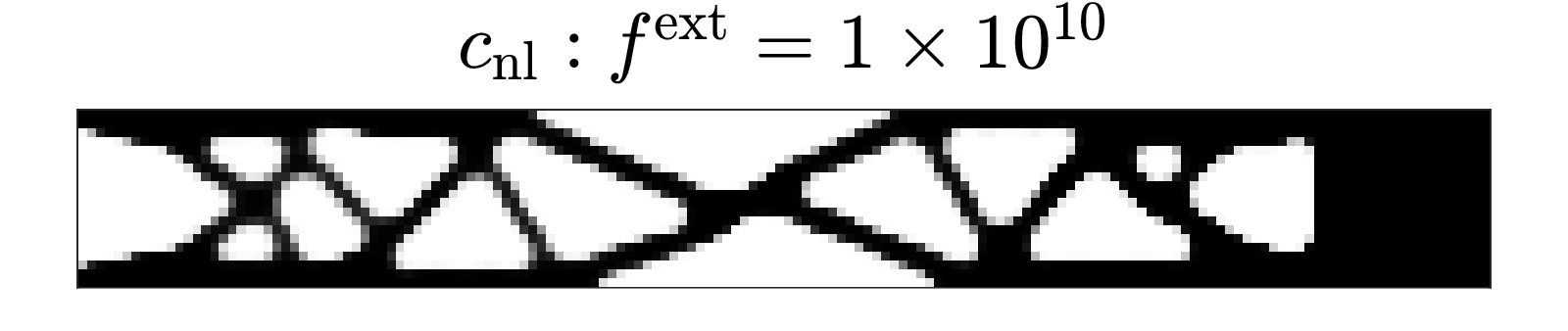}\\
 \includegraphics[width=0.5\textwidth]{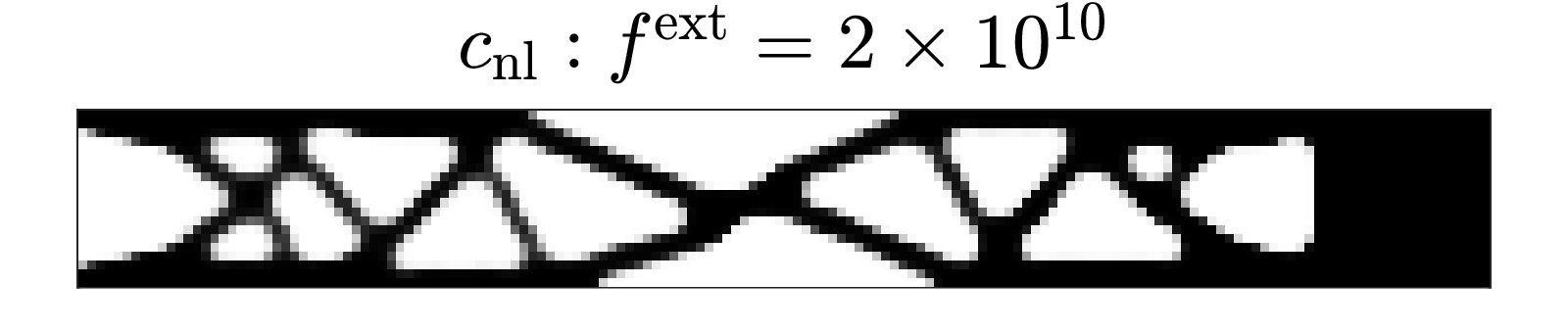}\\
 \caption{\textcolor{blue}{Optimal layouts obtained from the nonlinear optimization formulation Eq.~\eqref{eq:opt-rho_max} for different forcing amplitudes with $\gamma_{\mathrm{target}} = 5 \times 10^{-5}$.}}
 \label{fig:layouts_change_fext}
 \end{figure}

\begin{figure}[!ht]
 \centering
 \includegraphics[width=0.5\textwidth]{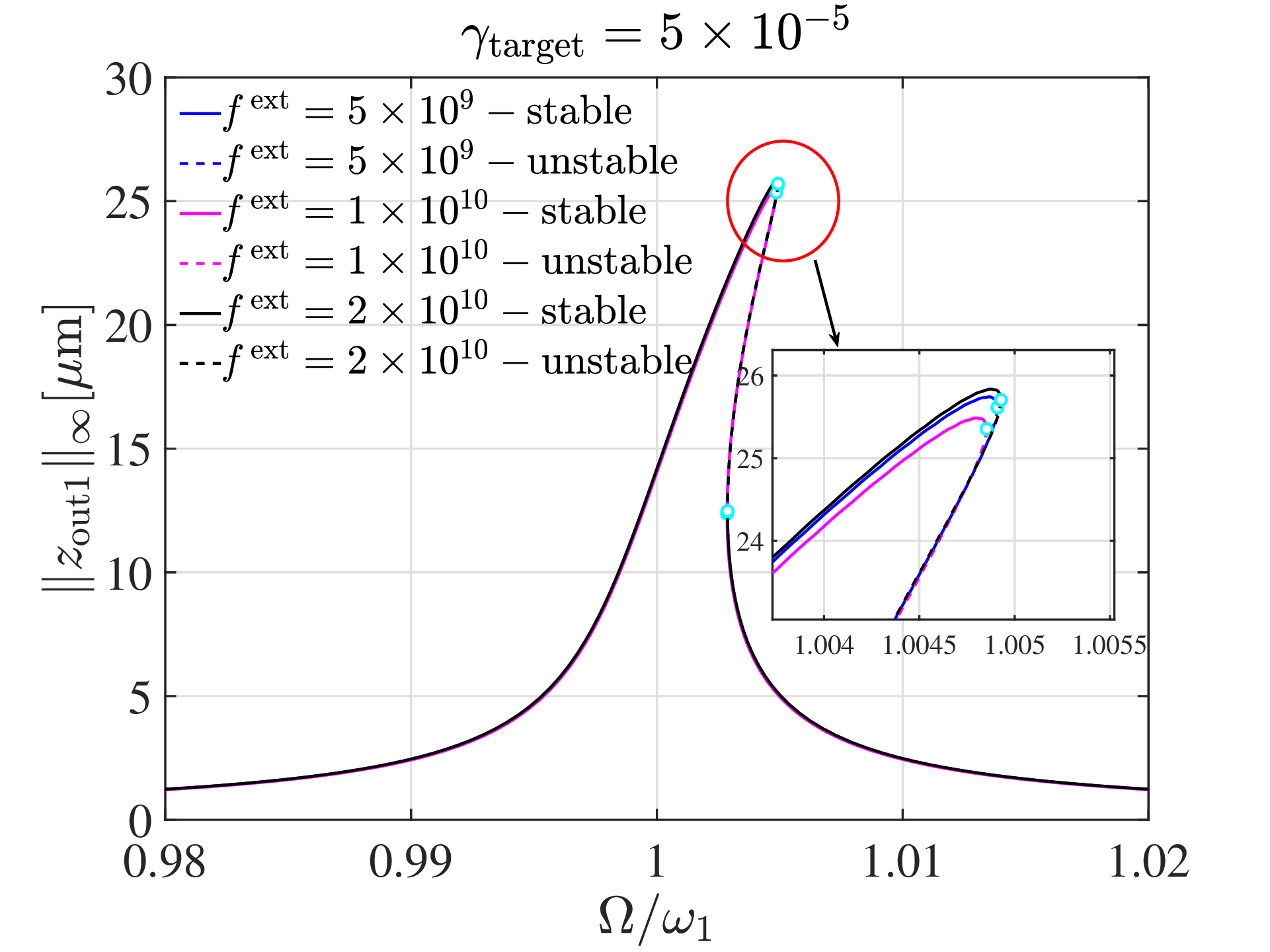}\\
 \caption{\textcolor{blue}{Comparison of the FRCs of the optimal layouts shown in Fig.~\ref{fig:layouts_change_fext} under forcing amplitude: $f^{\mathrm{ext}} = 2 \times 10^{10}$.}}
 \label{fig:FRC_change_fext}
 \end{figure}

\textcolor{blue}{The optimal layouts for different forcing amplitudes are shown in Fig.~\ref{fig:layouts_change_fext}. It can be observed that the layouts obtained for $f^{\mathrm{ext}} = 1 \times 10^{10}$ and $f^{\mathrm{ext}} = 2 \times 10^{10}$ are very similar to that obtained for $f^{\mathrm{ext}} = 5 \times 10^{9}$. Furthermore, the corresponding FRCs of these three optimized layouts under the same forcing are presented in Fig.~\ref{fig:FRC_change_fext}, which show only minor differences. These results indicate that, within the range of forcing amplitudes for which the third-order SSM reduction remains convergent, the optimized structures are robust to the change forcing amplitude.}

 \subsubsection{\textcolor{blue}{Influence of material stiffness on optimization}}

\textcolor{blue}{To further test the robustness of the optimized structures, we introduce a perturbation in the elastic modulus, changing it from $E = 148 \times 10^{9}$ Pa to $E' = 0.95 E$, and examine the resulting changes in the optimal layouts and their dynamic responses.}

\begin{figure}[!ht]
 \centering
 \includegraphics[width=0.5\textwidth]{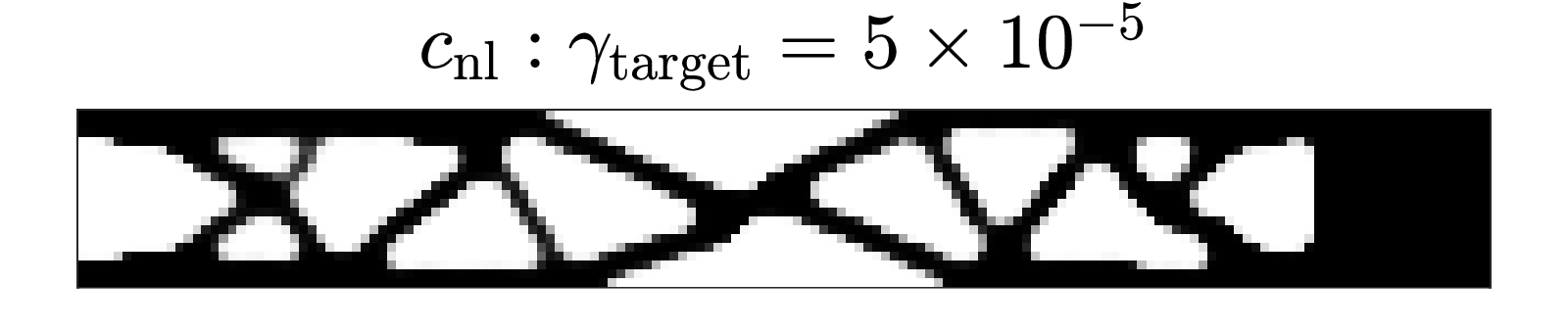}\\
 \includegraphics[width=0.5\textwidth]{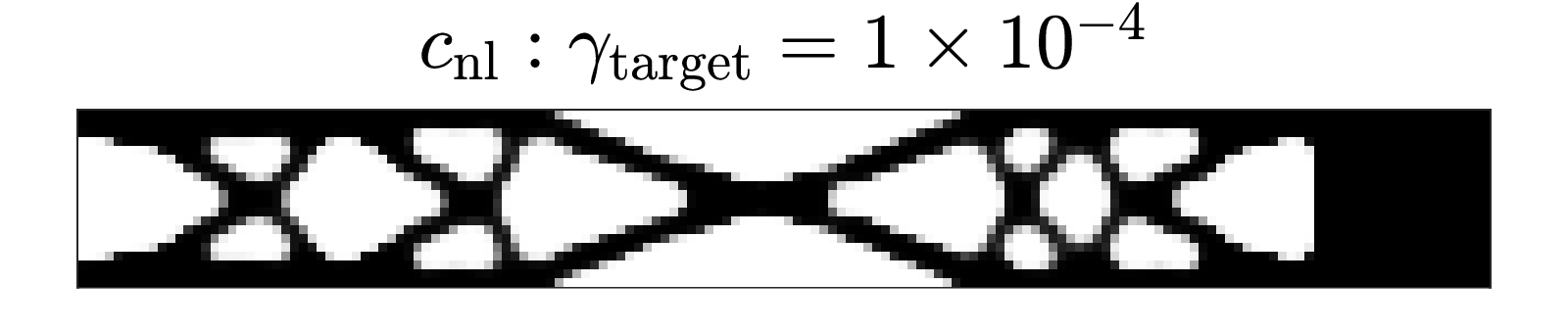}\\
 \includegraphics[width=0.5\textwidth]{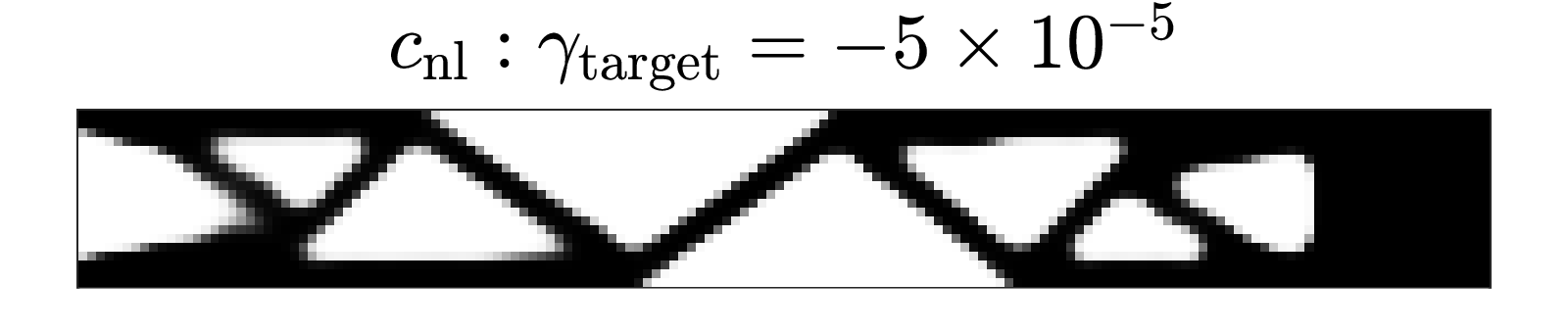}\\
 \includegraphics[width=0.5\textwidth]{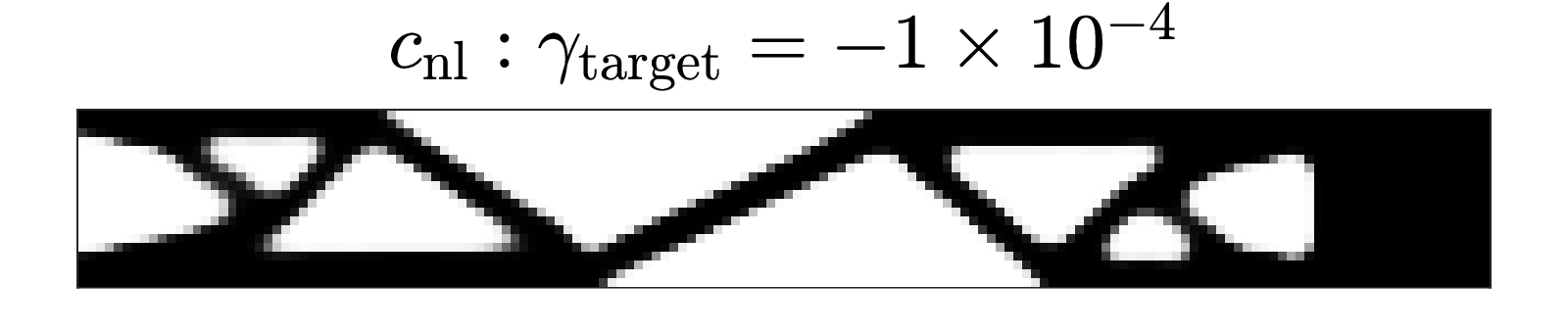}\\
 \caption{\textcolor{blue}{Optimal layouts obtained from the nonlinear optimization formulation~\eqref{eq:opt-rho_max} with a perturbed elastic modulus $E' = 0.95E$. Each panel corresponds to a nonlinear optimal layout for a distinct value of $\gamma_{\mathrm{target}}$.}}
 \label{fig:layouts_ex1_Epert}
 \end{figure}

\begin{figure}[!ht]
 \centering
 \includegraphics[width=0.5\textwidth]{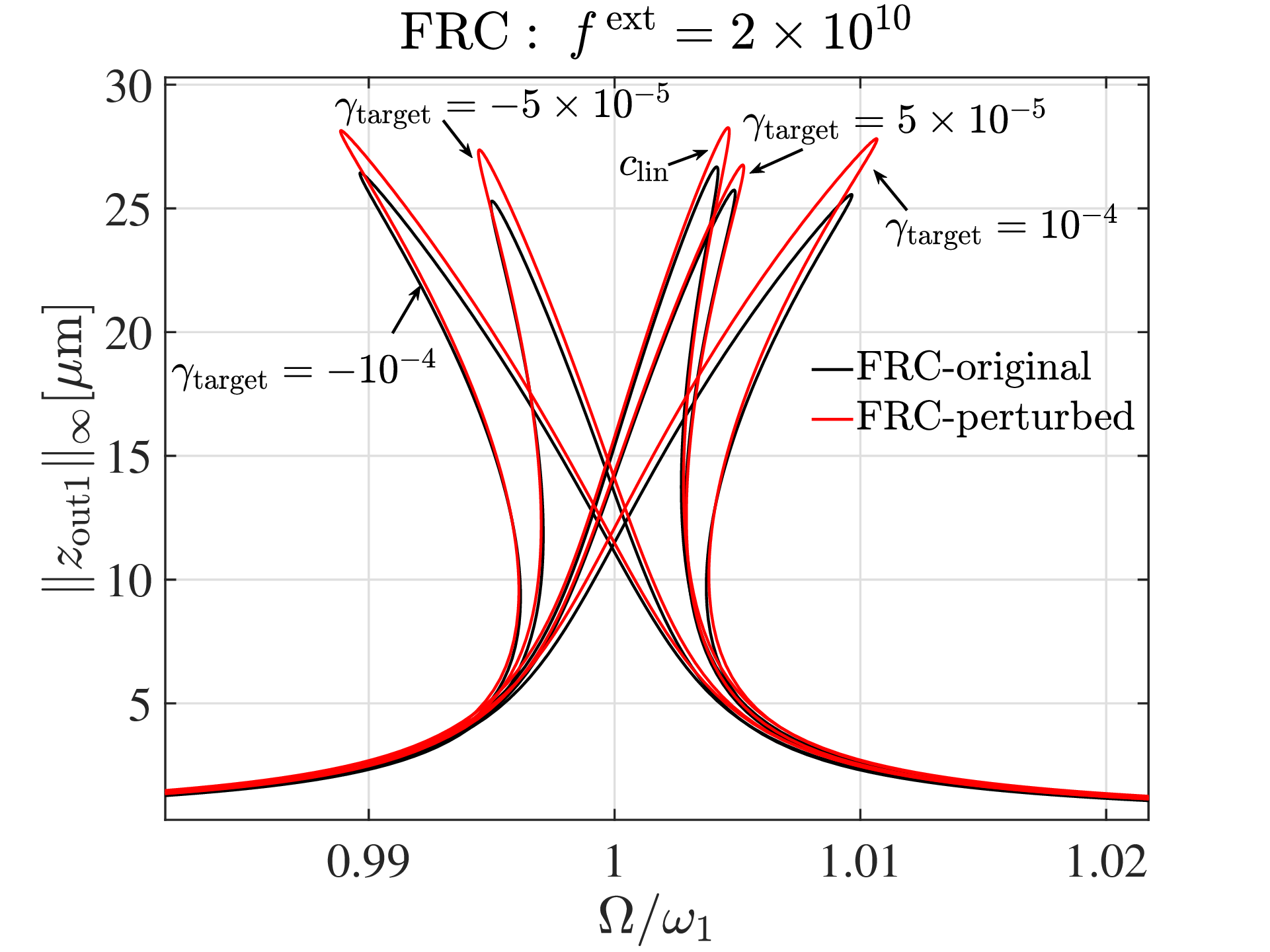}\\
 \caption{\textcolor{blue}{Comparison of the FRCs of the optimal layouts shown in Fig.~\ref{fig:layouts_linear_nonlinear} and Fig.~\ref{fig:layouts_ex1_Epert} under forcing amplitude: $f^{\mathrm{ext}} = 2 \times 10^{10}$.}}
 \label{fig:FRC_change_E_ex1}
 \end{figure}

\textcolor{blue}{As seen in Fig.~\ref{fig:layouts_ex1_Epert}, the optimal layouts exhibit some local changes after a perturbation of the elastic modulus, while the overall layouts are robust to such a perturbation. Further, the FRCs of the optimal layouts obtained with the perturbed elastic modulus are compared with those of the layouts obtained with the original modulus in Fig.~\ref{fig:FRC_change_E_ex1}. We observe that although the perturbation alters the response amplitudes, their respective nonlinear characteristics, especially the hardening/softening behavior, are mostly preserved.}

\textcolor{blue}{These results indicate that the optimized results are not overly sensitive to moderate stiffness variations. Therefore, the proposed optimization framework yields designs with good robustness against typical fabrication uncertainties encountered in MEMS applications.}

\subsection{\textcolor{blue}{Test of mode collisions and master mode consistency}} 
\label{appA3: test_mode_collisions}

\begin{figure}[!ht]
\centering
\includegraphics[width=0.45\textwidth]{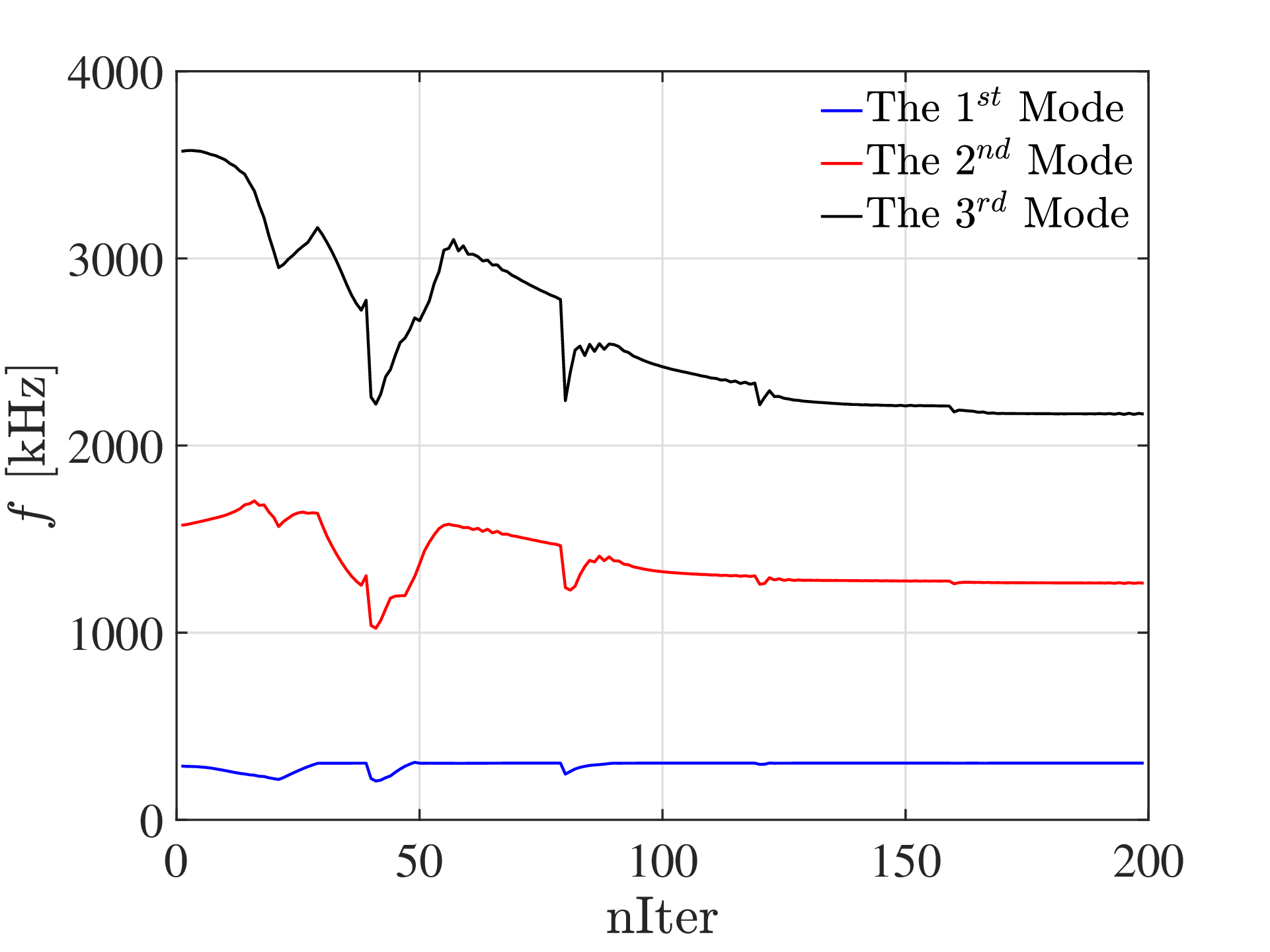}
\caption{\textcolor{blue}{The evolution of the eigenfrequencies of the first three modes for the optimization with $\gamma_{\mathrm{target}} = 5 \times 10^{-5}$ in Fig.~\ref{fig:layouts_linear_nonlinear}.}}
\label{fig:om2nIter_Case1}
\end{figure}

\textcolor{blue}{To detect possible modal collisions and internal resonances, we present the evolution of the first three eigenfrequencies for the optimization with $\gamma_{\mathrm{target}} = 5 \times 10^{-5}$ in Fig.~\ref{fig:layouts_linear_nonlinear}. As shown in Fig.~\ref{fig:om2nIter_Case1}, the eigenfrequencies of the three modes remain well separated and do not overlap. Indeed, we consistently have $\omega_X>3\omega_Y$ throughout the iteration process thanks to the constraints imposed on natural frequencies. Therefore, no internal resonance involving the master mode and the second mode occurs. Potential internal resonances with higher-order modes are not checked here because $1:m$ internal resonances with $m\geq4$ are not relevant for systems with only quadratic and cubic nonlinearities.}

\begin{figure}[!ht]
\centering
\includegraphics[width=0.45\textwidth]{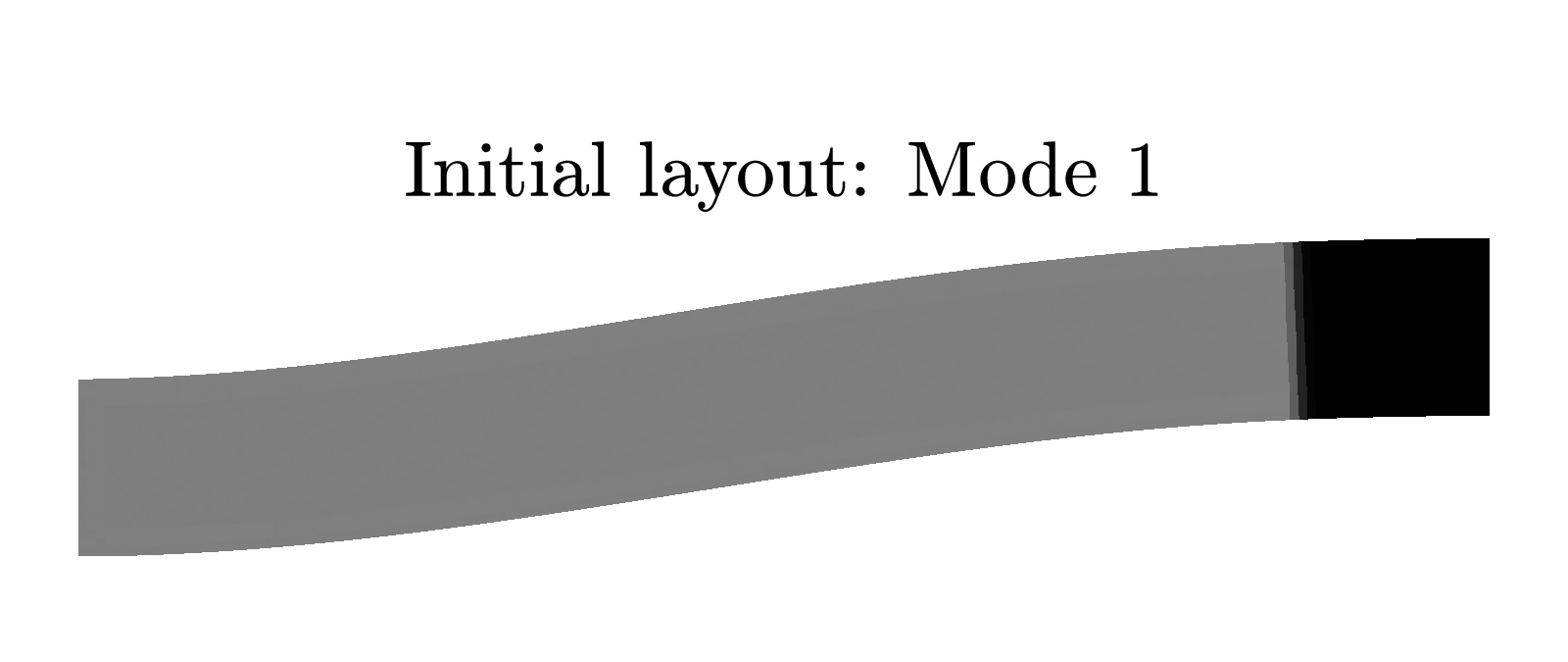}
\includegraphics[width=0.45\textwidth]{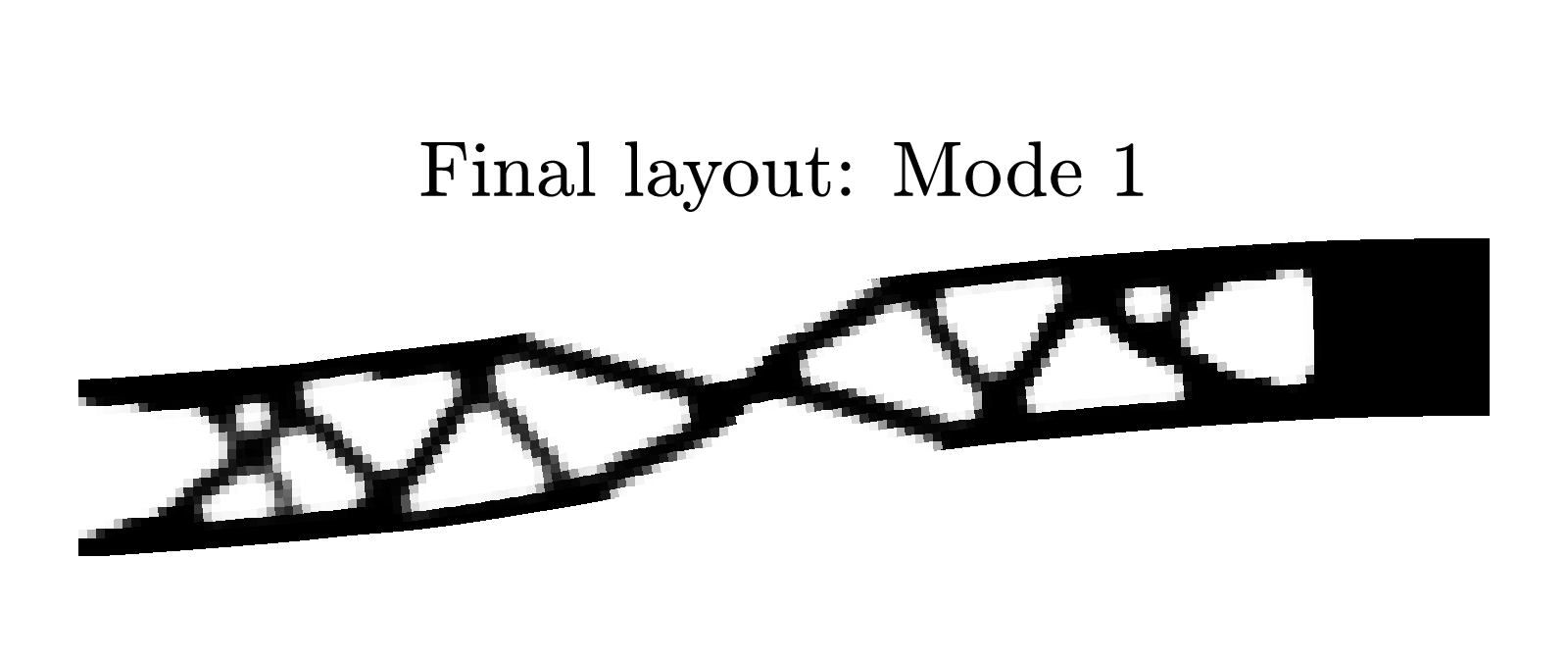} 
\caption{\textcolor{blue}{The linear modal shape of the first modes of the initial and optimized structures for the optimization with $\gamma_{\mathrm{target}} = 5 \times 10^{-5}$ in Fig.~\ref{fig:layouts_linear_nonlinear}.}}
\label{fig:modes_Case1}
\end{figure}

\textcolor{blue}{To check the consistency of master modes during iterations, we compare the linear modal shape of the first modes of the initial and optimized structures for the optimization with $\gamma_{\mathrm{target}} = 5 \times 10^{-5}$ in Fig.~\ref{fig:layouts_linear_nonlinear}. As shown in Fig.~\ref{fig:modes_Case1}, the initial and optimized structures exhibit similar modal shapes, indicating that the target master mode is tracked successfully in the optimization process.}

\section{\textcolor{blue}{Supplementary materials for the example in Sec.~\ref{ssec:example2}}}

\subsection{\textcolor{blue}{Test of mode collisions and master mode consistency}} 
\label{appD3: test_mode_collisions_ex2}

\begin{figure}[!ht]
\centering
\includegraphics[width=0.45\textwidth]{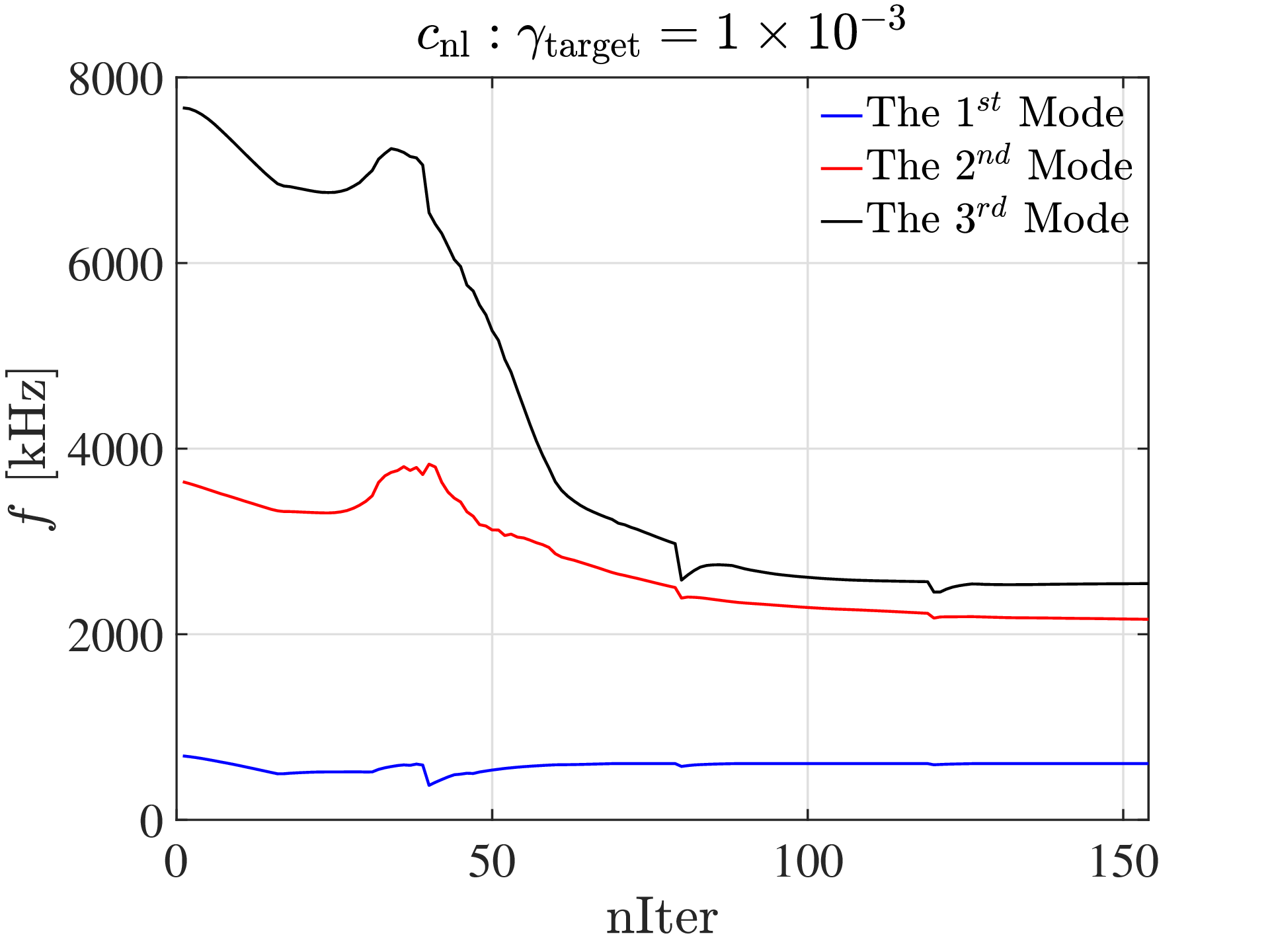}
\includegraphics[width=0.45\textwidth]{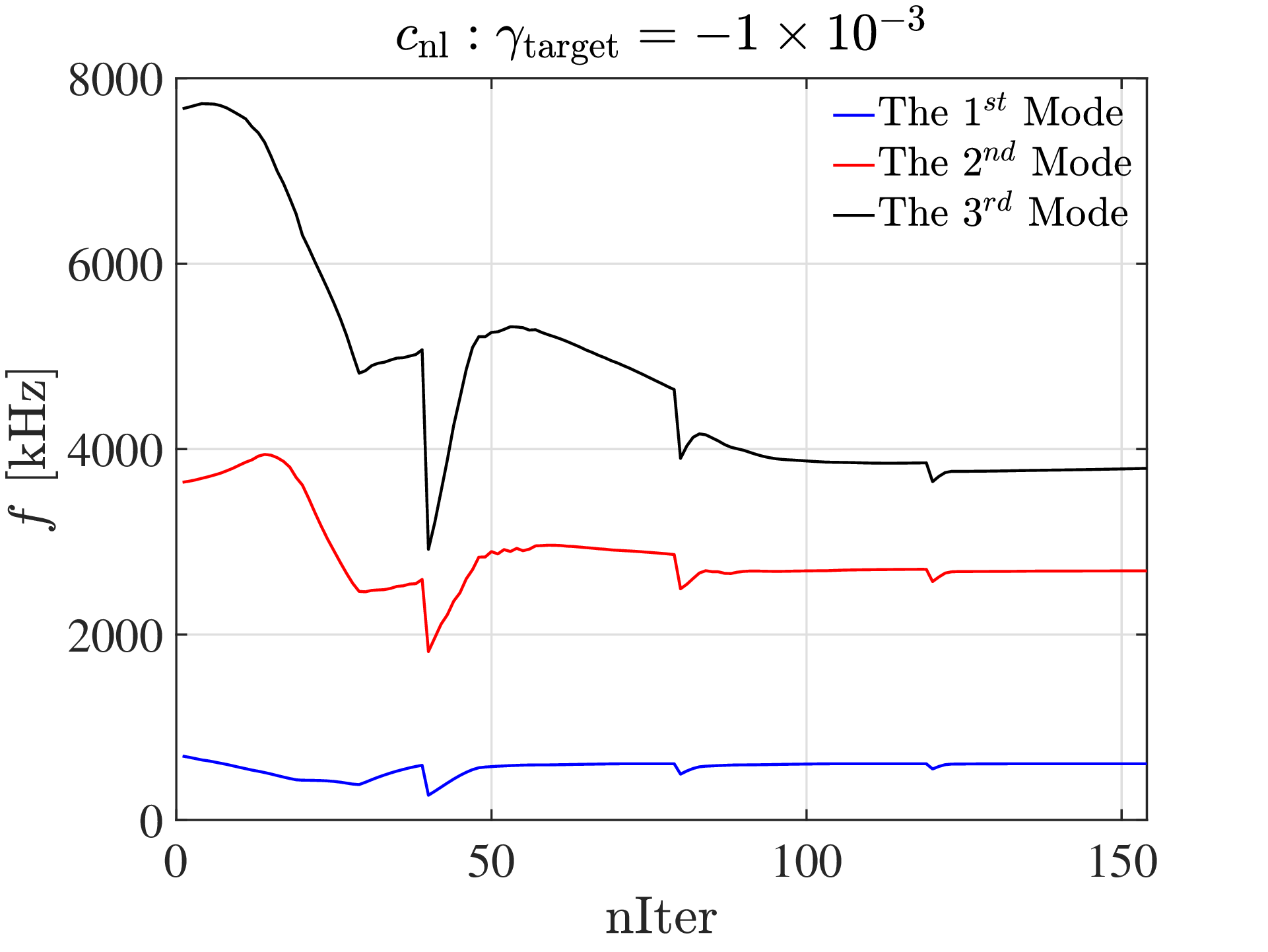}
\caption{\textcolor{blue}{The evolution of eigenfrequencies of the first three modes for the optimization $c_\text{nl}$ in Fig.~\ref{fig:optHarden-gamma0001} and Fig.~\ref{fig:optSoften-gamma_0001}.}}
\label{fig:om2nIter_Case2}
\end{figure}

\textcolor{blue}{Similar to Appendix~\ref{appA3: test_mode_collisions}, we track the evolution of the eigenfrequencies of the first three modes for the optimization $c_\text{nl}$ in Fig.~\ref{fig:optHarden-gamma0001} and Fig.~\ref{fig:optSoften-gamma_0001} to check for modal collisions and internal resonances. As shown in Fig.~\ref{fig:om2nIter_Case2}, although the eigenfrequencies of the second and third modes approach each other during the iterations, no modal collision occurs. Moreover, since $\omega_X>3\omega_Y$ throughout the iteration process, the master mode does not experience internal resonance with the second mode.}

\begin{figure}[!ht]
\centering
\includegraphics[width=0.3\textwidth]{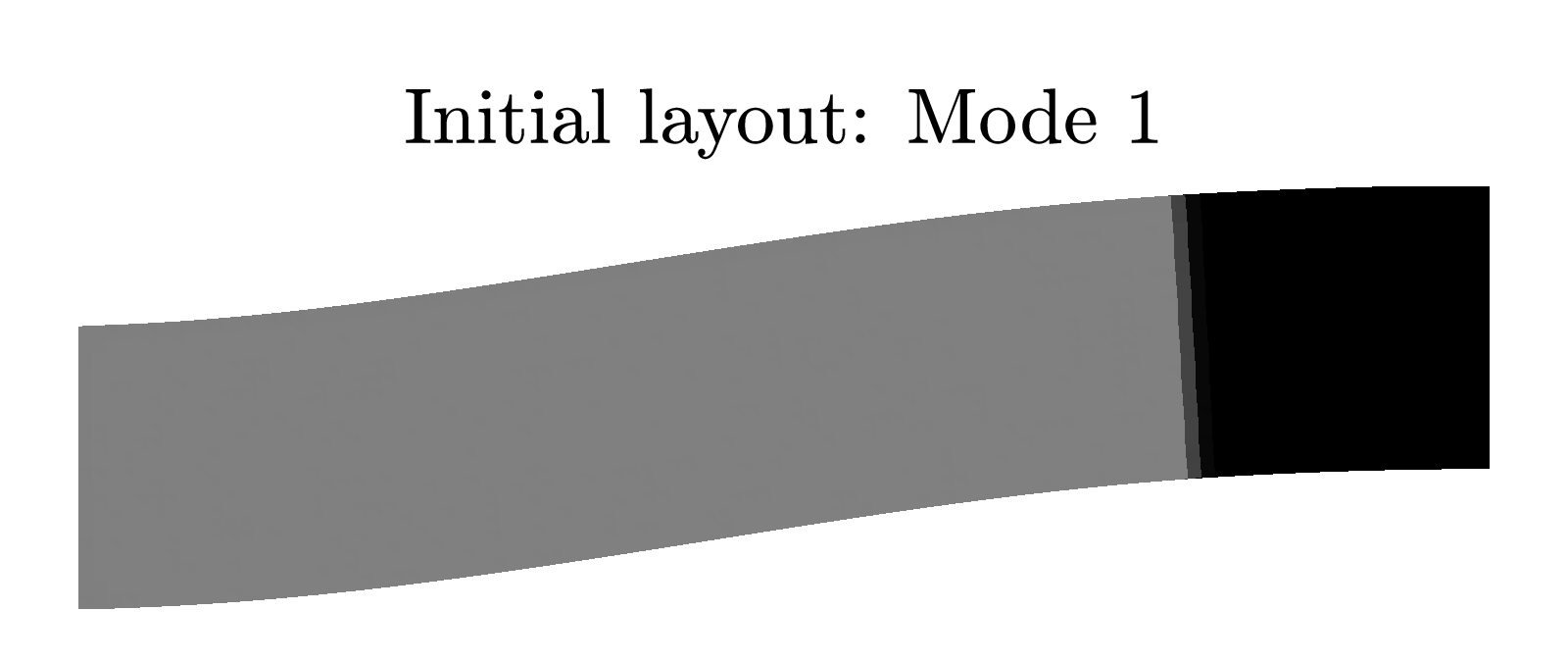}
\includegraphics[width=0.3\textwidth]{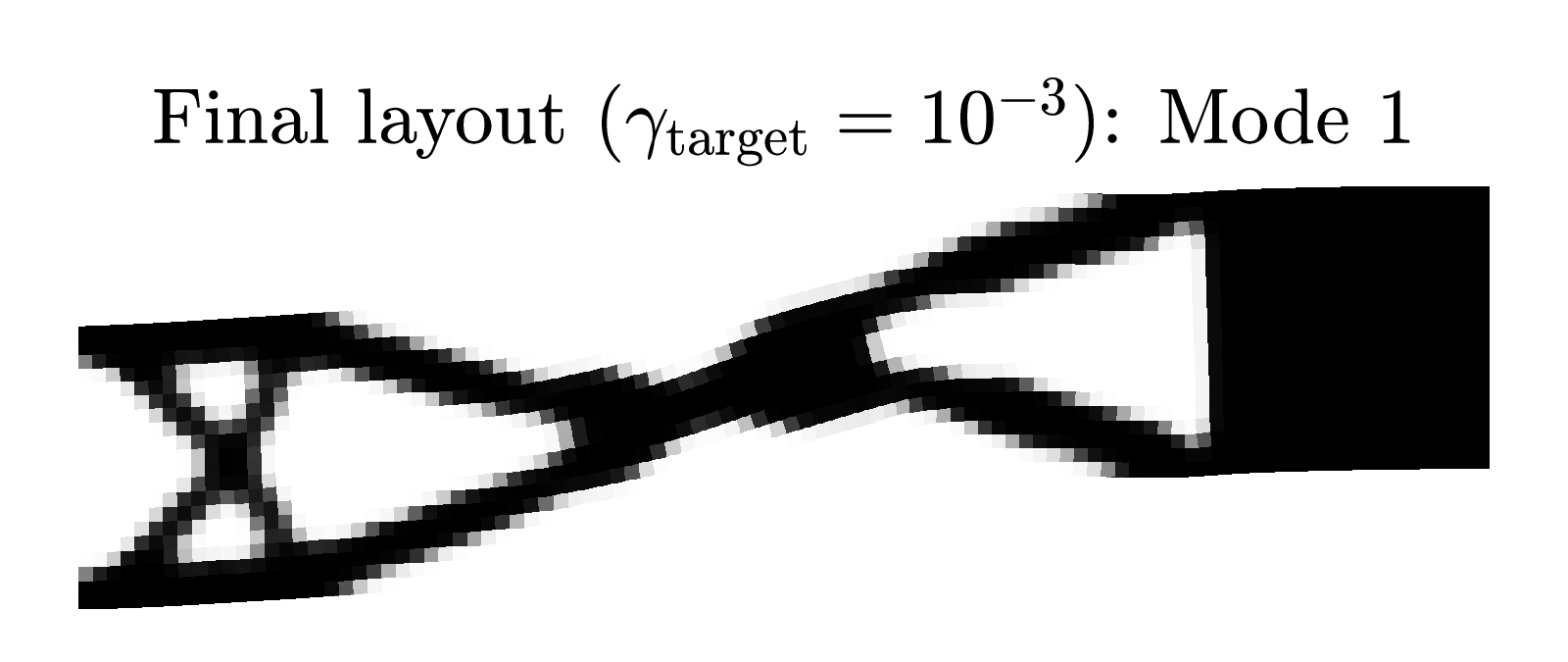} 
\includegraphics[width=0.3\textwidth]{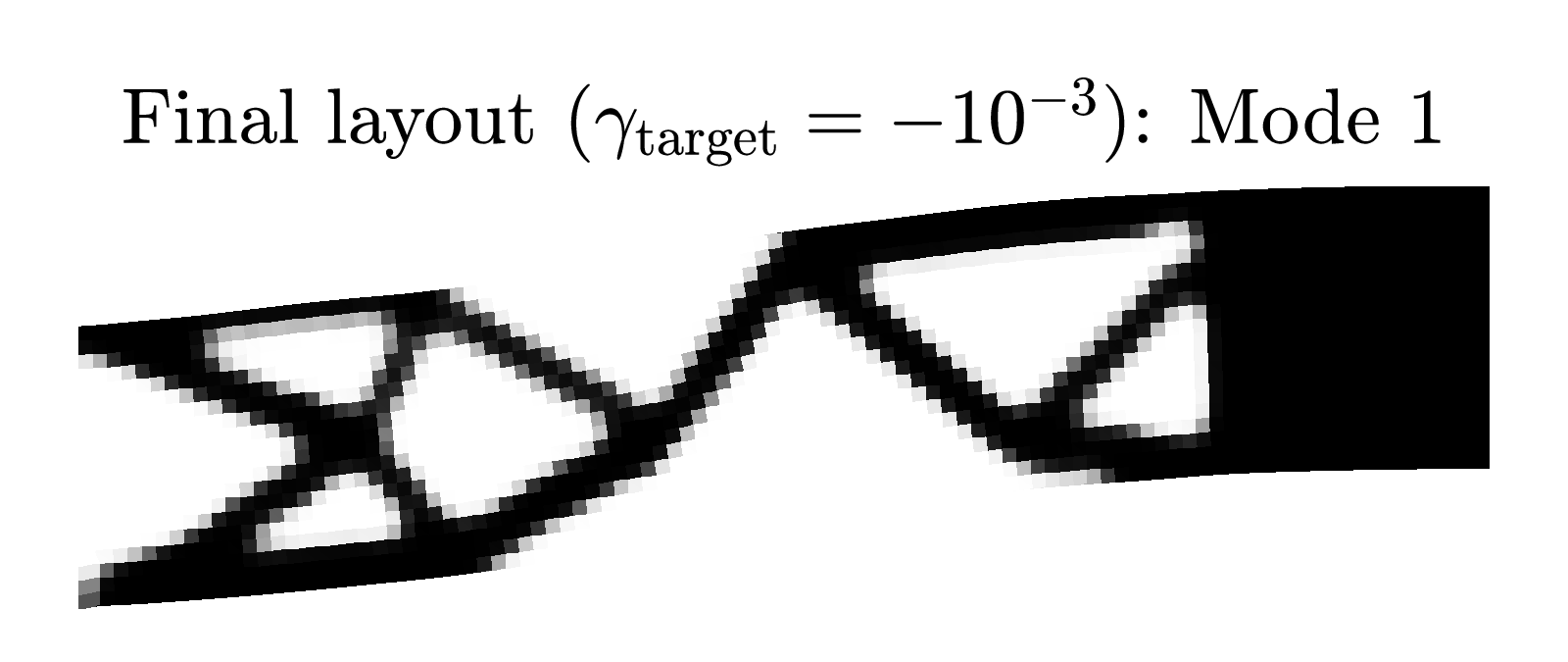} 
\caption{\textcolor{blue}{The linear modal shape of the first modes of the initial and optimized structures for the optimization $c_\text{nl}$ in Fig.~\ref{fig:optHarden-gamma0001} and Fig.~\ref{fig:optSoften-gamma_0001}.}}
\label{fig:modes_Case2}
\end{figure}

\textcolor{blue}{We compare the linear modal shape of the first modes of the initial and optimized structures for the optimization $c_\text{nl}$ in Fig.~\ref{fig:optHarden-gamma0001} and Fig.~\ref{fig:optSoften-gamma_0001} to check the consistency of master modes during iterations. As shown in Fig.~\ref{fig:modes_Case2}, the initial and optimized structures exhibit similar modal shapes, indicating that the target master mode is tracked successfully.}


\section{\textcolor{blue}{Supplementary materials for the example in Sec.~\ref{ssec: control_SN}}}

\subsection{The convergence validation of SSM-based reduction} 
\label{appA: SSM_order_test_ex3}

\begin{figure}[!ht]
\centering
\includegraphics[width=0.3\textwidth]{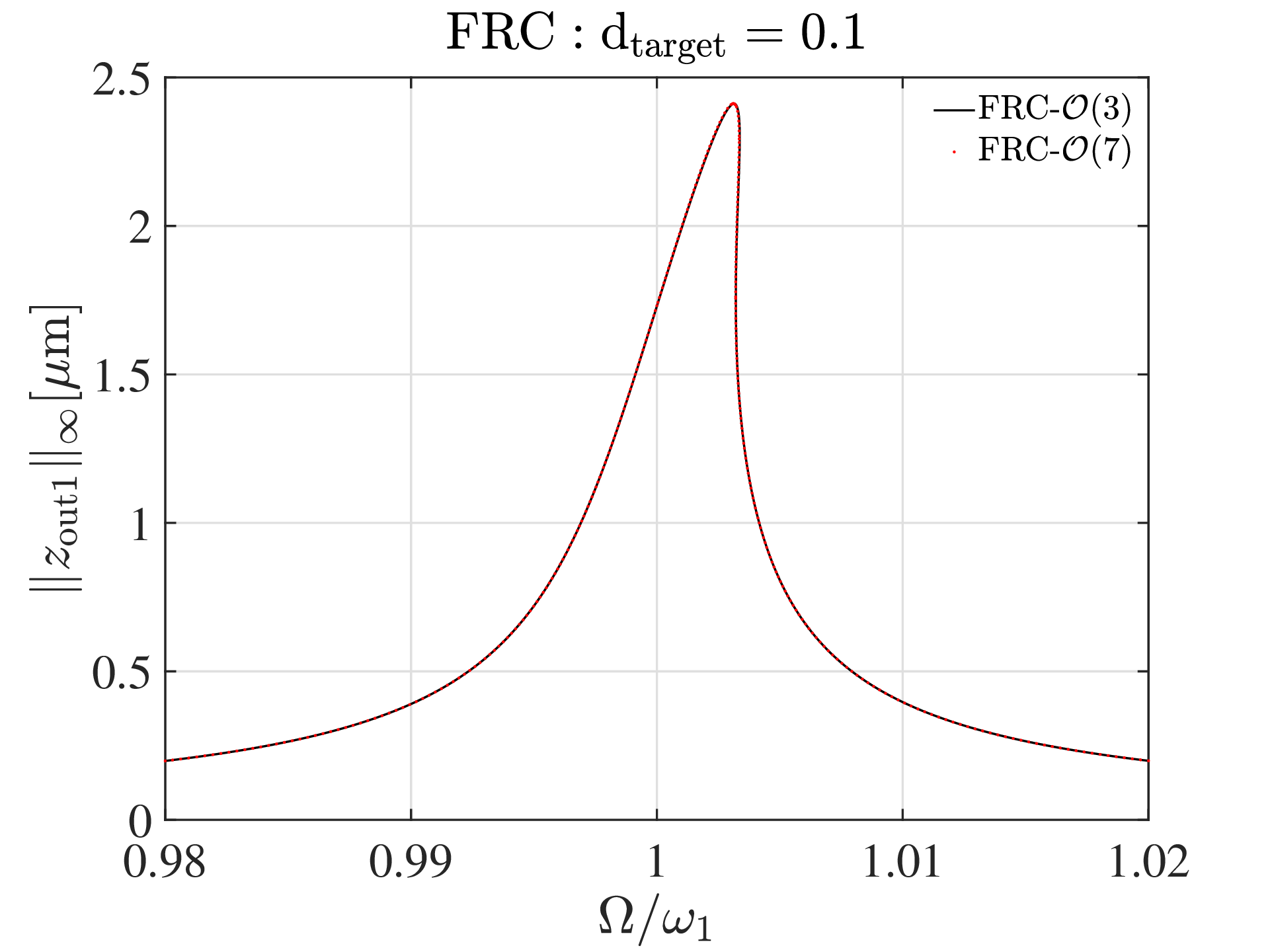} 
\includegraphics[width=0.3\textwidth]{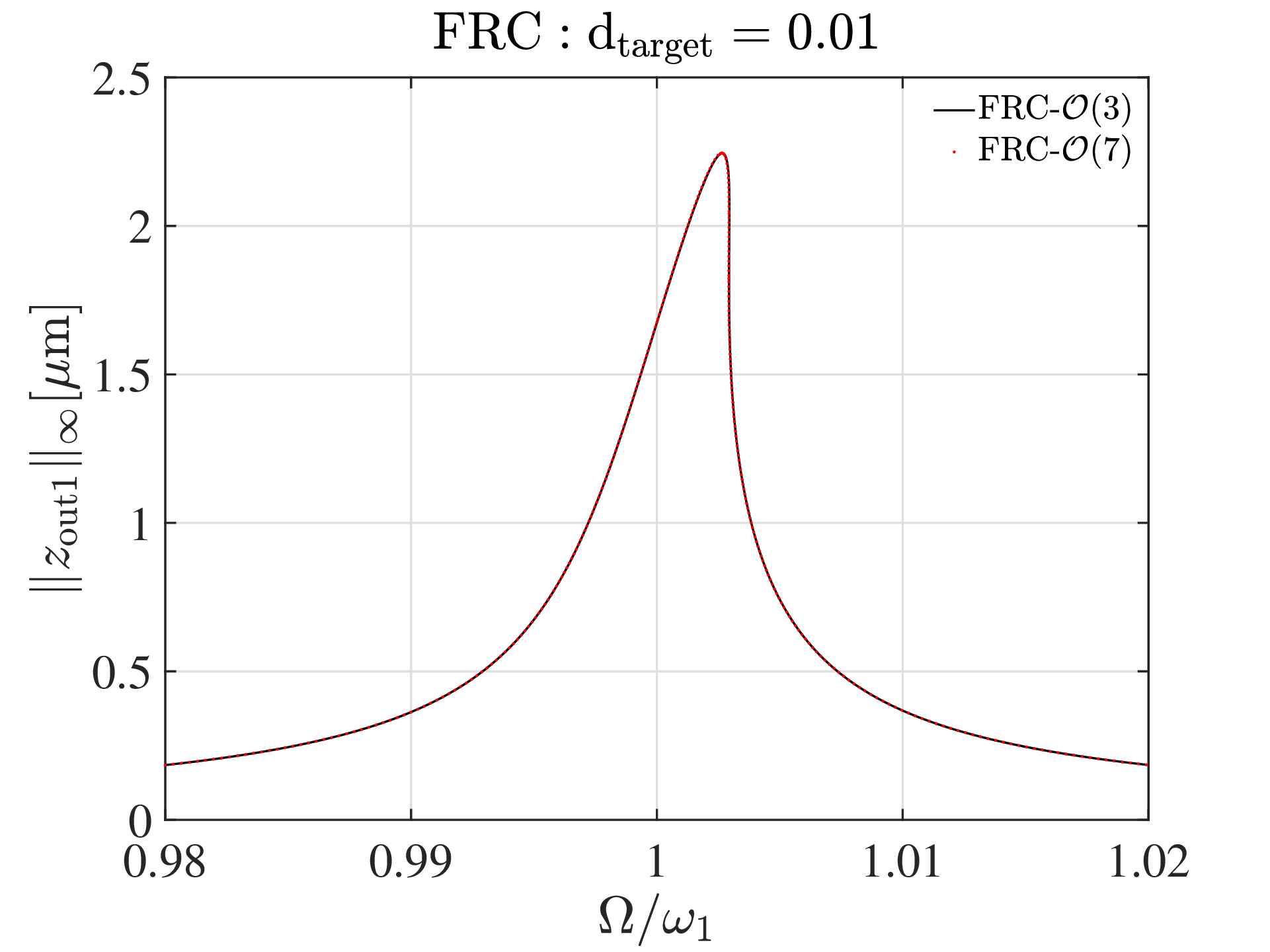} 
\includegraphics[width=0.3\textwidth]{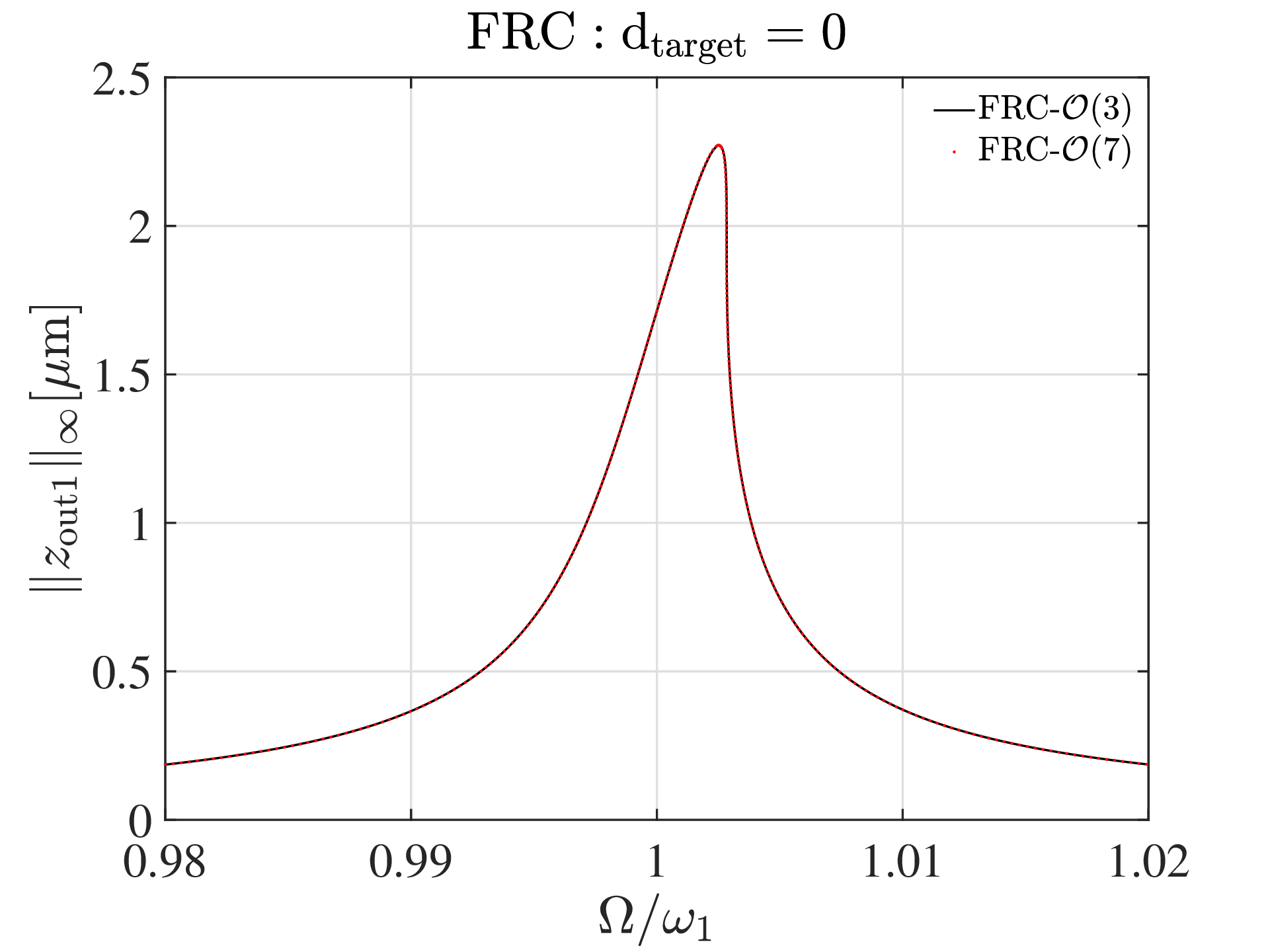} 
\caption{Comparison of FRCs computed using third-order and seventh-order SSM-reduced models for the optimized structures with \textcolor{blue}{$d_{\mathrm{target}}$ = 0.1 (left panel), 0.01 (middle panel), and 0 (right panel)}.}
\label{fig:FRC_O3_SN_O7}
\end{figure}

To validate the convergence of the SSM-based reduction in the SN bifurcation control example presented in Sec.~\ref{ssec: control_SN}, we compute the FRCs of the optimized structures using both third-order and seventh-order reduced models. As shown in Fig.\ref{fig:FRC_O3_SN_O7}, consistent agreement is observed across all three target values of \textcolor{blue}{$d_{\mathrm{target}}$ (0.1, 0.01, and 0)}, indicating that the third-order SSM approximation provides sufficient accuracy for capturing the key nonlinear response features in this study.

\subsection{\textcolor{blue}{Test of mode collisions and master mode consistency}} 
\label{appE3: test_mode_collisions_ex3}

\begin{figure}[!ht]
\centering
\includegraphics[width=0.45\textwidth]{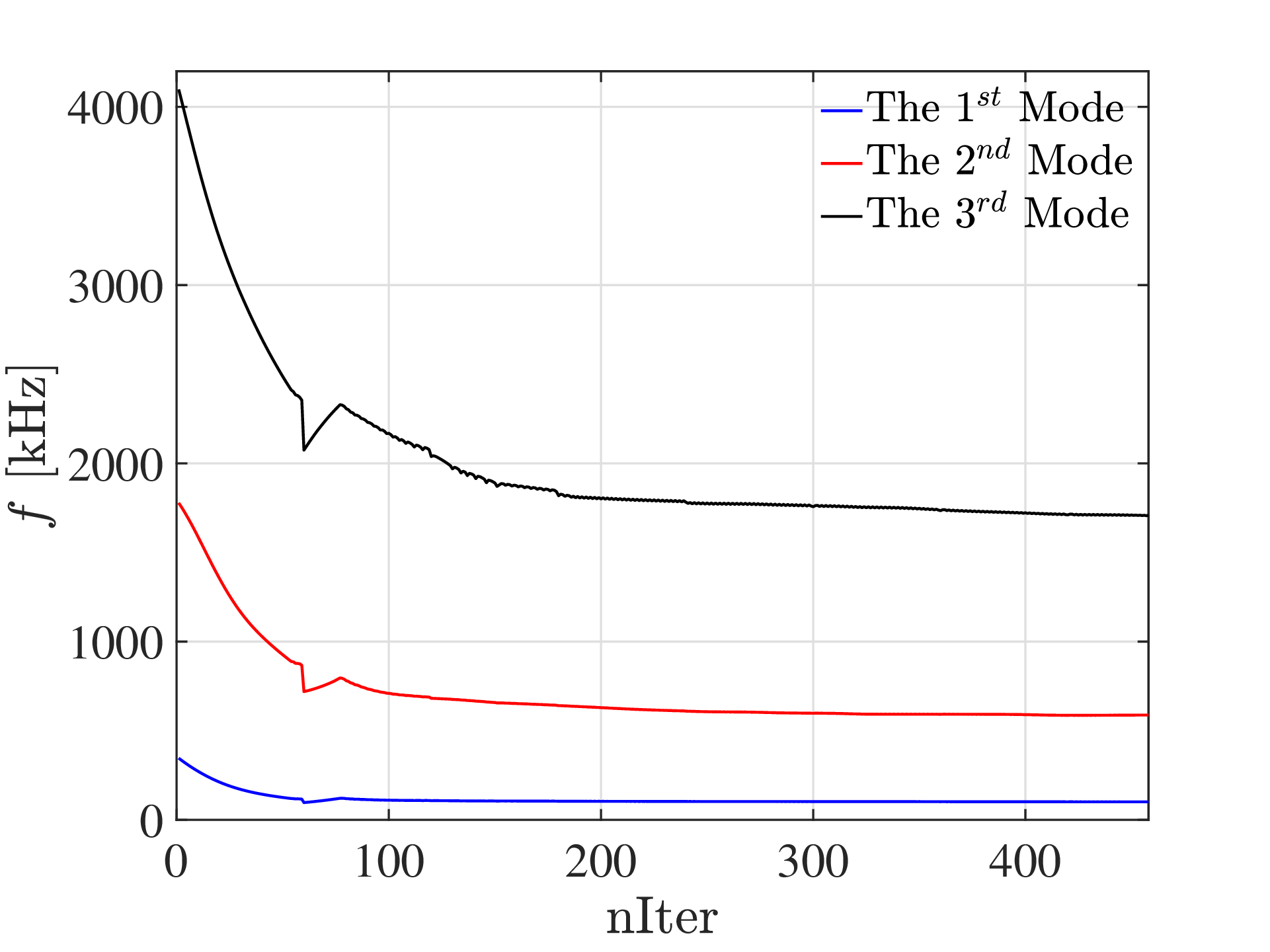} 
\caption{\textcolor{blue}{The evolution of the eigenfrequencies of the first three modes for the optimization with $d_{\mathrm{target}} = 0$ in Fig.~\ref{fig:optSN-layoutFRC}.}}
\label{fig:om2nIter_Case3}
\end{figure}

\textcolor{blue}{To detect possible modal collisions and internal resonances, we present the evolution of the first three eigenfrequencies during the optimization with $d_{\mathrm{target}} = 0$ in Fig.~\ref{fig:optSN-layoutFRC}. As shown in Fig.~\ref{fig:om2nIter_Case3}, the eigenfrequencies of the three modes remain well separated and do not overlap. Moreover, since $ \omega_X>3\omega_Y$ throughout the iteration process, no internal resonance involving the master mode and the second mode occurs.}

\begin{figure}[!ht]
\centering
\includegraphics[width=0.4\textwidth]{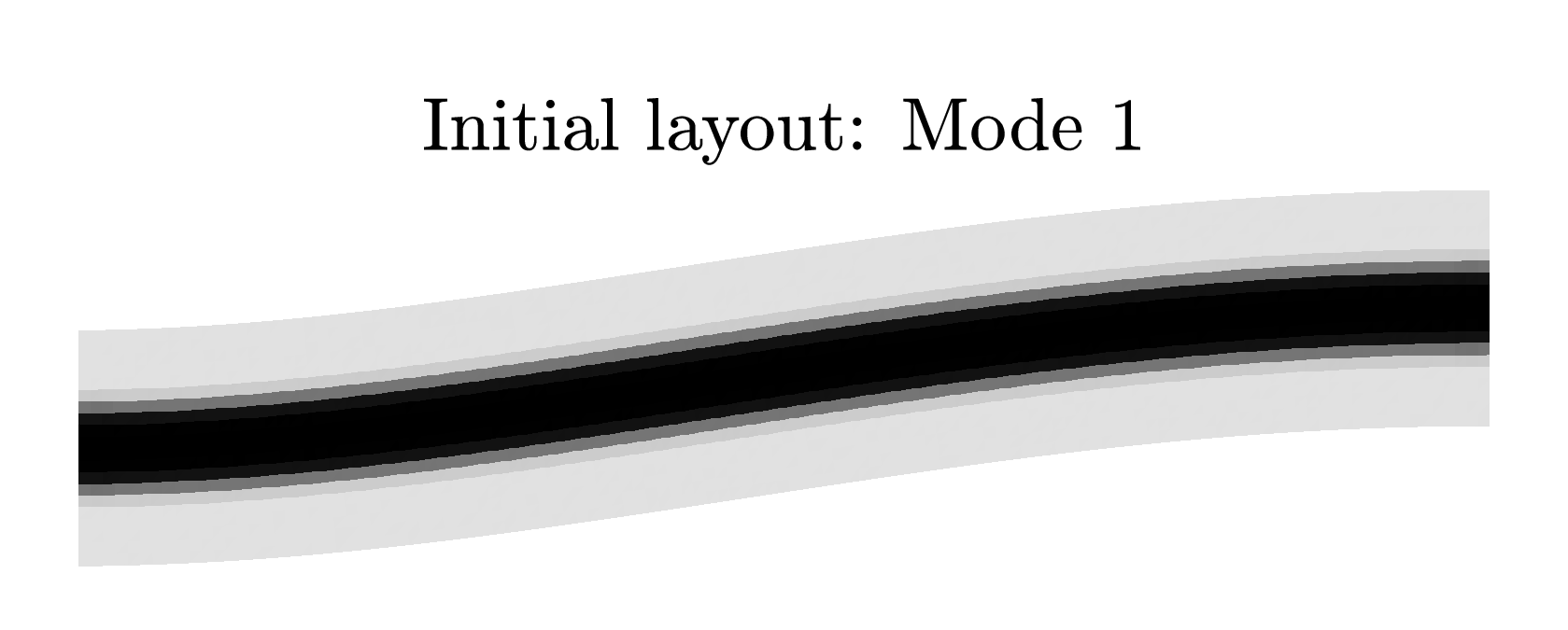}
\includegraphics[width=0.4\textwidth]{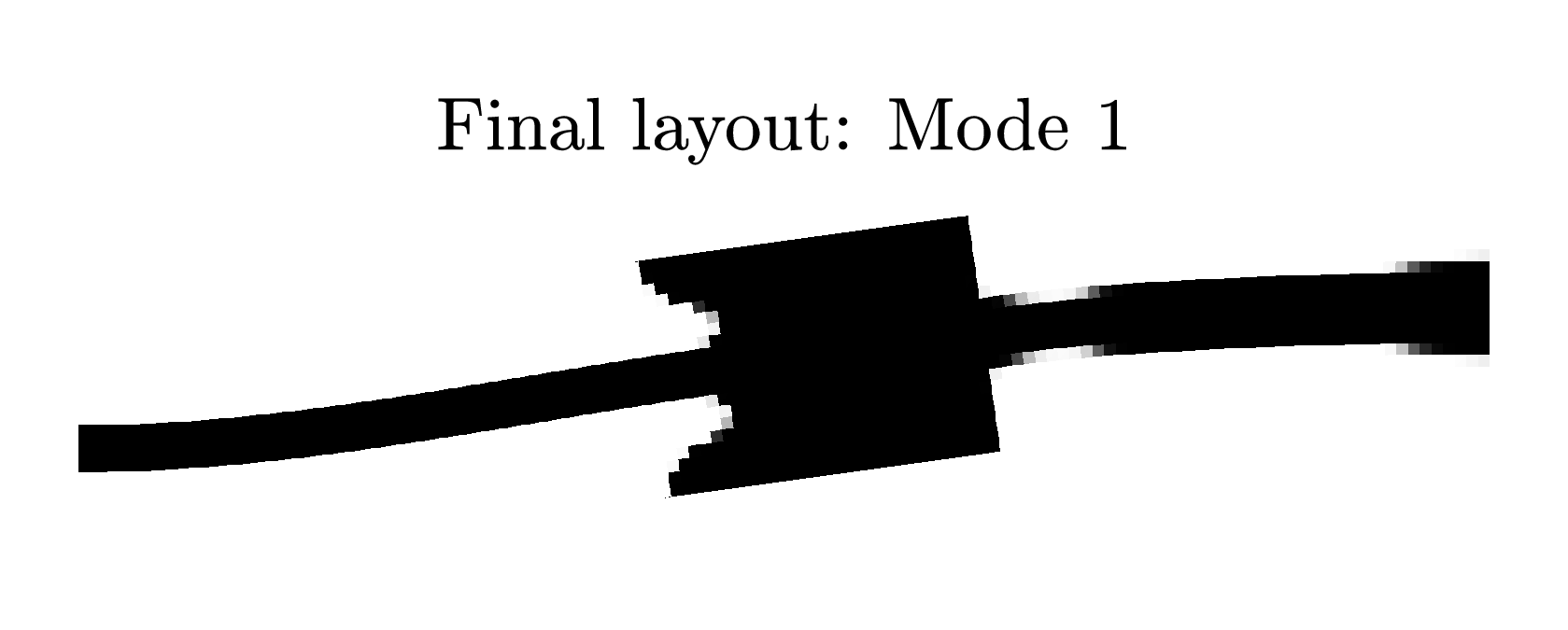} 
\caption{\textcolor{blue}{The linear modal shape of the first modes of the initial and optimized structures for the optimization with $d_{\mathrm{target}} = 0$ in Fig.~\ref{fig:optSN-layoutFRC}.}}
\label{fig:modes_Case3}
\end{figure}

\textcolor{blue}{To check the consistency of master modes during iterations, we compare the modal shape of the first modes of the initial and optimized structures for the optimization with $d_{\mathrm{target}} = 0$ in Fig.~\ref{fig:optSN-layoutFRC}. As shown in Fig.~\ref{fig:modes_Case3}, the initial and optimized structures exhibit similar modal shapes, indicating that the target master mode is tracked successfully..}

\bibliographystyle{ieeetr}       
\bibliography{ref}   


\end{document}